\documentclass[
showkeys,
 amsmath,amssymb,
 aps,
Prb,
]{revtex4-2}

\usepackage{graphicx}
\usepackage{dcolumn}
\usepackage{bm}
\usepackage{color}
\usepackage{subfigure}
\usepackage{csquotes}

\begin{document}

\preprint{APS/123-QED}

\title{Creating compact localized modes for robust sound transport via singular flat band engineering}

\author{Emanuele Riva}
\author{Federico Bellinzoni}
\author{Francesco Braghin}
 \affiliation{Department of Mechanical Engineering, Politecnico di Milano, 20156 (Italy)}

\date{\today}

\begin{abstract}
We experimentally demonstrate the emergence of flat-band-induced compact-localized modes in acoustic Kagome lattices. Compact localized states populate singular dispersion bands characterized by band crossing, where a quadratic and a flat-band dispersion coalesce into a singularity. These conditions enable intriguing wave phenomena when the Hilbert–Schmidt quantum distance, measuring the strength of the singularity, is nonzero. We report numerically and experimentally the formation of compact localized states (CLS), extremely localized in space and protected by dispersion flatness. In our system of coupled acoustic waveguides, sound waves are confined to propagate within tightly localized sites positioned both at the boundaries and within the interior of the lattice, achieving broadband and sustained confinement over time.
This framework opens new avenues for the manipulation and transport of information through sound waves, with potential application in mechanics and acoustics, including communication, signal processing, and sound isolation. This work also expands the exploration of flat-band lattice physics within the realm of acoustics.

\end{abstract}

\keywords{Singular flat-band, Kagome lattice, Acoustic metamaterial, Topological acoustics. }

\maketitle


\section*{Introduction}
Kagome lattices, originally investigated in condensed matter physics and photonics, have attracted growing interest due to their unique dispersion characteristics and nontrivial topological signature, propelling the quest for new and unprecedented dynamic behaviors. These properties, coupled with unique dispersion features, showcase how intricate geometrical configurations enable precise control over wave propagation, enhancing transport capabilities, back-scattering immunity, and robustness against defects \cite{mann2021broadband,kumar2024slow,ni2023topological}. A notable example in the realm of acoustics consists of precisely engineered Kagome lattices drawing on analogs inspired by the quantum Hall \cite{chen2019mechanical}, quantum spin Hall \cite{chen2018elastic}, and quantum valley Hall effects \cite{riva2018tunable,azizi2024omnidirectional,ni2017topological,ma2019topological}, where complex geometries featuring spatial or temporal symmetry breaks are employed to nucleate edge states and interface modes.

Among the most intriguing properties of kagome lattices is the emergence of flat dispersion bands, or flat-bands, which exhibit a vanishing group velocity across all momenta, completely suppressing the energy transport. These flat-band systems have drawn significant interest across various fields, from superconductivity in twisted bilayer graphene \cite{cao2018unconventional,cao2018correlated,balents2020superconductivity} to enhanced wave-matter interactions in photonics \cite{gersen2005real,baba2008slow,krauss2007slow} and their counterparts in phononic crystals \cite{gardezi2020acoustic,deng2020magic,yves2022moire,rosendo2020flat}. Within Kagome-decorated lattice waveguides, flat-band dispersion is driven by destructive interference among neighboring lattice sites, producing compact localized states (CLS) \cite{bergman2008band,rhim2021singular,maimaiti2017compact}. This interference confines the wave energy to a small region, with amplitudes dropping to zero outside this localized zone. Unlike other forms of wave localization, such as topological boundary modes \cite{xia2021experimental,riva2020edge,pantaleoni2024topological,rosa2023material,ni2019observation,chen2021creating,cheng2023revealing}, CLSs maintain strictly finite spatial confinement, a feature observable across both singular and non-singular flat-band lattices.
In non-singular configurations, the flat band is isolated from other dispersive bands, thereby preserving distinct Bloch modes. In contrast, singular flat bands exhibit band crossings where two eigenstates coalesce at specific points in momentum space \cite{rhim2019classification}. Despite the proximity of their Bloch functions, these states maintain a nonzero Hilbert-Schmidt quantum distance near the crossing point \cite{dodonov2000hilbert,rhim2020quantum}. Moreover, under singular conditions, CLSs alone are insufficient to form a complete set of Bloch functions across momentum space, leading to the appearance of additional states known as non-contractible loop states (NLS). NLSs are localized along one spatial direction but extend infinitely along the other and, according to the bulk-edge correspondence, NLSs emerge as boundary modes highly localized at the edges of the lattice, often forming closed loop-like profiles. 

Another similar yet fundamentally different compact localization mechanism relies on bound states in the continuum (BICs) \cite{hsu2016bound}. Notably, both CLSs and BICs result from interference effects, leading to confinement within specific spatial regions without radiating energy toward other modes. However, CLSs manifest in flat-band systems, where destructive interference between neighboring sites generates strictly localized modes that confine energy to a finite region within the lattice. In contrast, BICs are localized at discrete points within the spectrum of extended states and require precise symmetry or parameter tuning to avoid coupling with radiative modes, allowing for robust energy confinement within the continuum. While CLS are observed in various geometrically frustrated lattice structures, such as Kagome lattices, BICs may occur in photonic \cite{marinica2008bound,zhen2014topological}, acoustic \cite{huang2024acoustic,marti2024observation}, and mechanical systems \cite{rahman2024elastic,rahman2022bound} that allow for symmetry-protected interference. This distinction highlights the potential of CLS for engineering flat-band lattices that achieve energy localization independently of system boundary conditions, as opposed to BICs, which typically depend on external symmetry constraints. In other words, despite being trivial from the standpoint of topological band theory \cite{graf2013bulk,hasan2010colloquium}, flat-band systems introduce new fascinating wave phenomena driven by the unique geometry and resulting interference effects \cite{rhim2019classification}, aspects that remain relatively unexplored in acoustics.

Motivated by these intriguing properties, we numerically and experimentally investigate the dynamics of a carefully engineered acoustic Kagome lattice, which is intentionally designed to induce a singular flat band. We demonstrate that near the band-crossing point, the Hilbert-Schmidt quantum distance reaches a unitary value, providing a clear signature of the singularity. Through a combination of time-domain simulations and experiments, we explore the behavior of both compact localized states and robust boundary modes, revealing their role in the phonon transport capabilities of the lattice. In this framework, this paper presents the first experimental realization of flat-band-induced boundary modes in an acoustic system, where the wave energy is confined to propagate throughout intricate and localized profiles. These findings demonstrate how the unique properties of flat-band lattices enable practical applications by harnessing compact localization wave phenomena, with potential impacts on acoustic imaging, sensing, communication, and sound transport.

\section*{Results}
\begin{figure}[!t]
    \centering
    \subfigure[]{\includegraphics[width=0.74\textwidth]{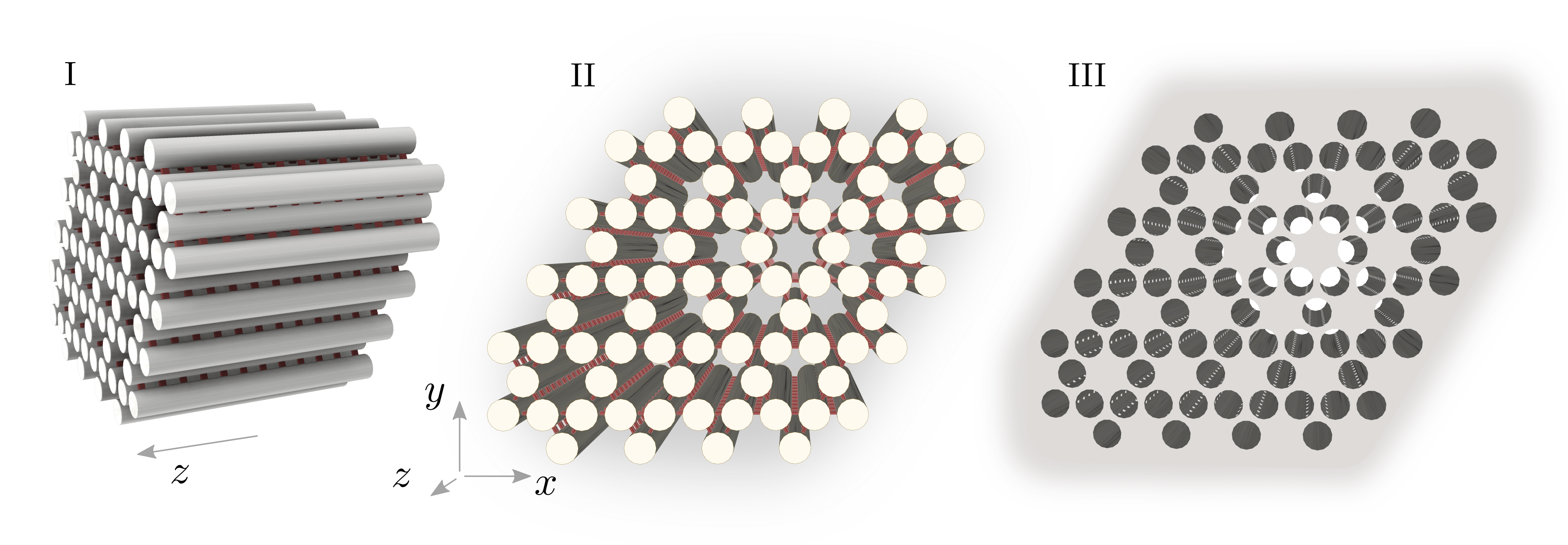}}\\
    \vspace{0.2cm}
    \subfigure[]{\includegraphics[width=0.62\textwidth]{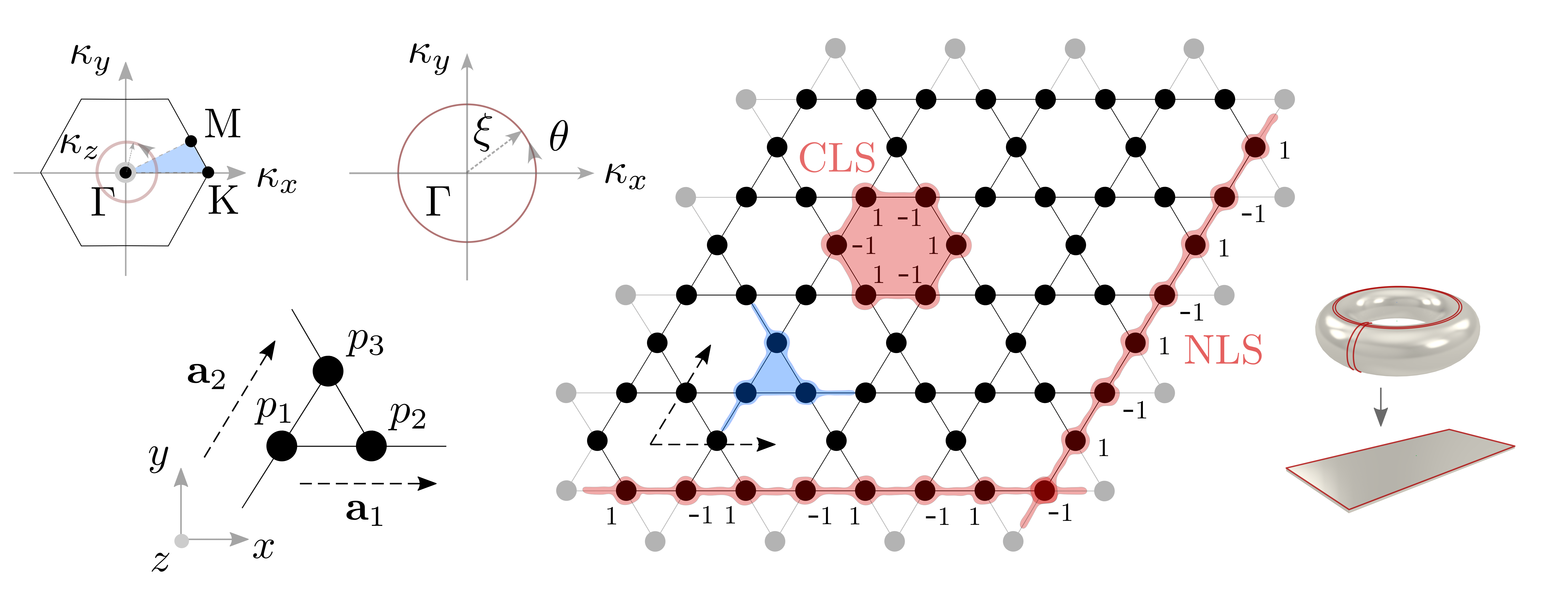}}\hspace{0.15cm}
    \subfigure[]{\includegraphics[width=0.33\textwidth]{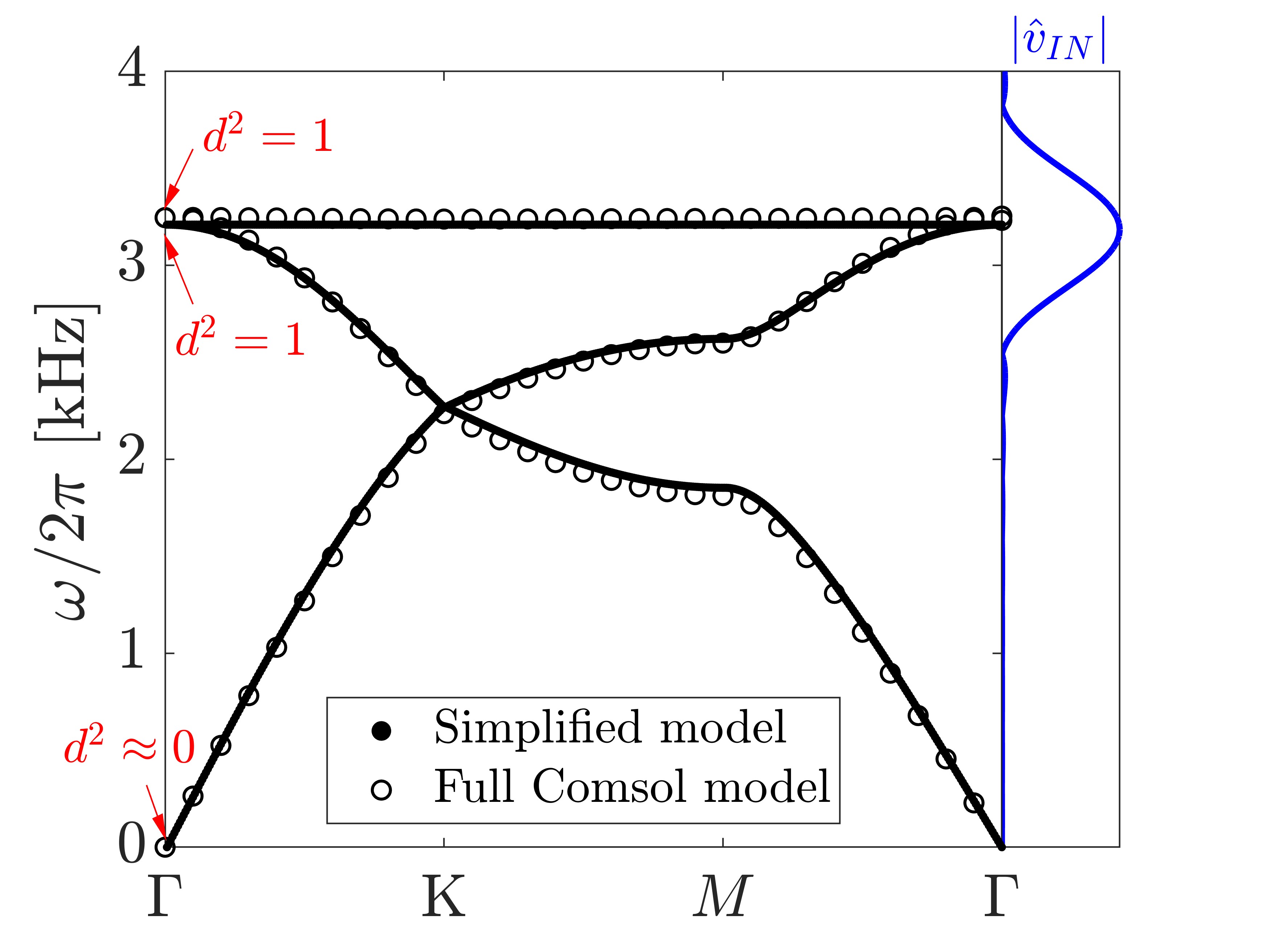}}\\
    \vspace{0.4cm}
    \hspace{-0.6cm}
    \subfigure[]{\includegraphics[width=0.26\textwidth]{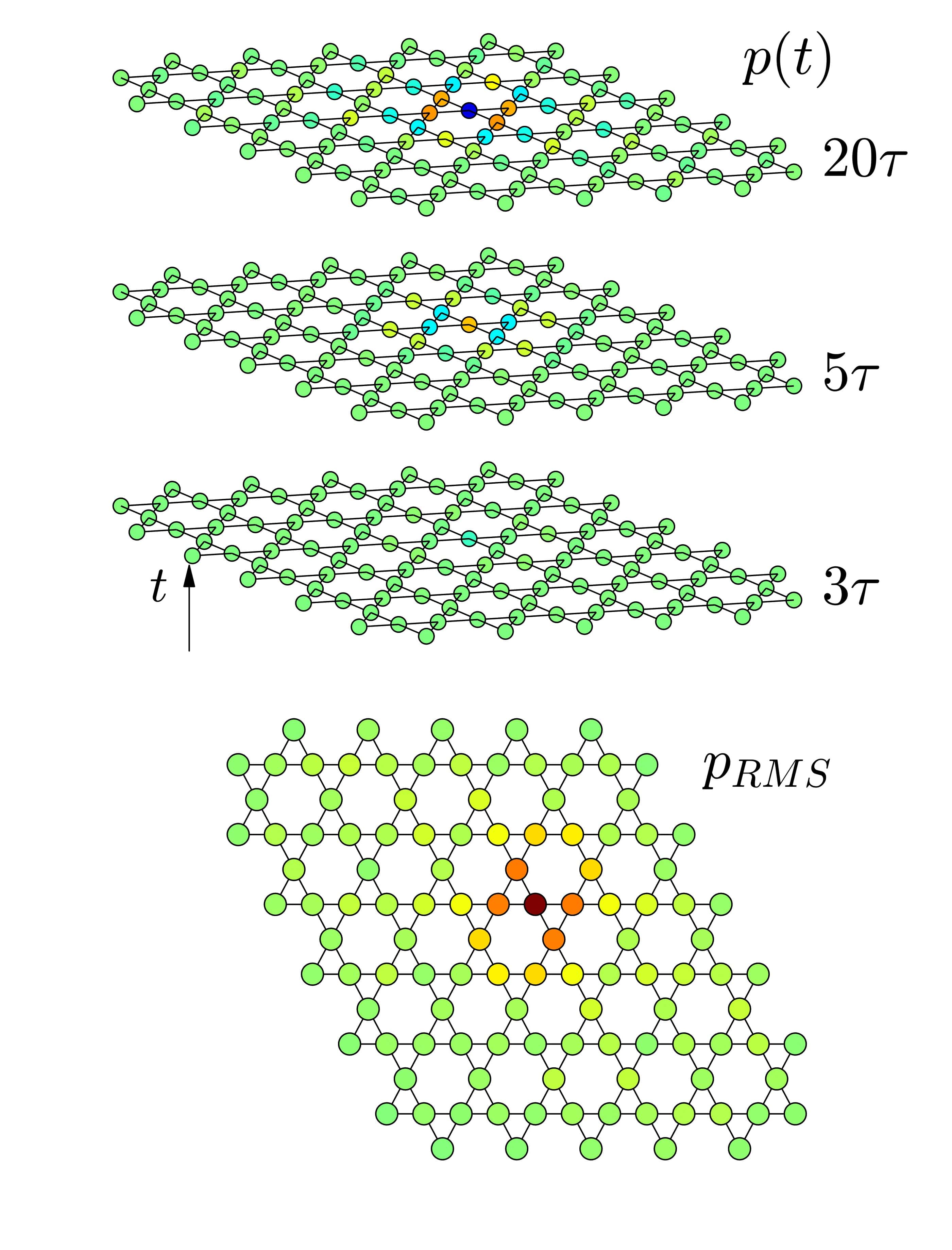}}\hspace{-0.22cm}
    \subfigure[]{\includegraphics[width=0.26\textwidth]{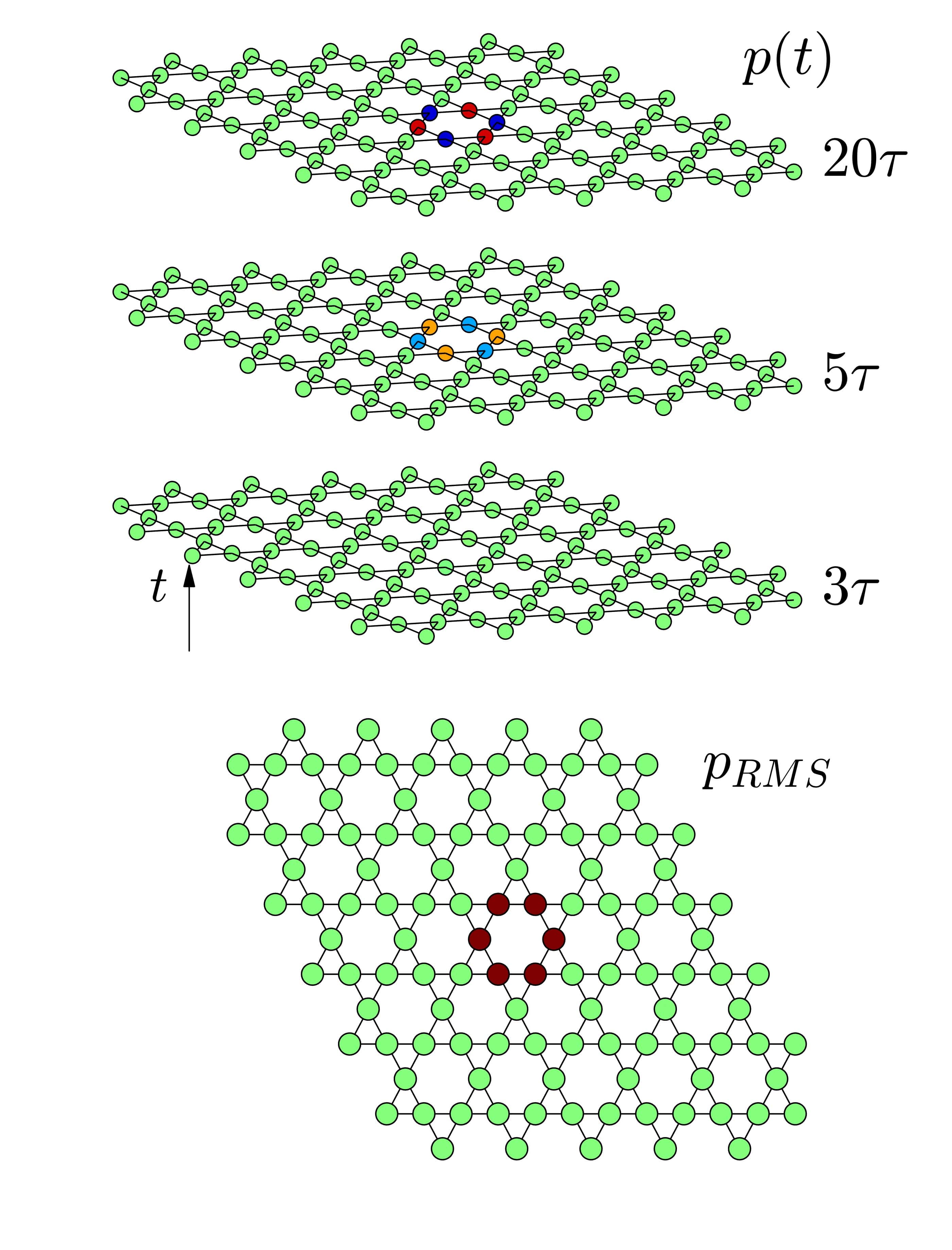}}\hspace{-0.22cm}
    \subfigure[]{\includegraphics[width=0.26\textwidth]{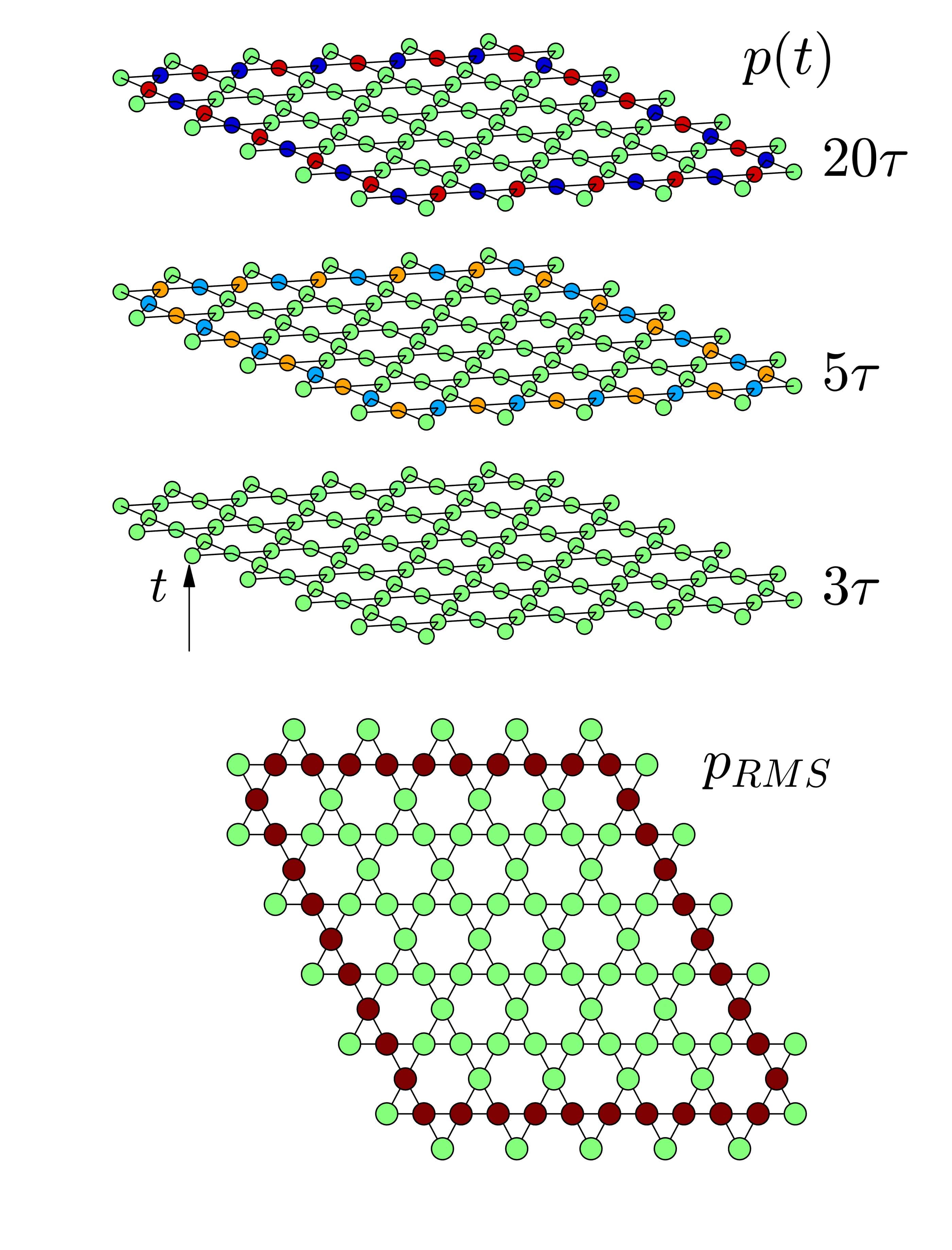}}\hspace{-0.22cm}
    \subfigure[]{\includegraphics[width=0.26\textwidth]{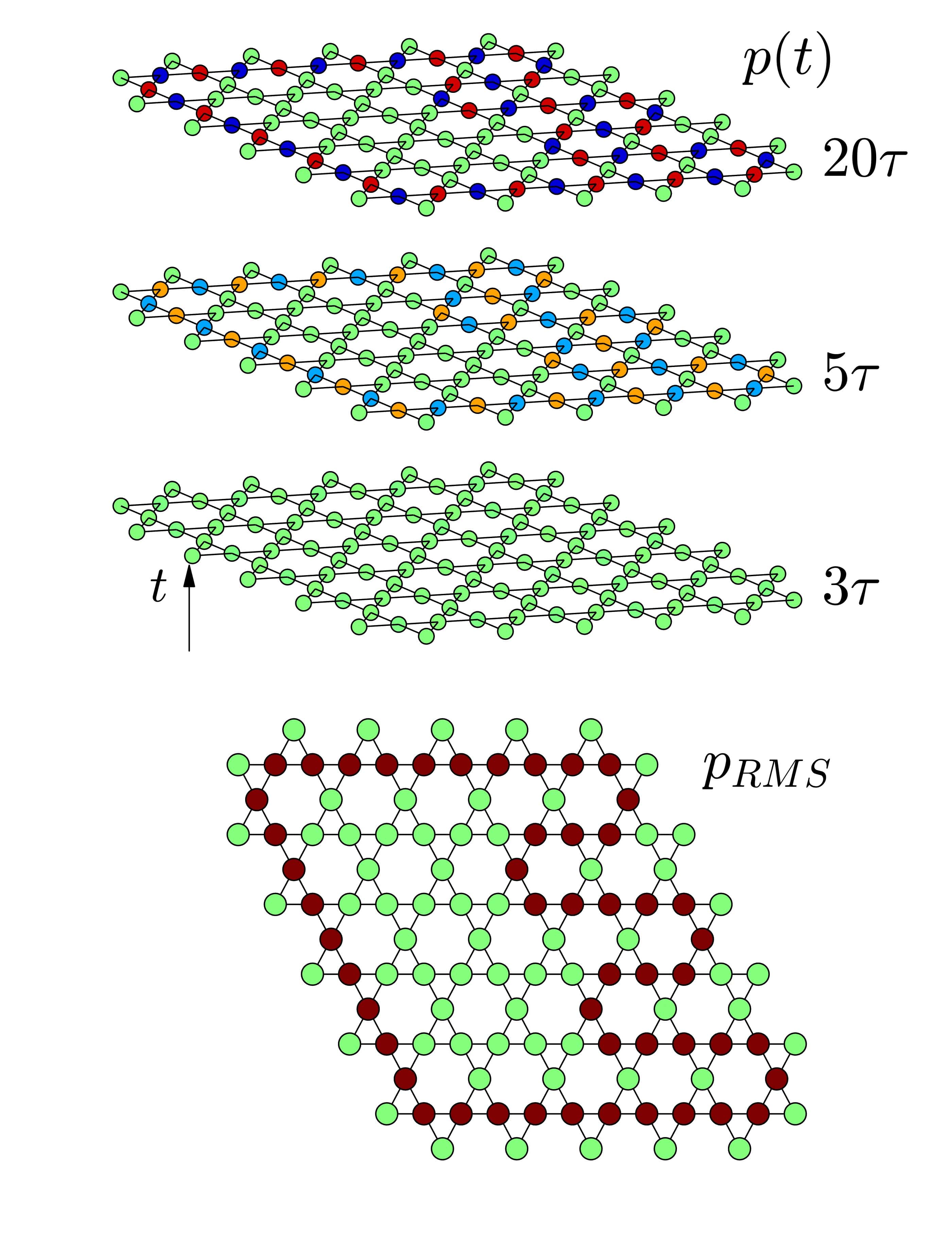}}
    \centering
    \caption{(a) Schematic of the Kagome lattice. (I-II) Graphical representation of the air-filled domain, where the pipes and the coupling elements are represented in white and red, respectively. (III) The external, rigid frame encloses the air-filled domain. (b) Lumped parameter model of the lattice, along with the corresponding Wigner-Seitz cells in direct and reciprocal spaces and the schematic of the polar coordinates used to encircle the $\Gamma$ point. The emergence of NLS and Boundary modes are schematically illustrated in the torus-plate transformation and further highlighted in red on the lattice. (c) Bloch diagram evaluated along the boundaries of the IBZ. The solid line is computed through the approximate formulation, while the white dots are evaluated with a COMSOL high-fidelity model. The integer numbers, denoted with $d^2$, represent the Hilbert-Schmidt quantum distance evaluated for the three corresponding dispersion branches (linear, quadratic, and flat band). The blue curve, reported alongside the dispersion, represents the spectral content injected in the lattice during time simulation. (d-g) Temporal evolution and RMS value of the pressure field for a 2D lattice forced through (d) point excitation, (e) 6 site CLS excitation, (f) boundary excitation, (g) "E" shape excitation. }
    \label{fig:01}
\end{figure}

\textbf{Theory.} We start the discussion by considering the acoustic system in Fig. \ref{fig:01}(a), which is made of a Kagome lattice distribution of air-filled waveguides. The air domain is illustrated in Fig. \ref{fig:01}(a)-I-II, while the external, rigid frame is shown in Fig. \ref{fig:01}(a)-III. The waveguides, represented with white cylinders, are connected to the nearest neighbors by way of link tubes with tailored hopping amplitude. These air-filled domains, in turn, are embedded in a sufficiently rigid external frame to avoid undesired fluid-structure interactions, hence, dramatically simplifying the discussion while retaining the dynamics of interest. 

To ease interpretation, the front view of the lattice is schematically illustrated in Fig. \ref{fig:01}(b), along with the corresponding Weigner-Seitz cell in the direct and reciprocal spaces. Due to in-plane periodicity, the pressure of neighboring unit cell elements is linked with exponential relations $p_j^{m+1,n}(x,y,z)=\hat p_j^{m,n}(z){\rm e}^{-{\rm i}\bm{\kappa}\cdot\bm{a}_1}$ and $p_j^{m,n+1}(x,y,z)=\hat p_j^{m,n}(z){\rm e}^{-{\rm i}\bm{\kappa}\cdot\bm{a}_2}$, where $\bm{\kappa}=\left(\kappa_x,\kappa_y\right)$ is the wavevector in the $x-y$ plane and $\textbf{a}_1=\left(a,0\right),\textbf{a}_2=\left(a\cos{\frac{\pi}{3}},a\sin{\frac{\pi}{3}}\right)$ are the direct lattice vectors. 
Then, complex exponential functions are used to describe wave propagation along the third direction $z$, i.e., $\hat p_j^{m,n}(z)=\hat p_j^{m,n}{\rm e}^{{\rm i}\left(\kappa_zz-\omega t\right)}$. 
This ansatz, along with the wave equation of the lattice, allows reducing the wave dynamics to the following linear eigenvalue problem (details are provided in the supplementary material \cite{SM}):
\begin{equation}
    \omega^2 I\left|\bm{p}\right>=H(\bm{\kappa},\kappa_z)\left|\bm{p}\right>
    \label{eq:evp}
\end{equation}
Eq. \ref{eq:evp} suitably describes the formation of Bloch modes populating the $x-y$ plane and propagating along $z$, whereby the eigenvector $\left|\bm{p}\right>$ accommodates the pressure amplitudes $p_j$ relative to $\omega$, being $j=1,2,3$. 

Now, the emergence of CLSs and robust boundary modes is hereafter studied in the $x-y$ plane by first assuming $\kappa_z=0$, which \textit{de facto} reduces the eigenvalue problem to that of a 2D Kagome lattice. This assumption will be later relaxed to show how the CLSs and ensuing boundary modes propagate along the direction $z$. Results are reported in Fig. \ref{fig:01}(c), where the dispersion is evaluated along the contours of the Irreducible Brillouin Zone (IBZ). To corroborate the fidelity of our simplified model, we compared the approximate formulation (black line) with the high-fidelity COMSOL model (white dots), demonstrating a very good agreement. Interestingly and consistently to prior studies on optical lattices \cite{rhim2021singular}, the dispersion exhibits a singular flat band at $\omega_{FB}=\sqrt{6}\omega_0$ where a quadratic dispersion intersects at the $\Gamma$ point. Here, $\omega_0=\sqrt{k_{eq}/m_{eq}}$ is a characteristic constant of the lattice and $k_{eq}$ and $m_{eq}$ are effective mass and stiffness parameters, which depend on the geometry of the waveguides and the acoustic links \cite{kinsler2000fundamentals,riva2023adiabatic}.  Under these settings, the eigenstates coalesce into a singularity in $\Gamma$, where the algebraic and geometric multiplicities differ (2 and 1, respectively). 
The strength of such a singularity, measurable through the Hilber-Schmidt quantum distance, is inherently connected to the dynamics that manifest in physical space and is responsible for the formation of compact localized and boundary modes. These aspects are discussed in what follows.

First and foremost, the emergence of CLSs is physically justified by the group velocity vanishing at the flat-band frequency and the geometrical characteristic of the lattice, which promote the flatness of the band. 
These features, along with the singularity at the $\Gamma$ point, reflect on eigenvectors with specially designed characteristics in $\left|\bm{p}\right>$ which, in turn, takes the following form (see the supplementary material for the derivation \cite{SM}):
\begin{equation}
    \left|\bm{p}\right>=\begin{pmatrix}
        {\rm e}^{\rm{i}\bm{\kappa}\cdot\bm{a}_2}-{\rm e}^{\rm{i}\bm{\kappa}\cdot\bm{a}_1}\\[3pt]
        1-{\rm e}^{\rm{i}\bm{\kappa}\cdot\bm{a}_2}\\[3pt]
        -1+{\rm e}^{\rm{i}\bm{\kappa}\cdot\bm{a}_1}
    \end{pmatrix}
\end{equation}
Note that the Bloch eigenstate $\left|\bm{p}\right>$ vanishes at the $\Gamma$ point. As such, normalization is performed to illustrate that the amplitude coefficients of $\left|\bm{p}\right>$ are discontinuous at $\Gamma$:
\begin{equation}
    \left|\bm{p_\kappa}\right>=\frac{\left|\bm{p}\right>}{\alpha_{\bm{\kappa}}}=\frac{1}{\sqrt{2\left(3-\cos{\left(\bm{\kappa}\cdot\bm{a}_1-\bm{\kappa}\cdot\bm{a}_2\right)-\cos{\left(\bm{\kappa}\cdot\bm{a}_2\right)}-\cos{\left(\bm{\kappa}\cdot\bm{a}_1\right)}}\right)}}\left|\bm{p}\right>
\end{equation}
with $\alpha_{\bm{\kappa}}^2=\langle\bm{p}|\bm{p}\rangle$. Notably, the components of $\left|\bm{p_\kappa}\right>$ are in the form $0/0$ near $\Gamma$, which implies singularity of the crossing point and that $\left|\bm{ p}_\kappa\right>$ depends on the path approaching $\Gamma$. For example:
\begin{equation}
\lim_{\bm{\kappa}\cdot\bm{a}_2\rightarrow0+}\left|\bm{p_\kappa}\right>=\begin{pmatrix}
    j/\sqrt{2}\\[4pt]
    -j/\sqrt{2}\\[4pt]
    0
\end{pmatrix}\;\;if\;\;\bm{\kappa}\cdot\bm{a}_1=0\hspace{1cm}\lim_{\bm{\kappa}\cdot\bm{a}_1\rightarrow0+}\left|\bm{p_\kappa}\right>=\begin{pmatrix}
    -j/\sqrt{2}\\[4pt]
    0\\[4pt]
    j/\sqrt{2}
\end{pmatrix}\;\;if\;\;\bm{\kappa}\cdot\bm{a}_2=0
\end{equation}
that is, orthogonal directions in momentum space reveal different eigenvectors close to the $\Gamma$ point. Such a singularity manifests in physical space in the form of compact localized states (CLS), non-contractible loop states (NLS), and robust boundary modes when a finite lattice is considered \cite{rhim2021singular}. The emergence of these eigenmodes is hereafter discussed starting from the existence of CLS in a system with periodic boundary conditions. 

Due to the flatness of the band, an infinite number of Bloch modes are degenerate at the frequency $\omega_{FB}$. This degeneracy enables the construction of compact localized states (CLS) by taking specific linear combinations of Bloch wave functions populating the flat-band:
\begin{equation}    \left|\chi_{\bm{R}}\right>=\sum_{\bm{\kappa}\in BZ}\displaystyle\alpha_{\bm \kappa}{\rm \displaystyle e}^{-{\rm i}\bm{\kappa}\cdot\bm{R}}\left| \bm{\psi}\left(\bm{\kappa}\right)\right>
\end{equation}
where the Bloch eigenstate is $\left| \bm{\psi}\left(\bm{\kappa}\right)\right>=\sum_{\bm{R}'}\sum_{j=1}^3{\rm \displaystyle e}^{{\rm i}\bm{\kappa}\cdot\bm{R}'}p_j\left(\bm{\kappa}\right)\left|\bm{R}',j\right>$. Here, $\bm{R}$ and $\bm{R}'$ represent lattice positions, and $p_j\left(\bm{\kappa}\right)$ denotes the amplitude on the $j^{th}$ site within each unit cell. As such, in a Kagome lattice that supports flat-band characteristics, the amplitude associated with each element in the $j^{th}$ unit cell element, evaluated by projecting the wave function $\left|\chi_{\bm{R}}\right>$ into physical space $\bm{R}'$, i.e., $\left<\bm{R}',j\right|\left.\chi_{\bm{R}}\right>$, is non-zero only within a finite set of lattice vectors:
\begin{equation}
\begin{pmatrix}
\left<\bm{R}',1\right|\left.\chi_{\bm{0}}\right>\\
\left<\bm{R}',2\right|\left.\chi_{\bm{0}}\right>\\
\left<\bm{R}',3\right|\left.\chi_{\bm{0}}\right>\\
\end{pmatrix}=
\begin{pmatrix}
\delta_{R,-a_2}-\delta_{R,-a_1}\\
\delta_{R,0}-\delta_{R,-a_2}\\
-\delta_{R,0}+\delta_{R,-a_1}\\
\end{pmatrix}
\label{eq:05}
\end{equation}
which confirms the localized nature of the flat-band modes. Constructing the CLS in this manner is equivalent to probing the dynamic response of the system at the flat-band frequency, a process that will be corroborated through numerical and experimental analysis. 

In addition to CLSs, the singular band structure necessitates the formation of non-contractible loop states (NLS) as additional states, required to complete the set of basis functions in momentum space \cite{rhim2021singular}. The mechanism behind the formation of these modes is illustrated schematically in Fig. \ref{fig:01}(b) with a “doughnut-plate” analogy, where red curves on the Brillouin torus $\left(\kappa_x,\kappa_y\right)\in \mathcal{T}^2=\left[0,2\pi\right]\times\left[0,2\pi\right]$ represent NLSs in an infinitely extended lattice with periodic boundary conditions. By “opening” the torus—i.e., removing periodic boundary conditions along one direction—NLSs form and manifest as boundary modes when open boundary conditions are applied along both directions in a finite Kagome lattice. Thus, two NLSs encircle the torus in the poloidal and toroidal directions, transforming into robust boundary modes when boundary conditions shift from periodic to open.
For a finite Kagome with $N$ unit cells, there exists $N-1$ linearly independent CLSs. However, two additional NLSs are required to complete the set of $N+1$ degenerate Bloch eigenstates at the flat-band frequency.
The singularity in the band structure directly leads to the emergence of NLSs, which in turn transform into robust edge modes as required by bulk-boundary correspondence. This correspondence links the presence of NLSs to the formation of boundary-localized modes in physical space, ensuring a complete basis for the flat-band states when transitioning from periodic to open boundary conditions \cite{rhim2021singular}.

Moreover, the presence of NLSs and the robustness of the flat-band modes are closely linked to the strength of the band singularity, which can be quantified using the Hilbert-Schmidt quantum distance $d^2\left(\bar p_1,\bar p_2\right)$ between two states $\left|\bar p_1\right>$ and $\left|\bar p_2\right>$ with momenta $\bm{\kappa}_1$ and $\bm{\kappa}_2$:
\begin{equation}
    d^2\left(\bm{p_{\kappa_1}},\bm{p_{\kappa_2}}\right)= 1 - \left|\langle \bm{p_{\kappa_1}}|\bm{p_{\kappa_2}}\rangle\right|^2
\end{equation}
where $0<d^2<1$. For non-singular crossings, $d^2\rightarrow 0$ drops to zero as $\bm{\kappa}_2\rightarrow\bm{\kappa}_1$, indicating that the quantum distance between states diminishes when the momenta converge. 
In contrast, singular flat-bands exhibit a persistent nonzero quantum distance even if $\bm{\kappa}_1$ and $\bm{\kappa}_2$ approach each other. 
To probe $d^2$ in the neighborhood of $\Gamma$, polar coordinates are used, with $\bm{\kappa}\cdot \bm{a}_1=\xi\cos{\theta_1}$ and $\bm{\kappa}\cdot \bm{a}_2=\xi\sin{\theta_2}$.
Here $\xi$ is arbitrarily small, fixed radius, and $\left|\bm{p_\kappa}\right>$ is evaluated as $\theta$ varies around $\Gamma$. The maximum values of $d^2$ for the three dispersion bands are displayed in Figs. \ref{fig:01}(c), where all possible permutations of the two arbitrary angles $\theta_1$  and $\theta_2$ are considered. In correspondence of the first band, $d^2$ remains very small. In contrast, $d^2=1$ for the quadratic and flat-band dispersion, respectively, with values oscillating between $0$ and $1$, indicating the presence of singularity. Additional numerical details on this matter are reported in the supplementary material \cite{SM}. In summary, the maximum quantum distance between Bloch eigenstates near the $\Gamma$ point serves as a bulk invariant. For this lattice configuration, the distance is $1$ for the flat and quadratic bands, indicating a robust singularity, while it remains $0$ for the first trivial band, confirming the absence of singularity.
\begin{figure}[!t]
    \centering
    \hspace{-0.6cm}\subfigure[]{\includegraphics[width=0.99\textwidth]{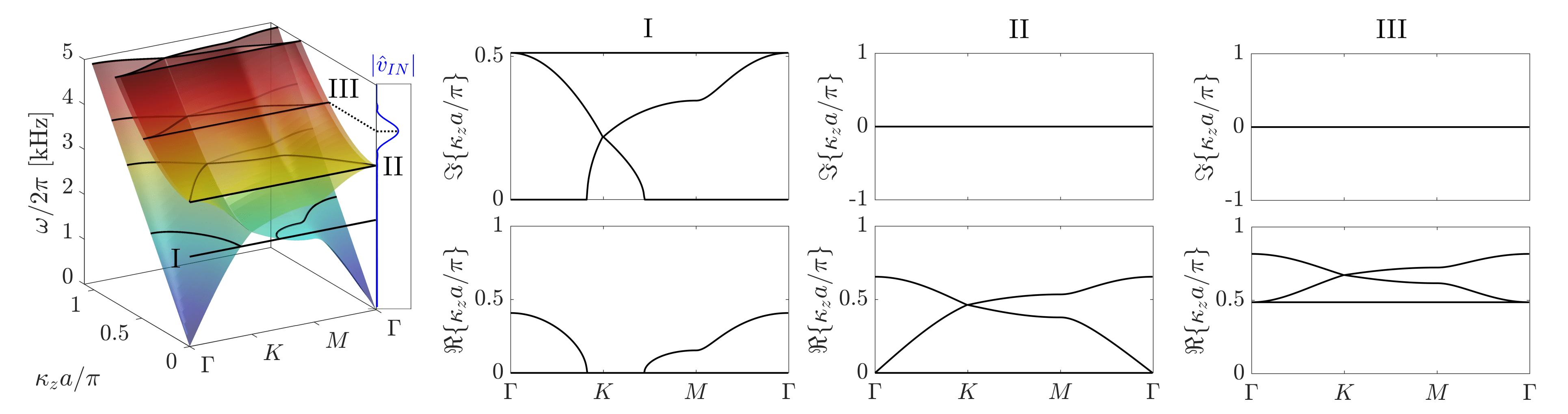}}\\
    \hspace{-0.6cm}\subfigure[]{\includegraphics[width=0.28\textwidth]{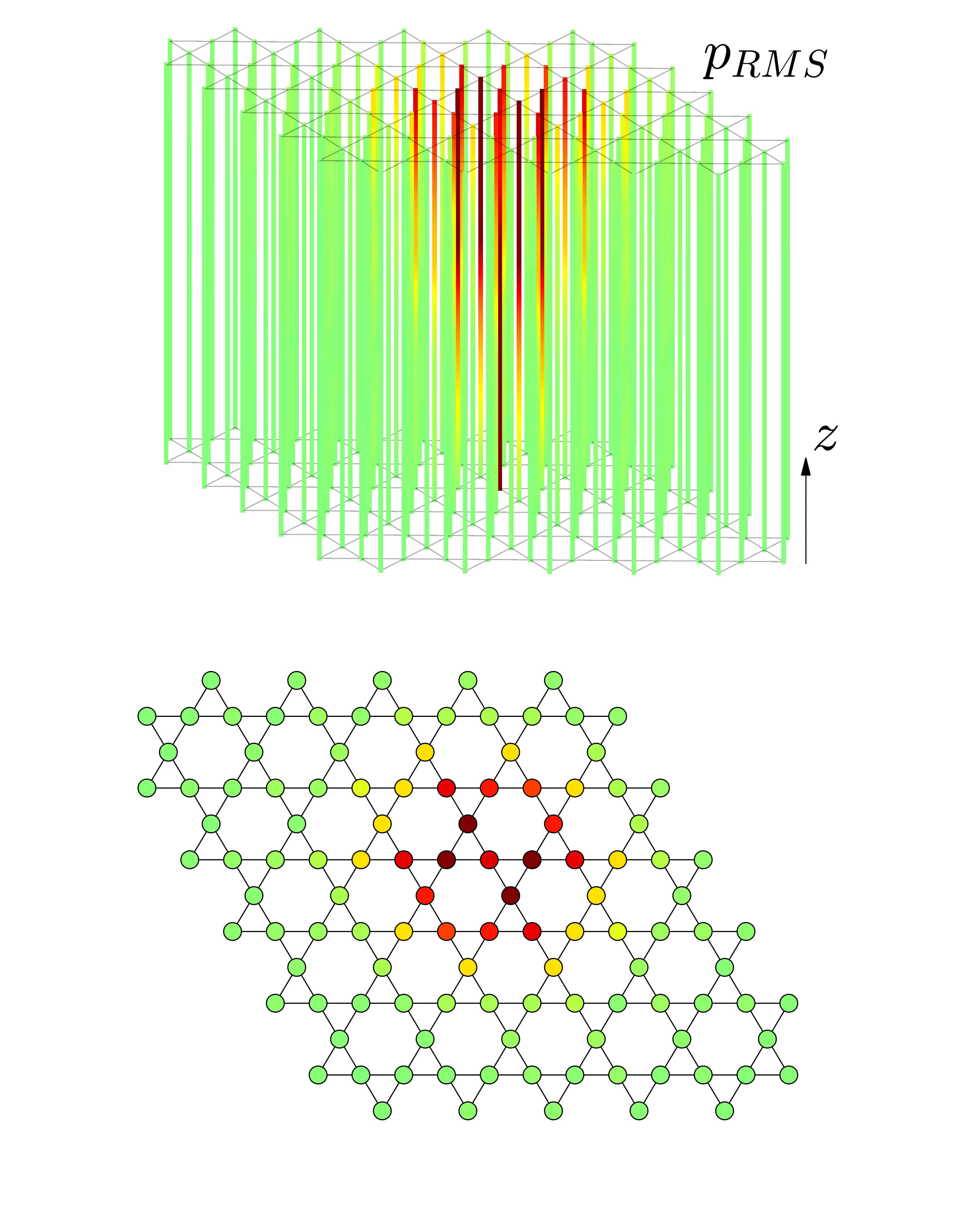}}\hspace{-0.75cm}
    \subfigure[]{\includegraphics[width=0.28\textwidth]{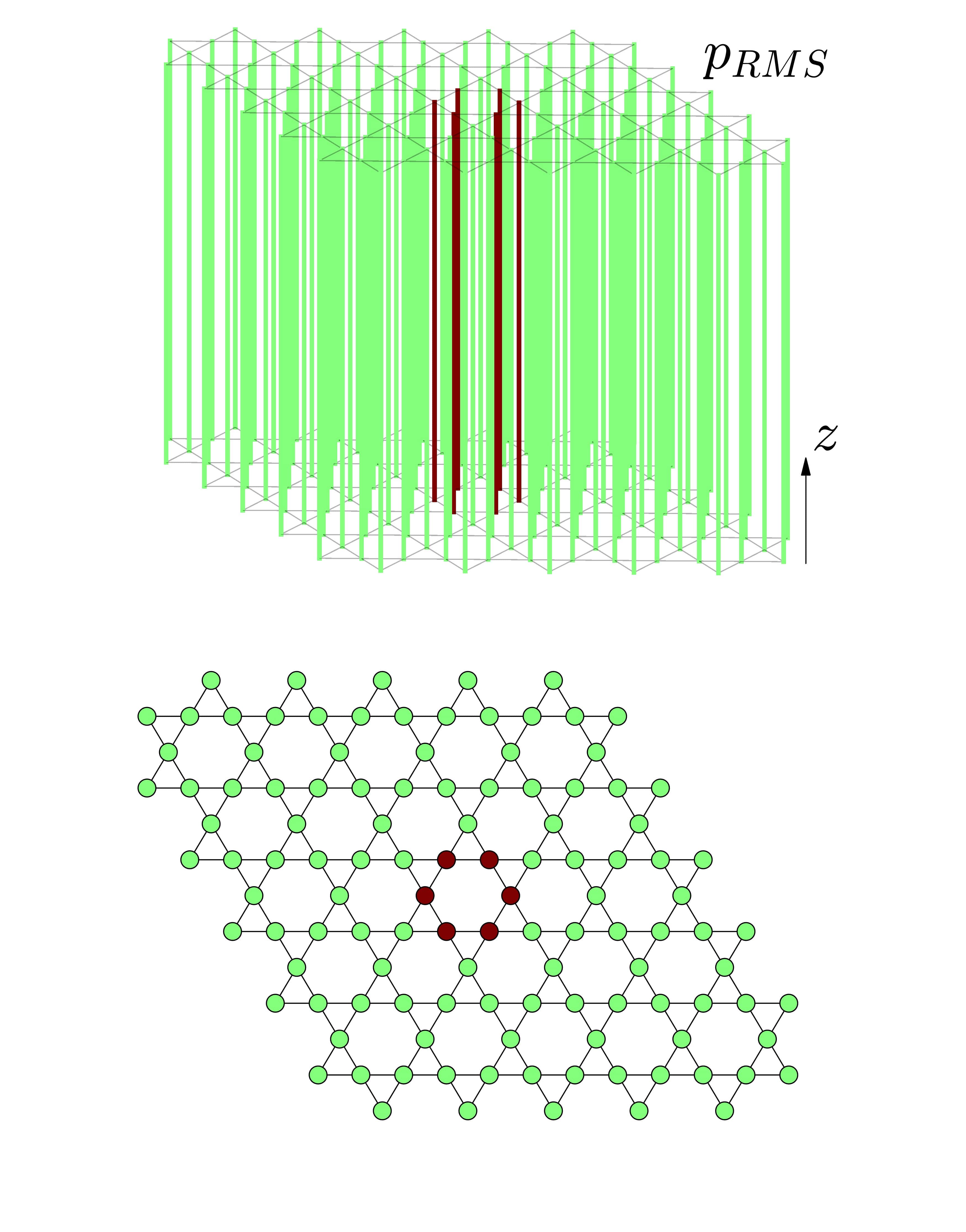}}\hspace{-0.75cm}
    \subfigure[]{\includegraphics[width=0.28\textwidth]{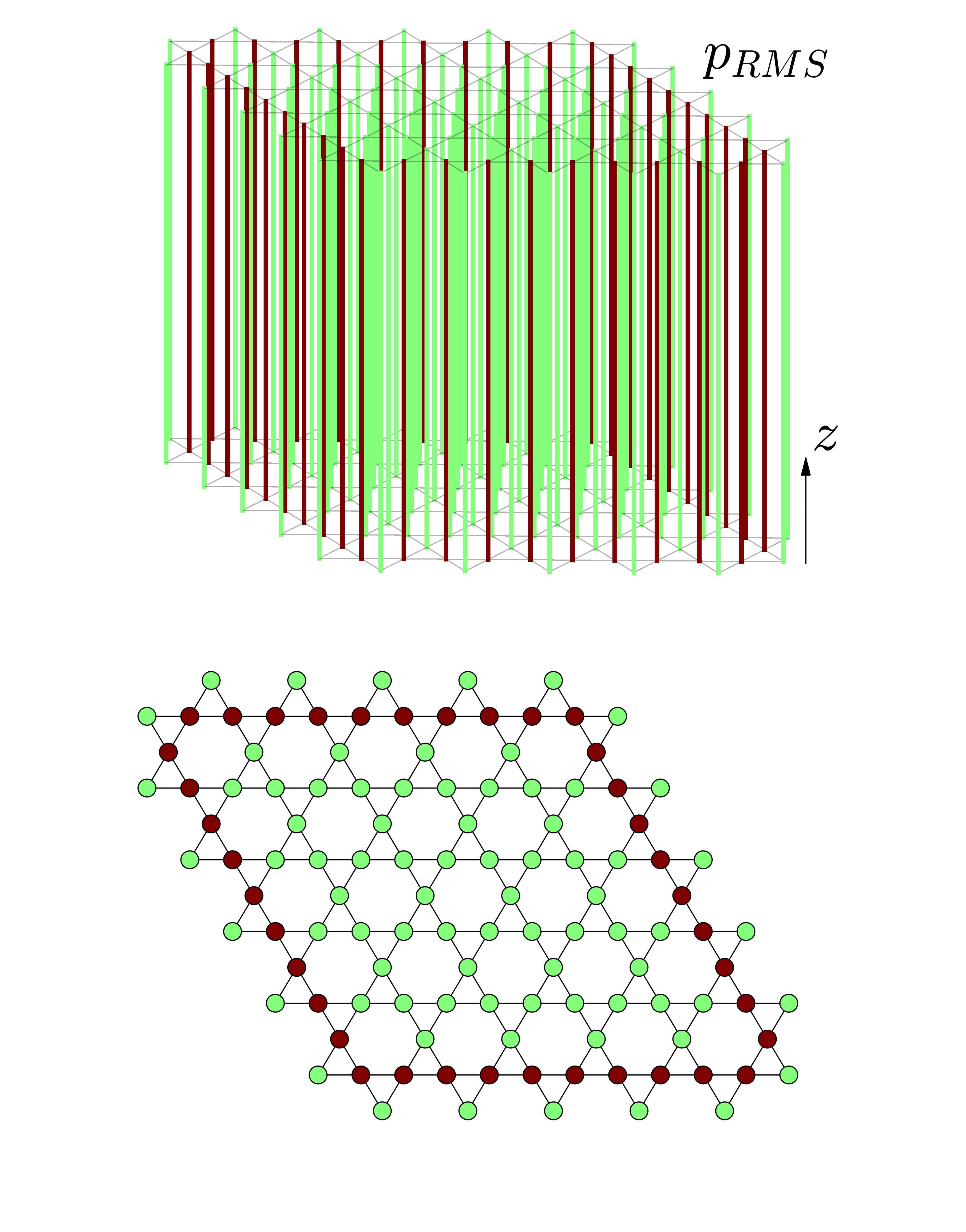}}\hspace{-0.75cm}
    \subfigure[]{\includegraphics[width=0.28\textwidth]{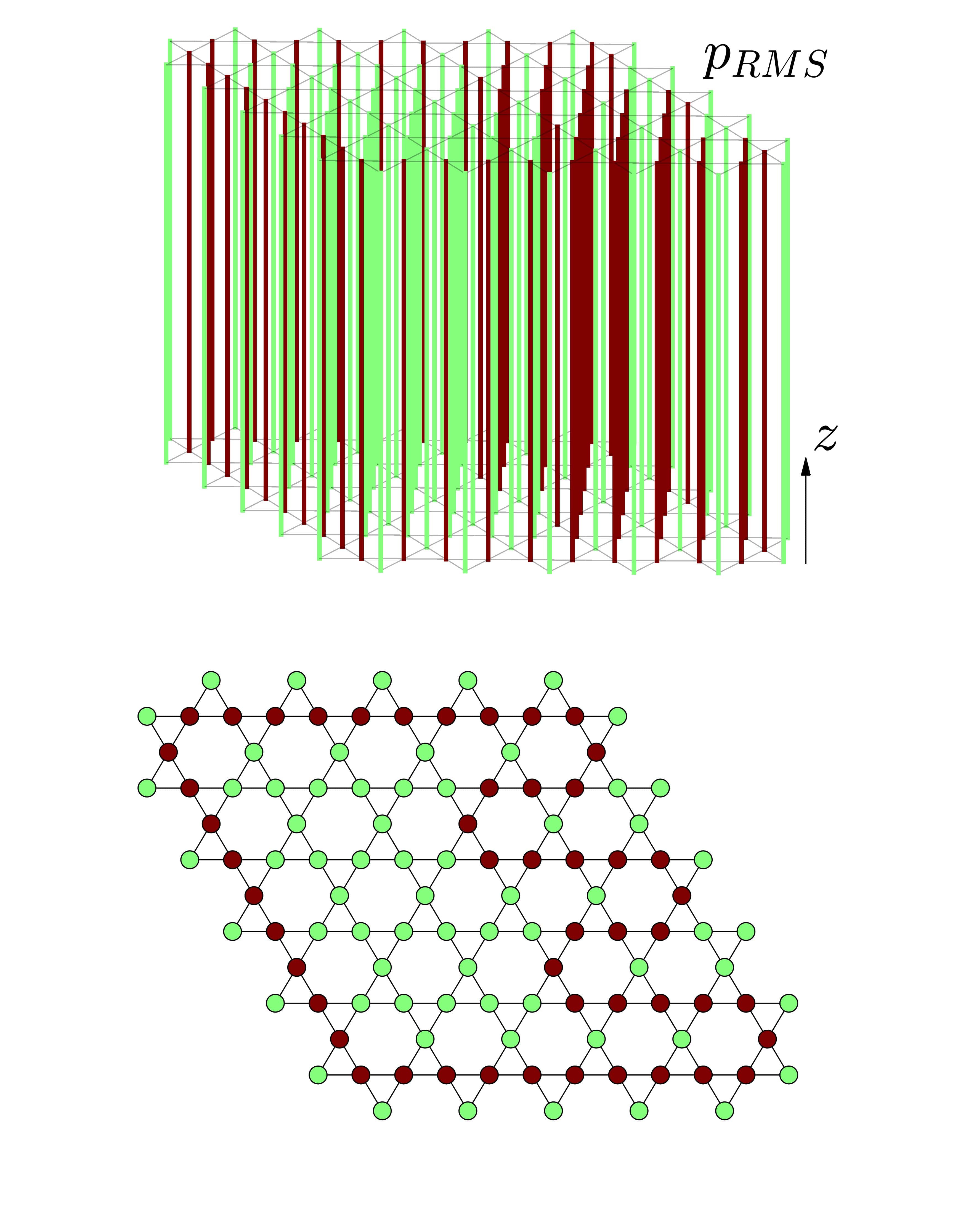}}
    \centering
    \caption{Numerical results for the 3D lattice waveguide. (a) 3D dispersion surface spanning the irreducible Brillouin zone and the lateral wavenumber $\kappa_z$. The blue curve represents the spectral content injected in the lattice during time simulation. (a)-I dispersion contours computed along the irreducible Brillouin zone (IBZ) at three selected values of input frequency equal to (I) $2$ kHz, (II) the flat-band frequency, and (III) $4$ kHz. To illustrate the propagative and evanescent nature of the wave, the imaginary and real parts are separated. (b-e) Numerical simulations showing the RMS value of the pressure field, normalized by its maximum in each plane at a given coordinate $z$, for different forcing conditions. (b) point excitation, (c) 6-site CLS excitation, (d) boundary excitation, (e) "E" shape excitation. }
    \label{fig:02}
\end{figure}

\textbf{Simulation.} To numerically investigate the presence of compact localized states (CLS) in a finite Kagome lattice, we consider a lattice made of $6\times 6$ unit cells. A velocity field is introduced as input on an internal lattice site, applying a 10-period tone burst centered at the flat-band frequency. The spectral content is reported alongside Fig. \ref{fig:01}(c), and corresponds to a narrowband excitation with central frequency of approximately $3.2$ kHz. 

More details about the simulation models and the numerical methods are reported both in the section \enquote{methods}. The resulting RMS pressure distributions are illustrated in Fig. \ref{fig:01}(d-g) for different excitation conditions. In Fig. \ref{fig:01}(d), a single-point excitation is applied, resulting in a dispersed energy distribution that spreads from the input site throughout the lattice. Instead, Fig. \ref{fig:01}(e) reports the response to an excitation configured in the form described in Eq. \ref{eq:05}. Here, the energy remains highly localized around the input site, without any propagation to the lattice boundaries, clearly demonstrating the formation of a compact localized state (CLS).
Additionally, we probe the robust boundary mode by configuring the input as a linear combination of CLSs that forms a closed loop around the boundary of the lattice. As shown in Fig. \ref{fig:01}(f), this boundary excitation results in energy confinement along the lattice edge, with no inward energy spread. This behavior is unique to singular flat-bands, as the boundary mode remains intact in time despite the addition or removal of CLS elements. This robustness is highlighted in Fig. \ref{fig:01}(g), where a similar boundary configuration forms a closed loop resembling the letter \enquote{E}, further underscoring the spatial confinement of these boundary states.
We also note that, in the absence of dissipation, the energy remains confined over time. This is justified by the matching of the excitation frequency, the flat-band frequency, and the selective excitation of a 6-site CLS. When the excitation frequency deviates from the flat-band frequency, the mode retains its compact localization. Yet, the oscillations do not persist over time (see supplementary material for further details, additional results, COMSOL validation, and animations \cite{SM}).

We now extend our analysis by relaxing the assumption of strictly in-plane dynamics to explore the effect of a nonzero wavenumber $\kappa_z$. Introducing a wavenumber $\kappa_z\neq0$ produces a shift in the dispersion properties while preserving the essential characteristics observed in the 2D case, including the flatness of the third band, the singularity at the $\Gamma$ point, and the presence of compact localized modes.
This is illustrated in Fig. \ref{fig:02}(a), where the dispersion surface spans the reciprocal space both in-plane and along $\kappa_z$. 
The figure shows how the Bloch modes, originally supported by the 2D lattice, propagate along the $z$ direction at frequencies above and below $\omega_{FB}$. The dispersion contours, computed along the irreducible Brillouin zone (IBZ) at three selected values of input frequency $\omega$, are highlighted in black, providing a clear picture of how the in-plane band flatness and singularity are preserved. These dispersion contours are reported alongside in Fig. \ref{fig:02}(a)-I-II-III for the distinct frequencies of $2$ kHz, $\omega_{FB}$, and $4$ kHz, showing that below $\omega_{FB}$, the wavenumber $\kappa_z$ takes a non-null imaginary part, which denotes an evanescent wave. Instead, frequencies above $\omega_{FB}$ can propagate freely. Hence, the dynamic behavior is tested through numerical simulation, where an input wave packet, centered at $4$ kHz is imposed. Note that a frequency above the flat band is necessary to excite propagating modes characterized by a non-null velocity along $z$ and a non-evanescent wavenumber $\kappa_z$. Results are reported in Fig. \ref{fig:02}(b) for point excitation, CLS excitation, and boundary mode excitation. As expected, point excitation leads to energy spread within the plane while propagating along $z$. In contrast, the CLS and boundary modes retain their compact localization within the plane as they propagate along the $z$ direction. For completeness, results for $2$ kHz excitation are reported in the supplementary material \cite{SM}.

\begin{figure}[!t]
    \centering
    \subfigure[]{\includegraphics[width=0.9\textwidth]{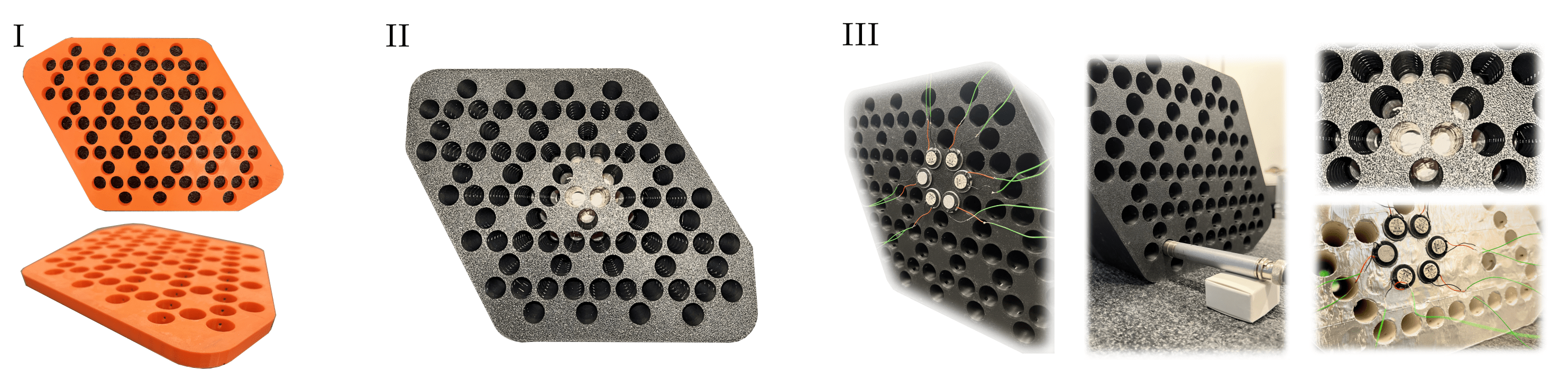}}\\
    \hspace{-0.85cm}\subfigure[]{\includegraphics[width=0.27\textwidth]{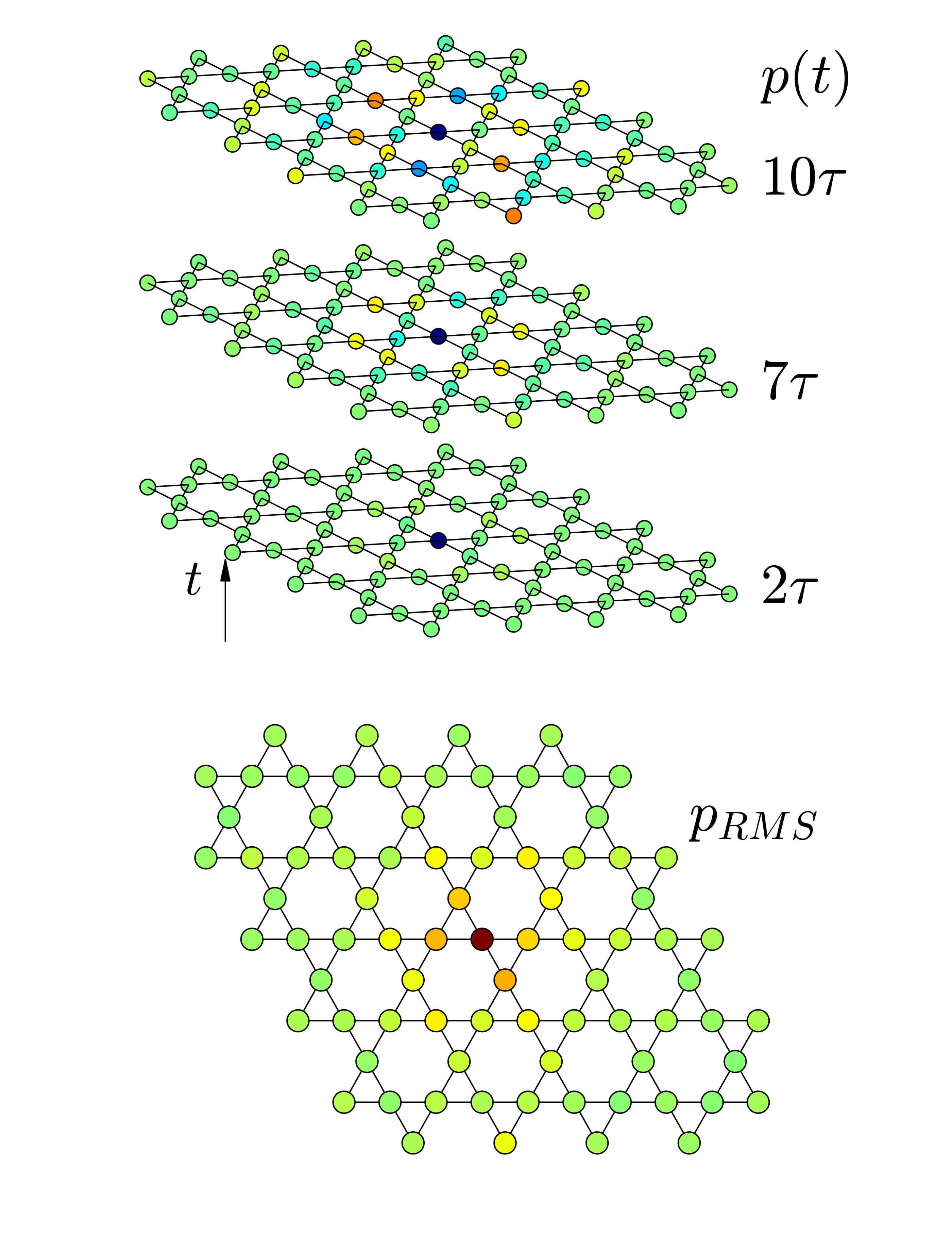}}\hspace{-0.30cm}
    \subfigure[]{\includegraphics[width=0.27\textwidth]{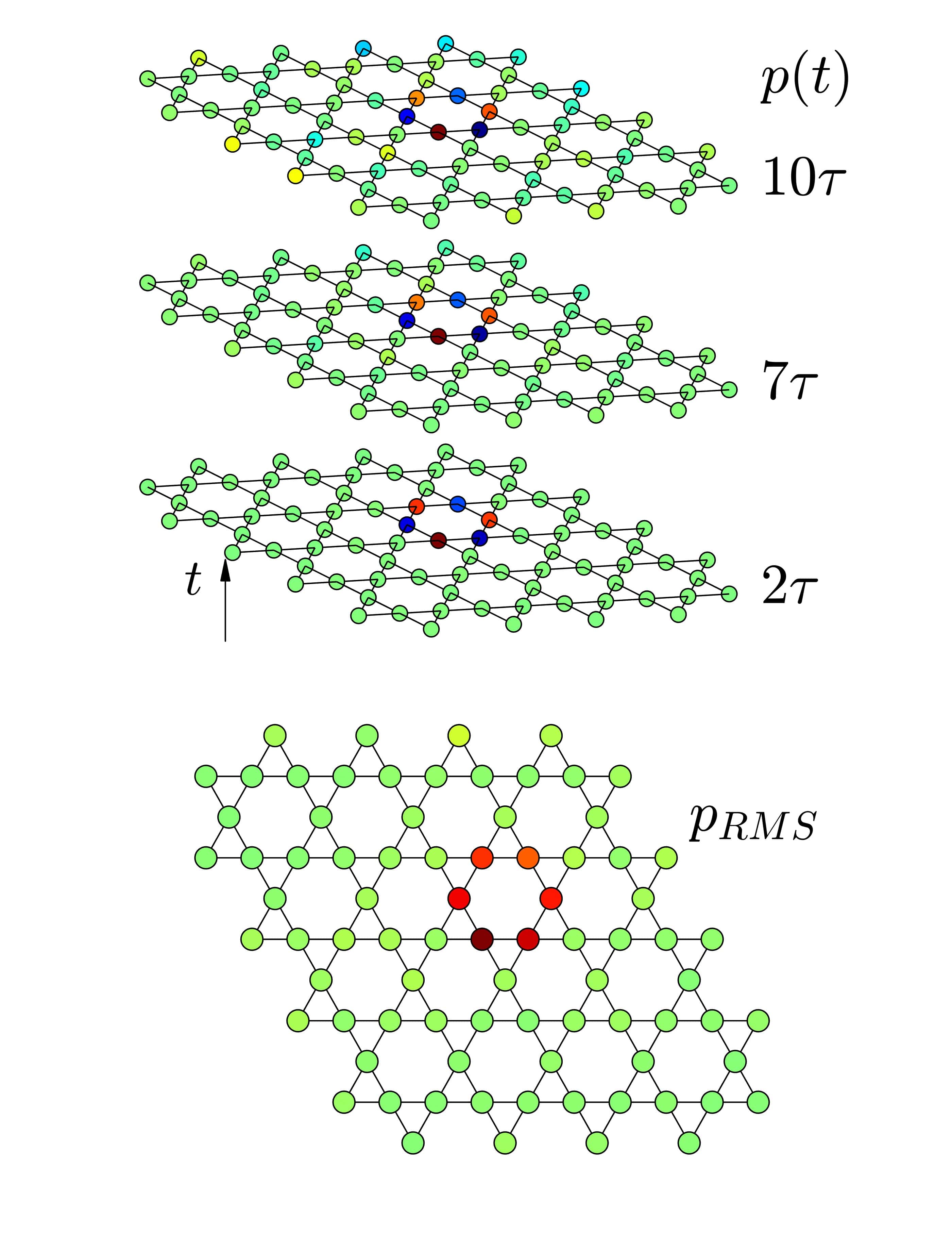}}\hspace{-0.30cm}
    \subfigure[]{\includegraphics[width=0.27\textwidth]{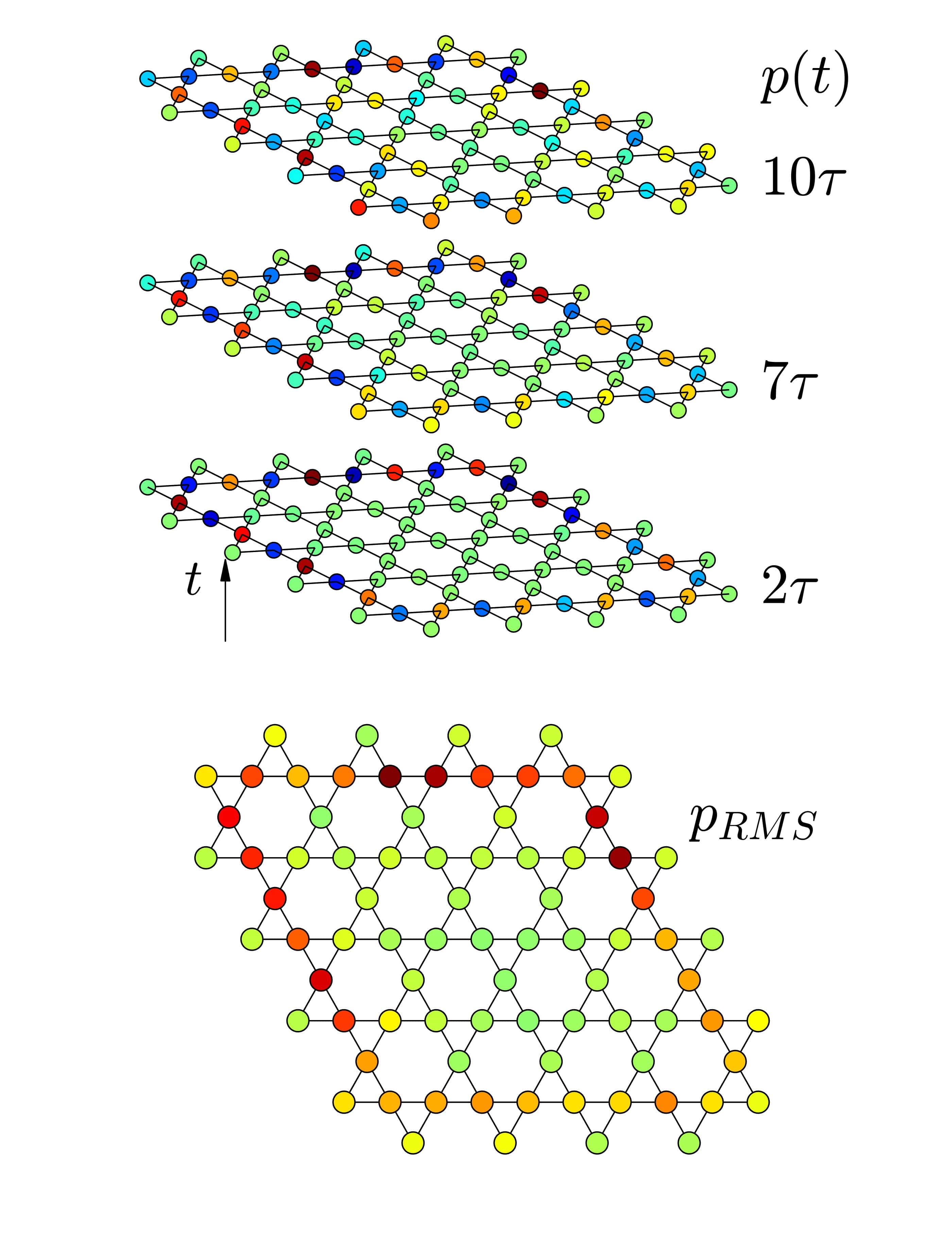}}\hspace{-0.30cm}
    \subfigure[]{\includegraphics[width=0.27\textwidth]{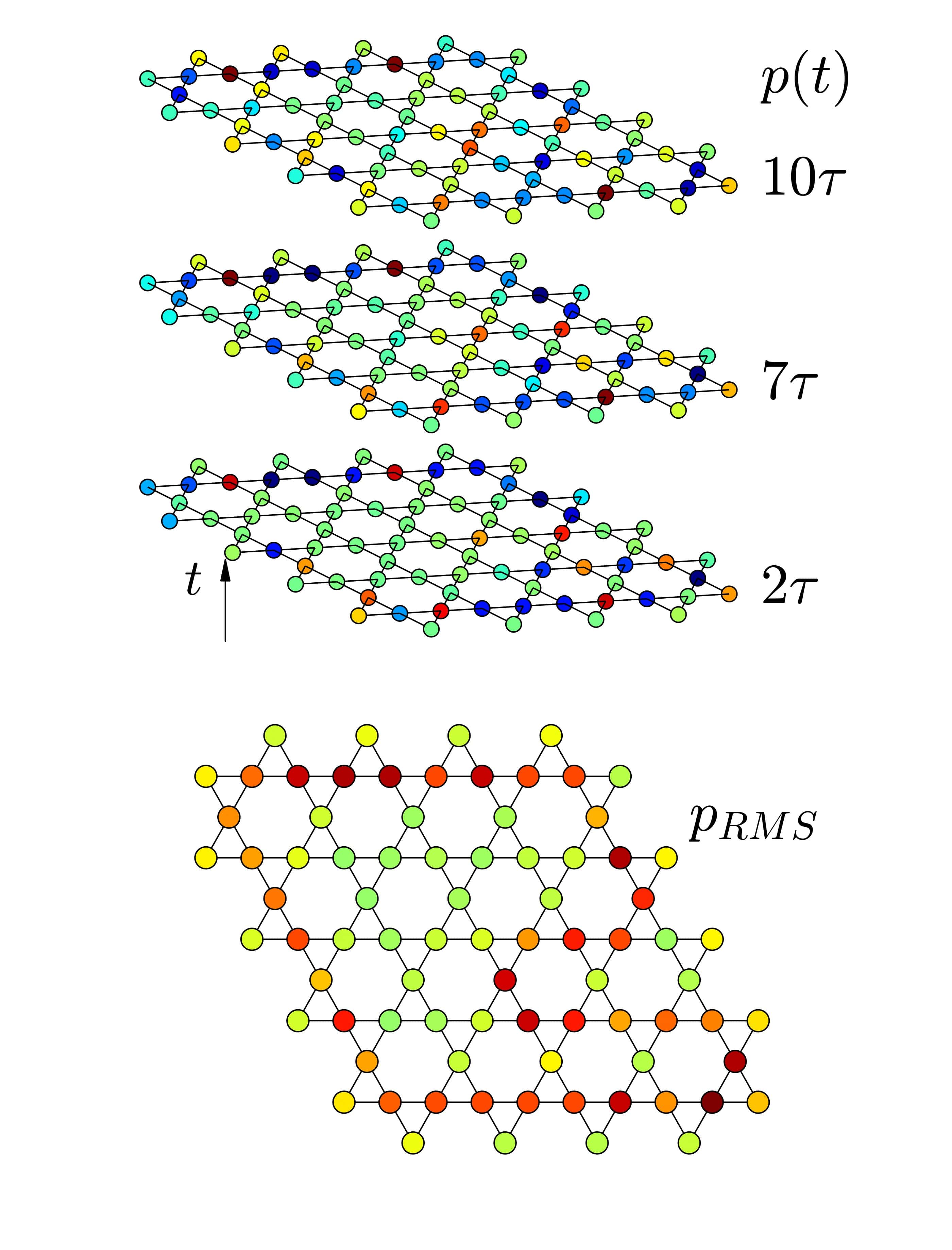}}\\
    \hspace{-0.85cm}\subfigure[]{\includegraphics[width=0.28\textwidth]{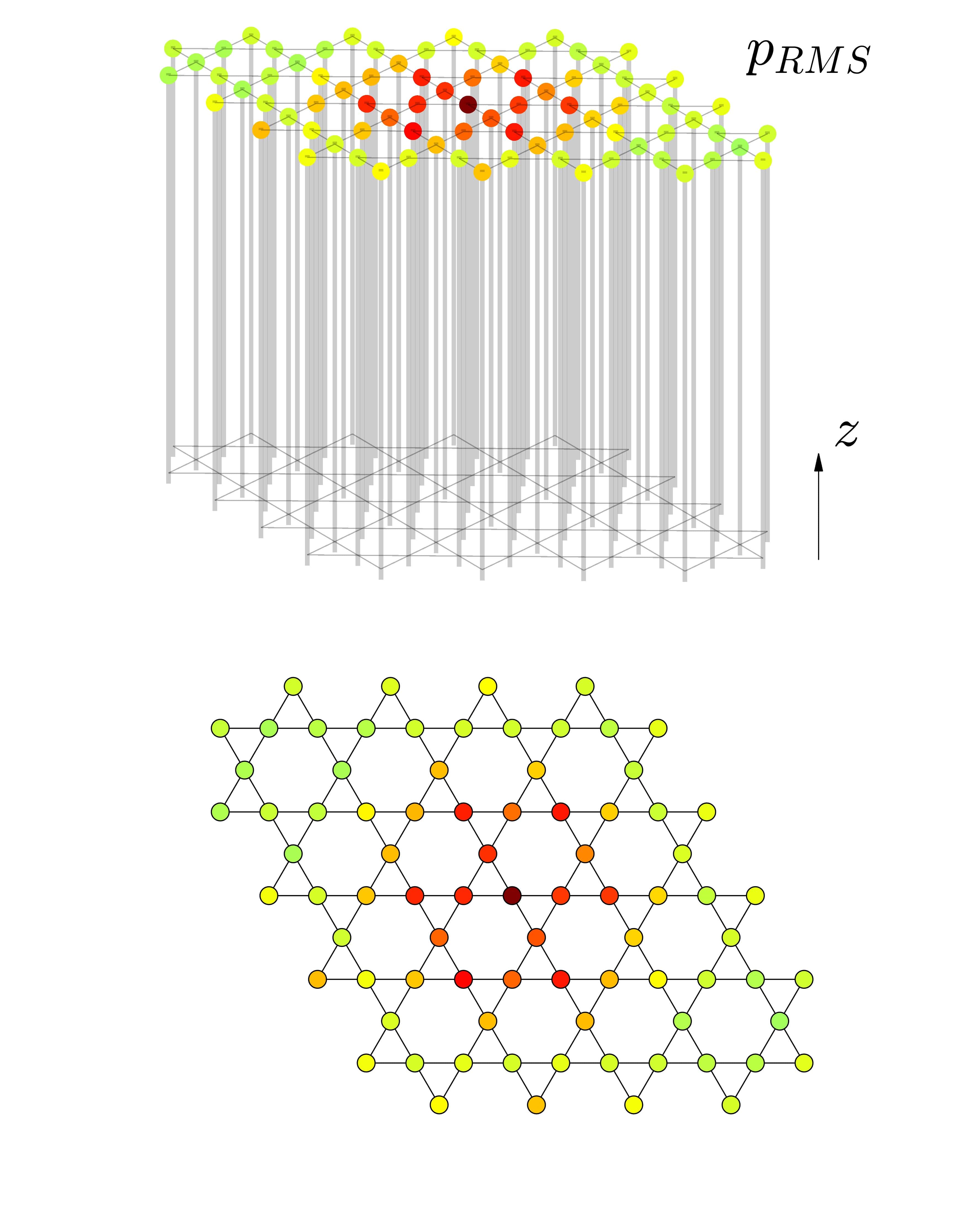}}\hspace{-0.50cm}
    \subfigure[]{\includegraphics[width=0.27\textwidth]{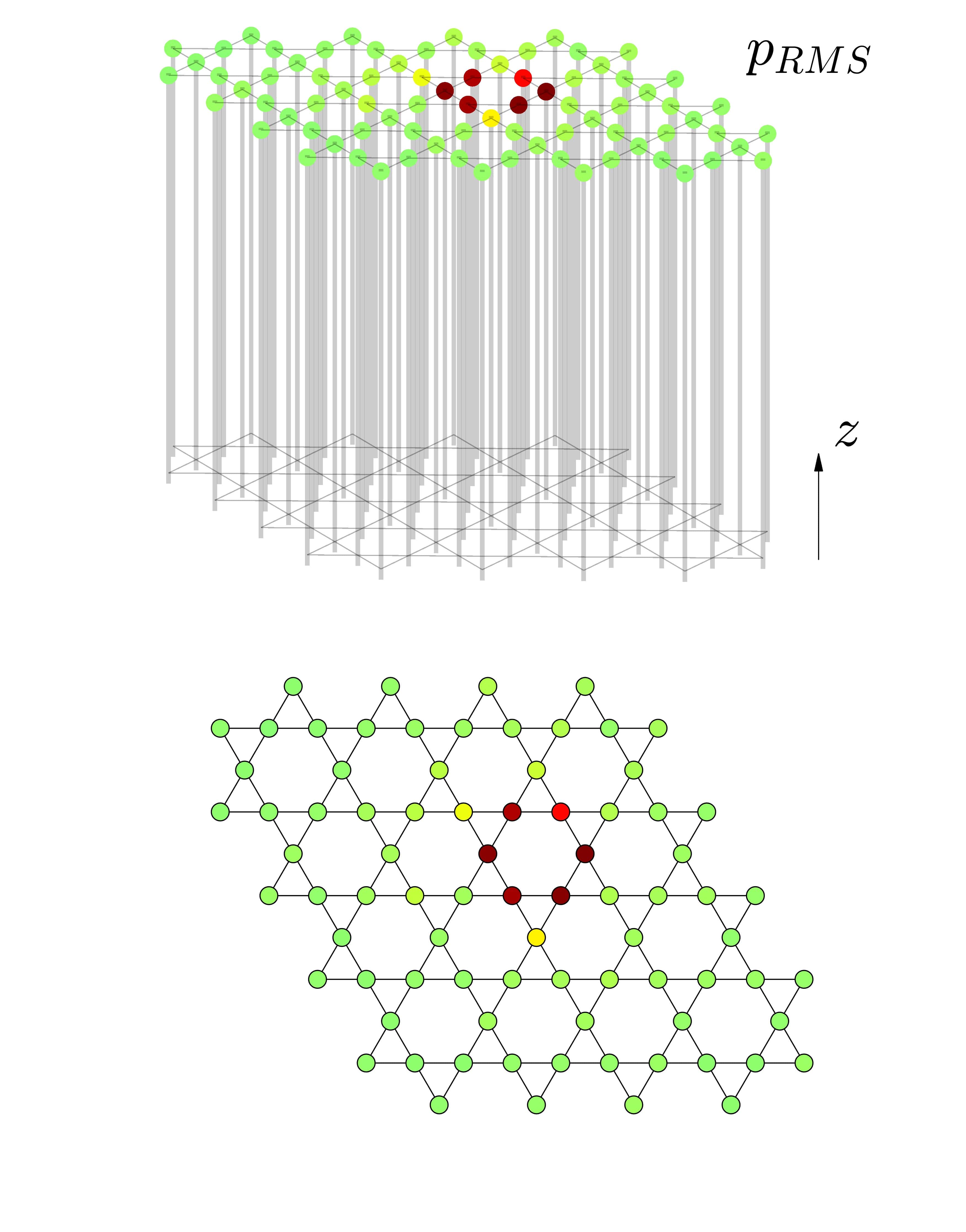}}\hspace{-0.50cm}
    \subfigure[]{\includegraphics[width=0.27\textwidth]{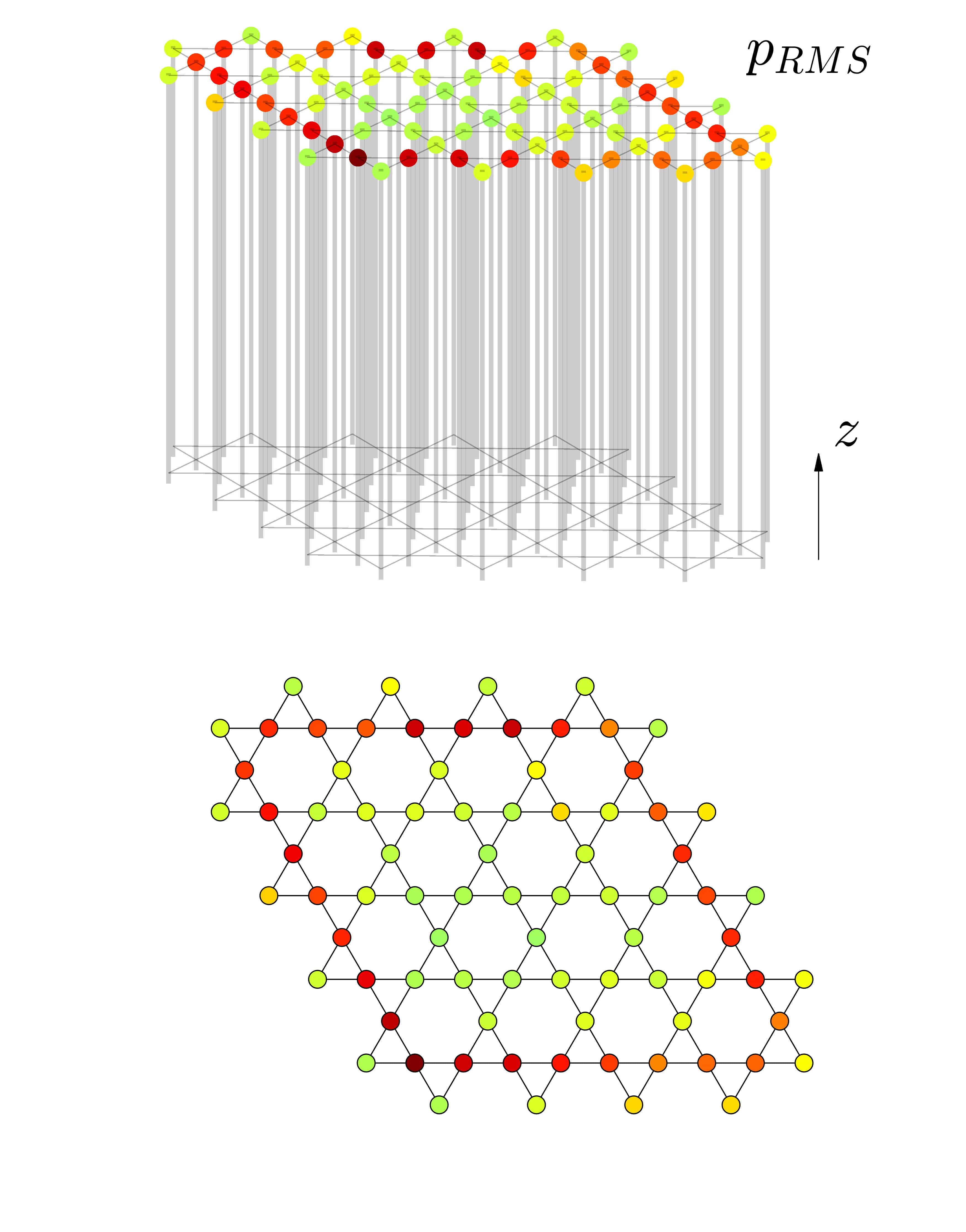}}\hspace{-0.50cm}
    \subfigure[]{\includegraphics[width=0.27\textwidth]{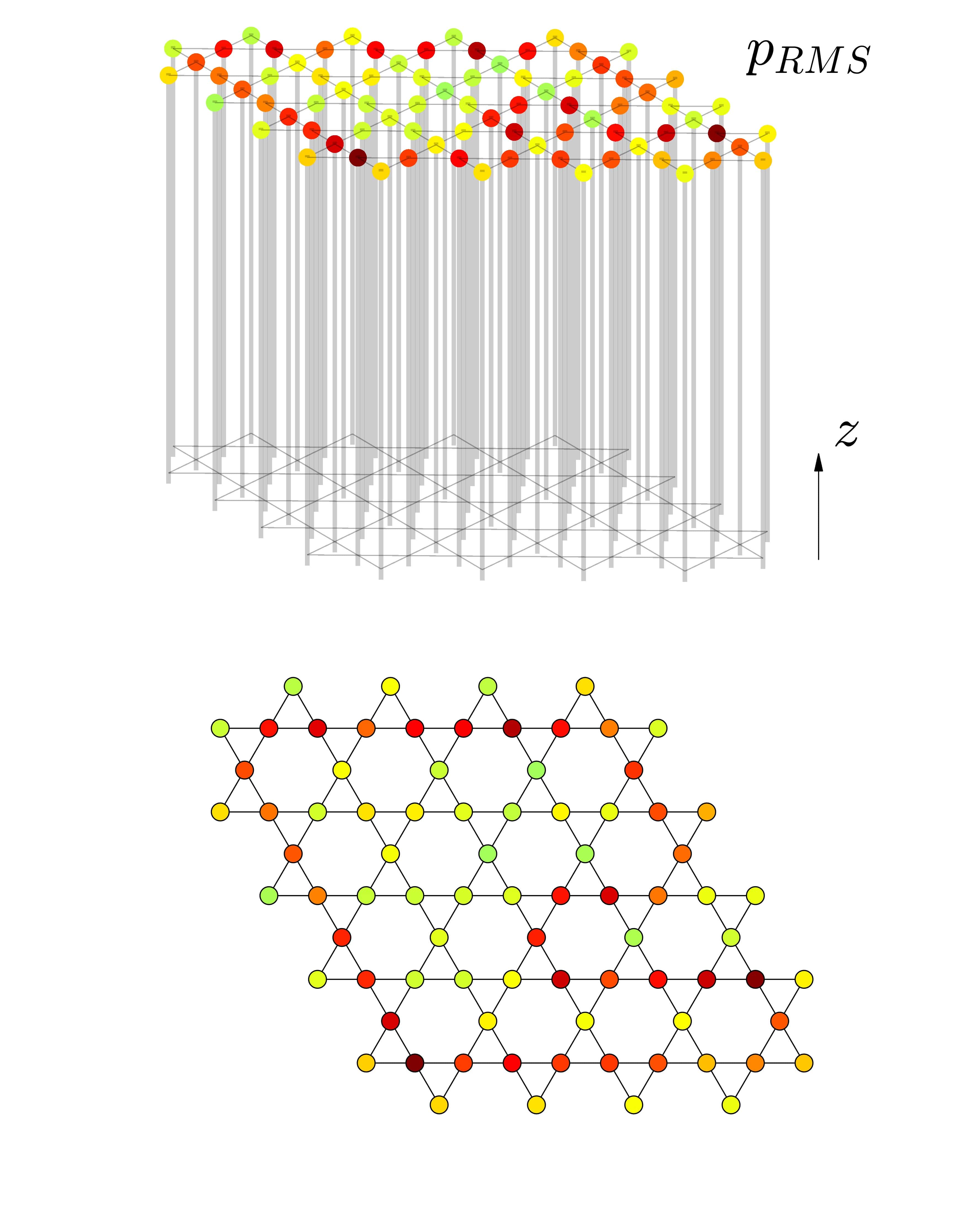}}\\
    \centering
    \caption{(a) Experimental setup and 3D printed prototypes. (I) 2D acoustic lattice. (II) front view of the 3D acoustic lattice. (III) Picture of the measurement site and 6-site CLS excitation. The waveguide elements not subjected to excitation and measurement are sealed with rigid tape. (b-e) RMS of the pressure field measured in the 2D lattice due to a (b) point excitation, (c) 6-site CLS excitation, (d) boundary excitation, and (e) Boundary excitation with removal of two CLSs. (f-i) RMS of the pressure field measured after propagation in the 3D lattice due to a (f) point excitation, (g) 6-site CLS excitation, (h) boundary excitation, and (i) boundary excitation with removal of two CLSs. }
    \label{fig:03}
\end{figure}

\textbf{Theory meets experiments.} We now focus on the experimental result, obtained through the setup displayed in Fig. \ref{fig:03}(a) (see \enquote{methods} for details on the manufacturing technology and the experimental methodologies). Briefly, the experimental campaign is performed on a $5\times5$ lattice. The 2D configuration is reported in Fig. \ref{fig:03}(a)-I, while the picture of the 3D lattice is illustrated in \ref{fig:03}(a)-II. An array of speakers is placed on one side of the sample, while the output pressure is measured on the opposite side (\ref{fig:03}(a)-III). Under these settings, the selective excitation of CLSs and boundary modes is achieved by phasing the electrical signal of consecutive array elements by $\pi$, i.e. by switching the polarity of the speakers. 
Results for the 2D lattice are shown in Fig. \ref{fig:03}(b-e) for an imposed input velocity field generated by a 10-period tone burst centered at the flat-band frequency. The response for a point excitation is reported in Fig. \ref{fig:03}(b). Notably, the energy spreads toward the lattice boundaries while, in contrast, CLS and boundary excitations (Fig. \ref{fig:03}(c-d)) generate pressure fields where the energy remains compactly localized, consistently with the numerical predictions. This behavior is further demonstrated in Fig. \ref{fig:03}(e), where the selective removal of two CLSs creates a closed loop encircling the region of removal, highlighting the robustness of the localized states.

The 3D configuration, instead, is probed with a $4$ kHz central excitation frequency, with sound pressure levels recorded along the z-axis after propagation. Unlike in 2D configurations, where localization was confined to a plane, the extended z-direction provides insight into the compact and boundary-localized modes across greater distances. Results are shown in Fig. \ref{fig:03}(f-i). Under point excitation (Fig. \ref{fig:03}(f)), the energy leaks in the plane during wave propagation. Instead, in the case of CLS or boundary-mode excitations (Fig. \ref{fig:03}(g-h)), the energy remains tightly localized, with minimal spreading along $x$ and $y$, highlighting the robust nature of these modes over propagation distances. Additionally, as illustrated in Fig. \ref{fig:03}(i), the selective removal of two CLSs forms a closed-loop region around the removed sites, preserving localization around this boundary without energy leaking into the surrounding lattice. This confirms that both CLS and boundary modes remain robust even with structural alterations, showing significant potential for controlled sound localization in three-dimensional acoustic devices. Corresponding animations that highlight the experimental response in each case are provided as supplementary files \cite{SM}. 

To conclude, this paper illustrates a numerical and experimental demonstration of flat-band phenomena in acoustic Kagome lattices, highlighting the unique wave dynamics of compact localized states (CLS) and robust boundary modes induced by the system’s singular flat band. We demonstrated how these compact localized states and boundary-confined modes emerge and propagate within Kagome lattices, with minimal energy spread in both 2D and 3D settings. The findings underscore the versatility of flat-band Kagome lattices for acoustic wave manipulation, suggesting new opportunities for sound-based information transport. This work also lays the foundation for further exploration of flat-band phenomena in acoustic and mechanical metamaterials, opening pathways to enhanced wave control technologies that leverage the unique properties of singular flat-band lattices.

\section*{Methods}
\textbf{Sample fabrication and measurements.}
The prototypes are 3D printed with a conventional Fused Layer Manufacturing (FLM) printer, with a tolerance of $\pm0.1$ mm, which is considered sufficiently accurate to guarantee the desired dynamic behavior and a low variability. The Kagome geometry is characterized by lattice constant $a=35$ mm and a waveguide diameter $d=14.5$ mm. The waveguides are separated by acoustic channels of cross-section area $A_c=5.45$ $\rm{mm}^2$. The 2D prototype has a thickness $s=12$ mm, while the 3D prototype is obtained by patterning direction $z$ with $N=16$ elements of thickness $d=12$ mm and 3D printed as a single sample for a total length of $L=192$ mm. Moreover, both 2D and 3D configurations are excited through an array of VISATON speakers of $16$ mm diameter and, in the case of CLS and boundary mode excitation, consecutive speakers are connected with opposite polarity. This allows to guarantee a selective excitation of the flat-band states. The input signal is provided as an arbitrary waveform through a KEYSIGHT 33500B signal generator. To avoid energy losses and non-ideal boundary conditions, the measurement is performed only in one lattice site, while the other waveguides are sealed with rigid tape to ensure hard boundaries. Hence, the experiment is repeated several times and the microphone is moved to different lattice sites. This allows combining the different time histories while preserving the same input phase. Finally, to obtain a constant gain measurement over the desired bandwidth a type 4188 Br\"uel and kj\ae~r microphone with $14$ mm diameter is employed. 

\textbf{Numerical simulations.}
The time simulations (Fig. \ref{fig:01} and Fig. \ref{fig:02} along with the supplementary results \cite{SM}) were carried out using a streamlined Matlab-based modeling approach. For the 2D configuration, a built-in Runge-Kutta integration scheme was applied, while the 3D configuration simulations employed finite difference time domain (FDTD) techniques, approximating partial derivatives through central difference. Both the dispersion characteristics and time-domain simulations were cross-validated against high-fidelity 3D COMSOL models (results are reported in the supplementary material \cite{SM}).

\newpage
\newcommand{\beginappendix}{%
	\setcounter{table}{0}
	\renewcommand{\thetable}{S\arabic{table}}%
	\setcounter{figure}{0}
	\renewcommand{\thefigure}{S\arabic{figure}}%
}
\beginappendix


\section*{Supplementary Note 1: modeling of the acoustic lattice}
\begin{figure*}[h]
	\centering
	\includegraphics[width=0.5\textwidth]{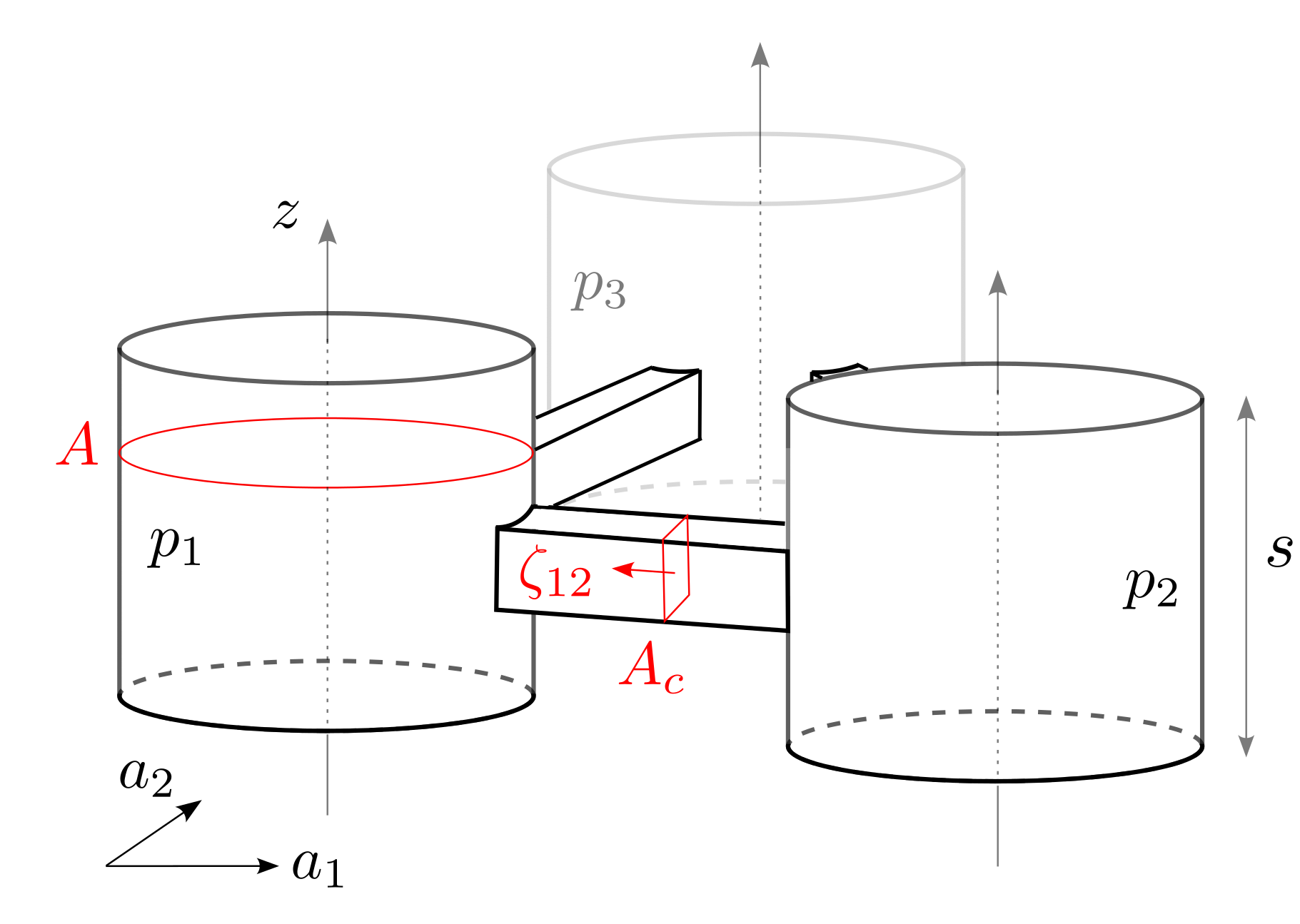}
	\caption{Schematic of the unit cell, along with the relevant quantities employed to describe the dynamics of interest.}
	\label{FigS01}    
\end{figure*}

The acoustic lattice illustrated in Fig. 1(a) is made of a Kagome tessellation of pipes extended along the $z$ direction and linked in the $x-y$ plane through air-filled connecting elements. A zoomed view of the unit cell geometry is reported in Fig. \ref{FigS01}. 
Following a spring-mass equivalence \cite{kinsler2000fundamentals,riva2023adiabatic}, the equation of motion for such a system writes:
\begin{equation}
	\begin{split}
		\frac{1}{c^2}\frac{\partial^2 \bm{p}^{m,n}}{\partial t^2}-\nabla^2 \bm{p}^{m,n}=\sum_{i\neq j}\frac{\partial \bm{G}_i}{\partial t}
	\end{split}
	\label{eq:S01}
\end{equation}
where $\bm{p}^{m,n}$ is the vector accommodating the pressure field amplitudes $p_j^{m,n}$ relative to each pipe in the unit cell located at coordinates $\bm{a}=(m\bm{a}_1,n\bm{a}_2)$. $\bm{G}_i$ is the vector accommodating the rate of input mass per unit volume $G_{ji}^{m,n}$ which, for the mass element between the $j^{th}$ and $i^{th}$ pipes, is defined as:
\begin{equation}
	G_{ji}^{(m,n)}=\rho\frac{A_c}{A}\frac{\partial\zeta_{ji}^{(m,n)}}{\partial t}
\end{equation}
$A$ and $A_c$ are the cross-section areas of the pipe and the connecting elements, respectively. $\rho$ is the density and $\zeta_{ji}^{(m,n)}$ is a kinematic variable used to approximate the displacement of the fluid inside the neck between the $i^{th}$ and $j^{th}$ pipes. 
The dynamic equilibrium inside the neck is:
\begin{equation}
	m_{eq}\ddot \zeta_{ji}^{(m,n)}+p_i^{(m,n)}A_c-p_j^{(m,n)}A_c=0
	\label{eq:S03}
\end{equation}
where the equivalent mass $m_{eq}$ is $m_{eq}=\rho A_cl'$, being $l'$ an effective length of the neck, which depends on the connection geometry \cite{kinsler2000fundamentals}. Combining Eqs. \ref{eq:S01}-\ref{eq:S03}, we get to:
\begin{equation}
	\begin{split}
		&\frac{\partial^2p_1^{m,n}}{\partial t^2}-c^2\frac{\partial^2p_1^{m,n}}{\partial z^2}=-4\omega_0^2p_1^{m,n}+\omega_0^2p_2^{m,n}+\omega_0^2p_3^{n,m}+\omega_0^2p_3^{m,n-1}+\omega_0^2p_2^{m-1,n}\\[5pt]
		&\frac{\partial^2p_2^{m,n}}{\partial t^2}-c^2\frac{\partial^2p_2^{m,n}}{\partial z^2}=-4\omega_0^2p_2^{m,n}+\omega_0^2p_1^{m,n}+\omega_0^2p_3^{m,n}+\omega_0^2p_1^{m+1,n}+\omega_0^2p_3^{m+1,n-1}\\[5pt]
		&\frac{\partial^2p_3^{m,n}}{\partial t^2}-c^2\frac{\partial^2p_3^{m,n}}{\partial z^2}=-4\omega_0^2p_3^{m,n}+\omega_0^2p_1^{m,n}+\omega_0^2p_2^{m,n}+\omega_0^2p_1^{m,n+1}+\omega_0^2p_2^{m-1,n+1}
	\end{split}
\end{equation}
where $\omega_0^2=\sqrt{k_{eq}/m_{eq}}$ and $k_{eq}=\rho A_c^2c^2/As$. $s$ is the length of the pipe along $z$. 
Due to the underlying periodicity of the lattice $p_j^{m+1,n}(x,y,z)=\hat{p}_j(z)^{m,n}{e}^{-{\rm i}\bm{\kappa}\cdot\bm{a}_1}$ and that $p_j^{m,n+1}(x,y,z)=\hat{p}_j^{m,n}(z){e}^{-{\rm i}\bm{\kappa}\cdot\bm{a}_2}$. Hence, 
\begin{equation}
	\begin{split}
		&\frac{\partial^2\hat{p}_1^{m,n}}{\partial t^2}-c^2\frac{\partial^2\hat{p}_1^{m,n}}{\partial z^2}=-4\omega_0^2\hat{p}_1^{m,n}+\omega_0^2\hat{p}_2^{m,n}\left(1+{\rm{e}^{\rm{i}\bm{\kappa}\cdot\bm{a_1}}}\right)+\omega_0^2\hat{p}_3^{n,m}\left(1+{\rm{e}^{\rm{i}\bm{\kappa}\cdot\bm{a_2}}}\right)\\[5pt]
		&\frac{\partial^2\hat{p}_2^{m,n}}{\partial t^2}-c^2\frac{\partial^2\hat{p}_2^{m,n}}{\partial z^2}=-4\omega_0^2\hat{p}_2^{m,n}+\omega_0^2\hat{p}_1^{m,n}\left(1+{\rm{e}^{-\rm{i}\bm{\kappa}\cdot\bm{a_1}}}\right)+\omega_0^2\hat{p}_3^{n,m}\left(1+{\rm{e}^{-\rm{i}\bm{\kappa}\cdot\bm{a_1}+\rm{i}\bm{\kappa}\cdot\bm{a_2}}}\right)\\[5pt]
		&\frac{\partial^2\hat{p}_3^{m,n}}{\partial t^2}-c^2\frac{\partial^2\hat{p}_3^{m,n}}{\partial z^2}=-4\omega_0^2\hat{p}_3^{m,n}+\omega_0^2\hat{p}_1^{m,n}\left(1+{\rm{e}^{-\rm{i}\bm{\kappa}\cdot\bm{a_2}}}\right)+\omega_0^2\hat{p}_2^{n,m}\left(1+{\rm{e}^{\rm{i}\bm{\kappa}\cdot\bm{a_1}-\rm{i}\bm{\kappa}\cdot\bm{a_2}}}\right)
	\end{split}
\end{equation}
wave propagation along the main pipe dimension is now enforced through the complex exponential term $\hat{p}_j^{m,n}(z)=\hat{p}_j^{m,n}{\rm e}^{{\rm i}\left(\kappa_zz-\omega t\right)}$, yielding the eigenvalue problem reported in the main text:
\begin{equation}
	\begin{split}
		\left(\omega^2 I - H\right)\left|\bm{p}\right>=0\hspace{1cm}
		H=\begin{pmatrix}
			4\omega_0^2&-\omega_0^2\left[1+{\rm{e}^{\rm{i}\bm{\kappa}\cdot\bm{a_1}}}\right]&-\omega_0^2\left[1+{\rm{e}^{\rm{i}\bm{\kappa}\cdot\bm{a_2}}}\right]\\[6pt]
			-\omega_0^2\left[1+{\rm{e}^{-\rm{i}\bm{\kappa}\cdot\bm{a_1}}}\right]&4\omega_0^2&-\omega_0^2\left[1+{\rm{e}^{\rm{i}\left(-\bm{\kappa}\cdot\bm{a_1}+\bm{\kappa}\cdot\bm{a_2}\right)}}\right]\\[6pt]
			-\omega_0^2\left[1+{\rm{e}^{-\rm{i}\bm{\kappa}\cdot\bm{a_2}}}\right]&-\omega_0^2\left[1+{\rm{e}^{\rm{i}\left(\bm{\kappa}\cdot\bm{a_1}-\bm{\kappa}\cdot\bm{a_2}\right)}}\right]&4\omega_0^2
		\end{pmatrix}+c^2\kappa_z^2I
	\end{split}
\end{equation}
where the in-plane dynamics is characterized by $\kappa_z=0$. In these settings, the dispersion exhibits a flat band at $\omega=\sqrt{6}\omega_0$, and the corresponding eigenvector can be conveniently evaluated by eliminating the $Q^{th}$ row from the $Q\times Q$ matrix $H_k=\omega^2 I - H$. As such, a more compact system of equation is derived, where the matrix elements $K_{k,qj}$ are $K_{k,qj}=H_{k,qj}H_{k,1Q}-H_{k,qQ}H_{k,1j}$ being $j=1,\hdots,Q-1$ the columns and $q=1,\hdots,Q-1$ the rows of the reduced matrix $K_k$. In the case at hand, $Q=3$ and:
\begin{equation}
	K_k=\begin{bmatrix}
		H_{k,21}H_{k,13}-H_{k,23}H_{k,11}&H_{k,22}   H_{k,13}-H_{k,23}H_{k,12}\\[7pt]
		H_{k,31}H_{k,13}-H_{k,33}H_{k,11}&H_{k,32}H_{k,13}-H_{k,23}H_{k,12}
	\end{bmatrix}
\end{equation}
where the matrix elements are:
\begin{equation}
	\begin{split}
		&K_{k,11}=\omega_0^4\left(-1+{\rm e}^{-{\rm i}\bm{\kappa}\cdot\bm{a}_1}+{\rm e}^{{\rm i}\bm{\kappa}\cdot\bm{a}_2}-{\rm e}^{-{\rm i}\bm{\kappa}\cdot\bm{a}_1+{\rm i}\bm{\kappa}\cdot\bm{a}_2}\right)\\[5pt]
		&K_{k,12}=\omega_0^4\left(1+{\rm e}^{{\rm i}\bm{\kappa}\cdot\bm{a}_2}-{\rm e}^{{\rm i}\bm{\kappa}\cdot\bm{a}_1}-{\rm e}^{-{\rm i}\bm{\kappa}\cdot\bm{a}_1+{\rm i}\bm{\kappa}\cdot\bm{a}_2}\right)\\[5pt]
		&K_{k,21}=\omega_0^4\left({\rm e}^{{\rm i}\bm{\kappa}\cdot\bm{a}_2}+{\rm e}^{-{\rm i}\bm{\kappa}\cdot\bm{a}_2}\right)\\[5pt]
		&K_{k,22}=\omega_0^4\left(-1-{\rm e}^{{\rm i}\bm{\kappa}\cdot\bm{a}_1}+{\rm e}^{{\rm i}\bm{\kappa}\cdot\bm{a}_2}+{\rm e}^{{\rm i}\bm{\kappa}\cdot\bm{a}_1+{\rm i}\bm{\kappa}\cdot\bm{a}_2}\right)
	\end{split}
\end{equation}
Note that a straightforward choice for the eigenvector is:
\begin{equation}
	\begin{split}
		&\hat{p}_1^{m,n}=K_{k,12}=\omega_0^4\left(1+{\rm e}^{{\rm i}\bm{\kappa}\cdot\bm{a}_2}-{\rm e}^{{\rm i}\bm{\kappa}\cdot\bm{a}_1}-{\rm e}^{-{\rm i}\bm{\kappa}\cdot\bm{a}_1+{\rm i}\bm{\kappa}\cdot\bm{a}_2}\right)\\[7pt]
		&\hat{p}_2^{m,n}=-K_{k,11}=\omega_0^4\left(1+{\rm e}^{-{\rm i}\bm{\kappa}\cdot\bm{a}_1}-{\rm e}^{{\rm i}\bm{\kappa}\cdot\bm{a}_2}+{\rm e}^{-{\rm i}\bm{\kappa}\cdot\bm{a}_1+{\rm i}\bm{\kappa}\cdot\bm{a}_2}\right)\\[5pt]
		&\hat{p}_3^{m,n}=-\frac{\hat{p}_1^{m,n}H_{k,31}+\hat{p}^{m,n}_2H_{k,32}}{H_{k,33}}=-\omega_0^4\left(2-{\rm e}^{{\rm i}\bm{\kappa}\cdot\bm{a}_1}-{\rm e}^{-{\rm i}\bm{\kappa}\cdot\bm{a}_1}\right)
	\end{split}
\end{equation}
hence:
\begin{equation}
	\left|\bm{p}\right>=\begin{pmatrix}
		1+{\rm e}^{\rm{i}\bm{\kappa}\cdot\bm{a}_2}-{\rm e}^{\rm{i}\bm{\kappa}\cdot\bm{a}_1}-{\rm e}^{-\rm{i}\bm{\kappa}\cdot\bm{a}_1+\rm{i}\bm{\kappa}\cdot\bm{a}_2}\\[3pt]
		1-{\rm e}^{-\rm{i}\bm{\kappa}\cdot\bm{a}_1}-{\rm e}^{\rm{i}\bm{\kappa}\cdot\bm{a}_2}+{\rm e}^{-\rm{i}\bm{\kappa}\cdot\bm{a}_1+\rm{i}\bm{\kappa}\cdot\bm{a}_2}\\[3pt]
		-2+{\rm e}^{\rm{i}\bm{\kappa}\cdot\bm{a}_1}+{\rm e}^{-\rm{i}\bm{\kappa}\cdot\bm{a}_1}
	\end{pmatrix}
	\label{eq:eigv}
\end{equation}
due to the linearity of the differential operator, the combination of eigenvectors is also an eigenvector. This allows us to get to the following more compact form:
\begin{equation}
	\left|\bm{p}\right>=\begin{pmatrix}
		{\rm e}^{\rm{i}\bm{\kappa}\cdot\bm{a}_2}-{\rm e}^{\rm{i}\bm{\kappa}\cdot\bm{a}_1}\\[3pt]
		1-{\rm e}^{\rm{i}\bm{\kappa}\cdot\bm{a}_2}\\[3pt]
		-1+{\rm e}^{\rm{i}\bm{\kappa}\cdot\bm{a}_1}
	\end{pmatrix}
\end{equation}
\section*{Supplementary Note 2: Hilbert-Schmidt quantum distance}
\begin{figure}[!h]
	\centering
	\subfigure[]{\includegraphics[width=0.21\textwidth]{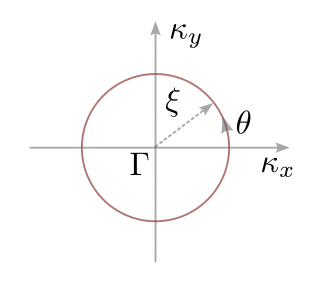}} 
	\subfigure[]{\includegraphics[width=0.24\textwidth]{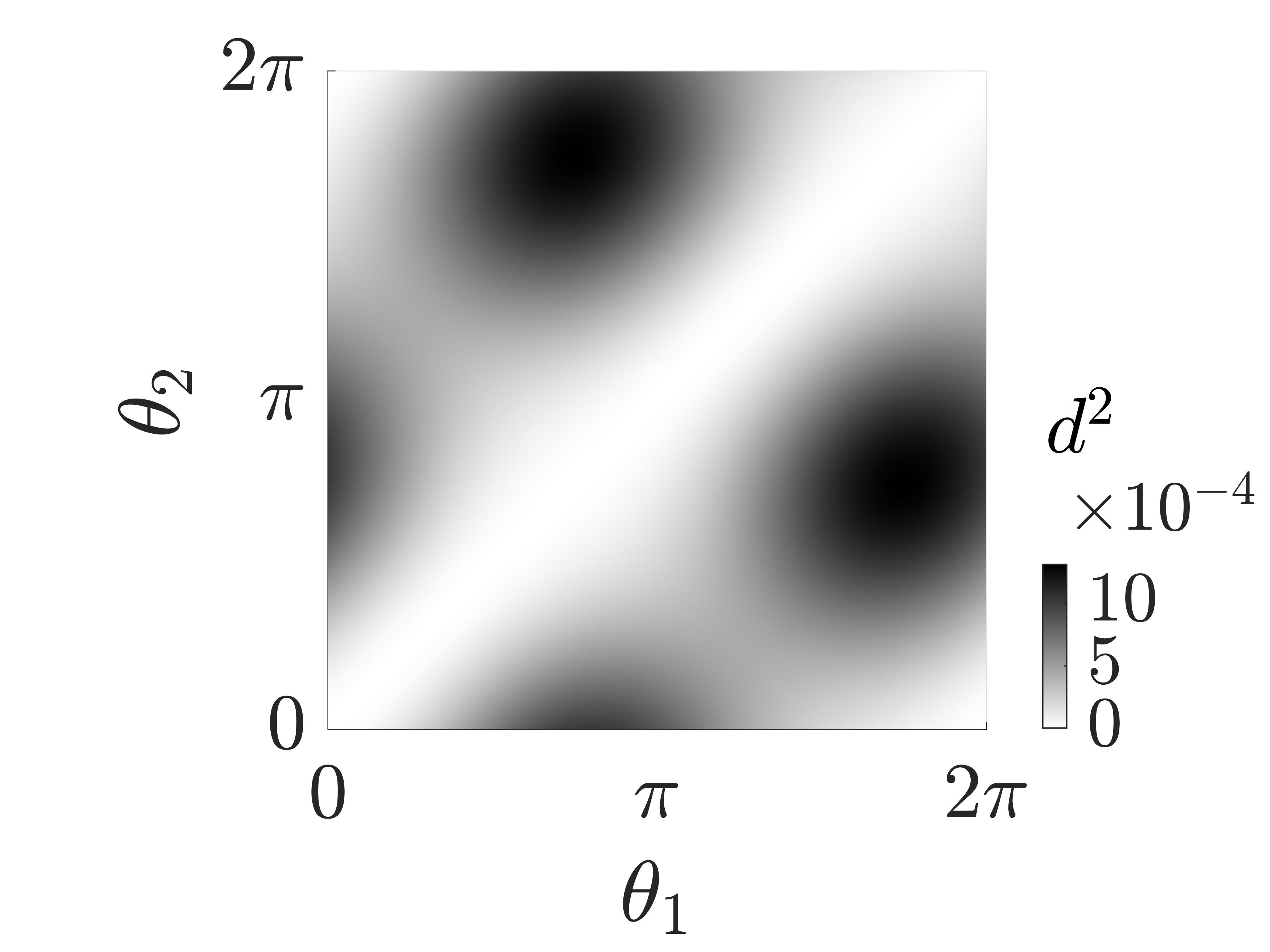}} 
	\subfigure[]{\includegraphics[width=0.24\textwidth]{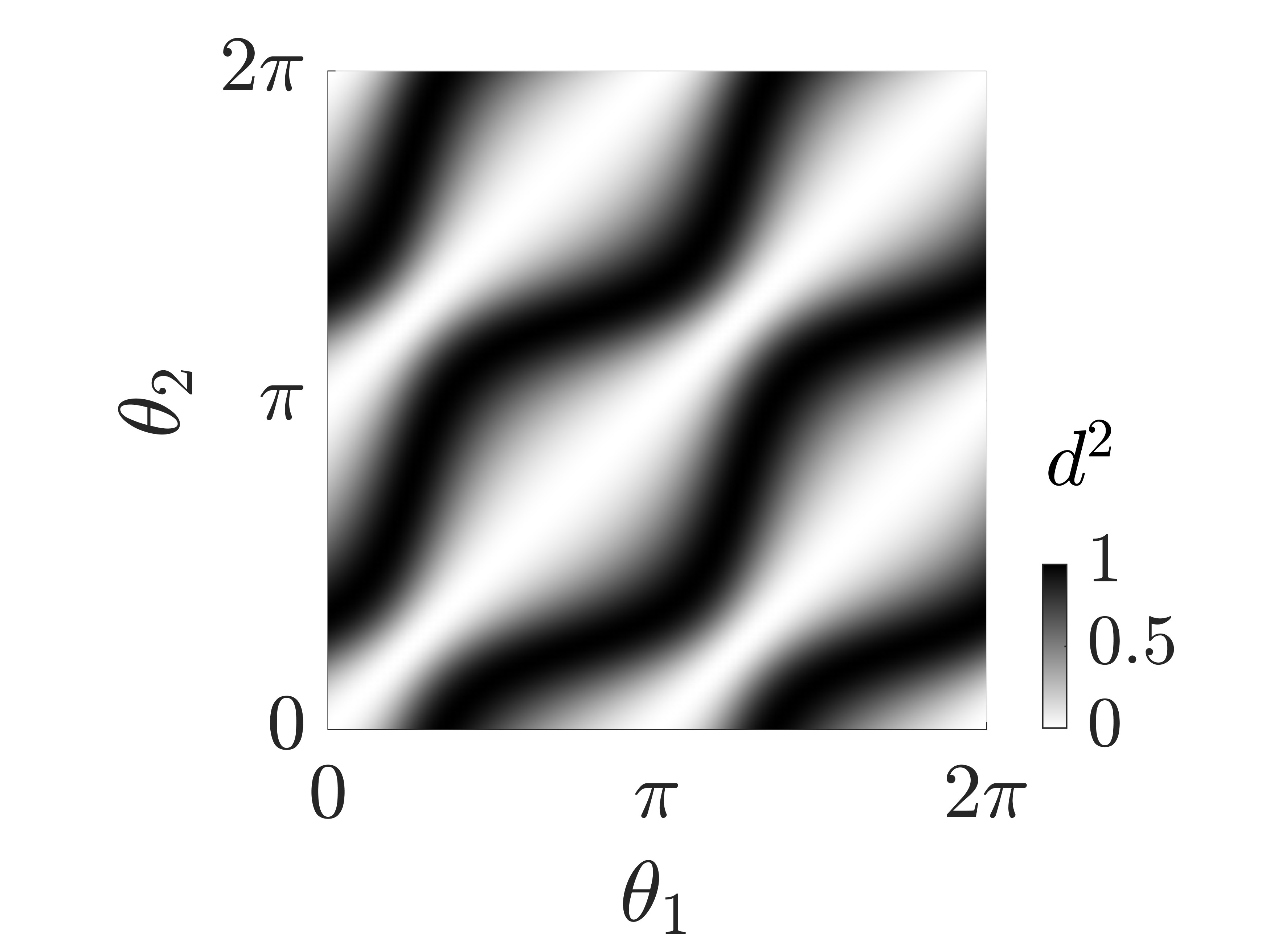}} 
	\subfigure[]{\includegraphics[width=0.24\textwidth]{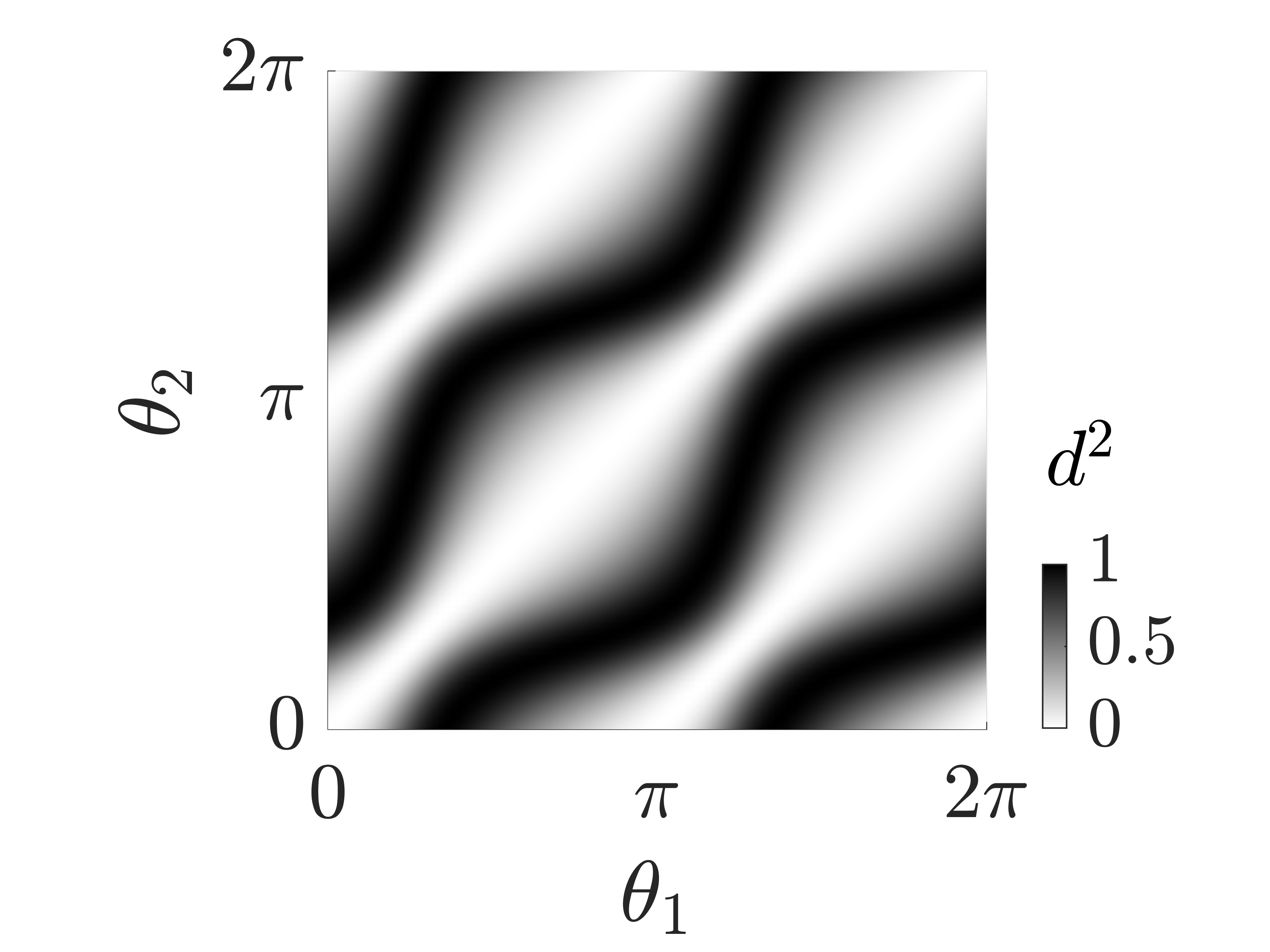}}\\     
	\centering
	\caption{(a) Schematic of the polar coordinates employed to encircle the $\Gamma$ point. $d^2$ for the (b) first band, (c) the second, quadratic band, and (d) the third, flat band. }
	\label{fig:S2}
\end{figure}

The Hilbert-Schmidt quantum distance is contextually evaluated as \cite{rhim2021singular}: 
\begin{equation}
	d^2\left(\bm{p_{\kappa_1}},\bm{p_{\kappa_2}}\right)= 1 - \left|\langle \bm{p_{\kappa_1}}|\bm{p_{\kappa_2}}\rangle\right|^2
\end{equation}
and quantifies the strength of the band singularity. Note that if the band crossing is singular, the eigenvector at the $\Gamma$ point depends upon the path approaching $\Gamma$. This leads to unitary values for $d^2$ near the singular point. 
As such, $d^2$ is probed in by encircling $\Gamma$ as shown in Fig. \ref{fig:S2}(a). Polar coordinates are conveniently used, with $\bm{\kappa}\cdot \bm{a}_1=\xi\cos{\theta_1}$ and $\bm{\kappa}\cdot \bm{a}_2=\xi\sin{\theta_2}$. Here, $\xi$ is arbitrarily small, fixed radius, and $\left|\bm{p_\kappa}\right>$ is evaluated as $\theta$ varies around $\Gamma$. 
Results are reported in Fig. \ref{fig:S2}(b-d). Notably, Fig. \ref{fig:S2}(b) displays $d^2$ for the first band, which is characterized by very small values $\approx 0$. In contrast, Fig. \ref{fig:S2}(c) and Fig. \ref{fig:S2}(d) display results for the second (quadratic) and third (flat) band near the $\Gamma$ point, showing that $d^2$ oscillates between $0$ and $1$.

\newpage
\section*{Supplementary Note 3: additional numerical results for the 2D lattice}
\begin{figure}[!h]
	\centering
	\hspace{-0.75cm}\subfigure[]{\includegraphics[width=0.27\textwidth]{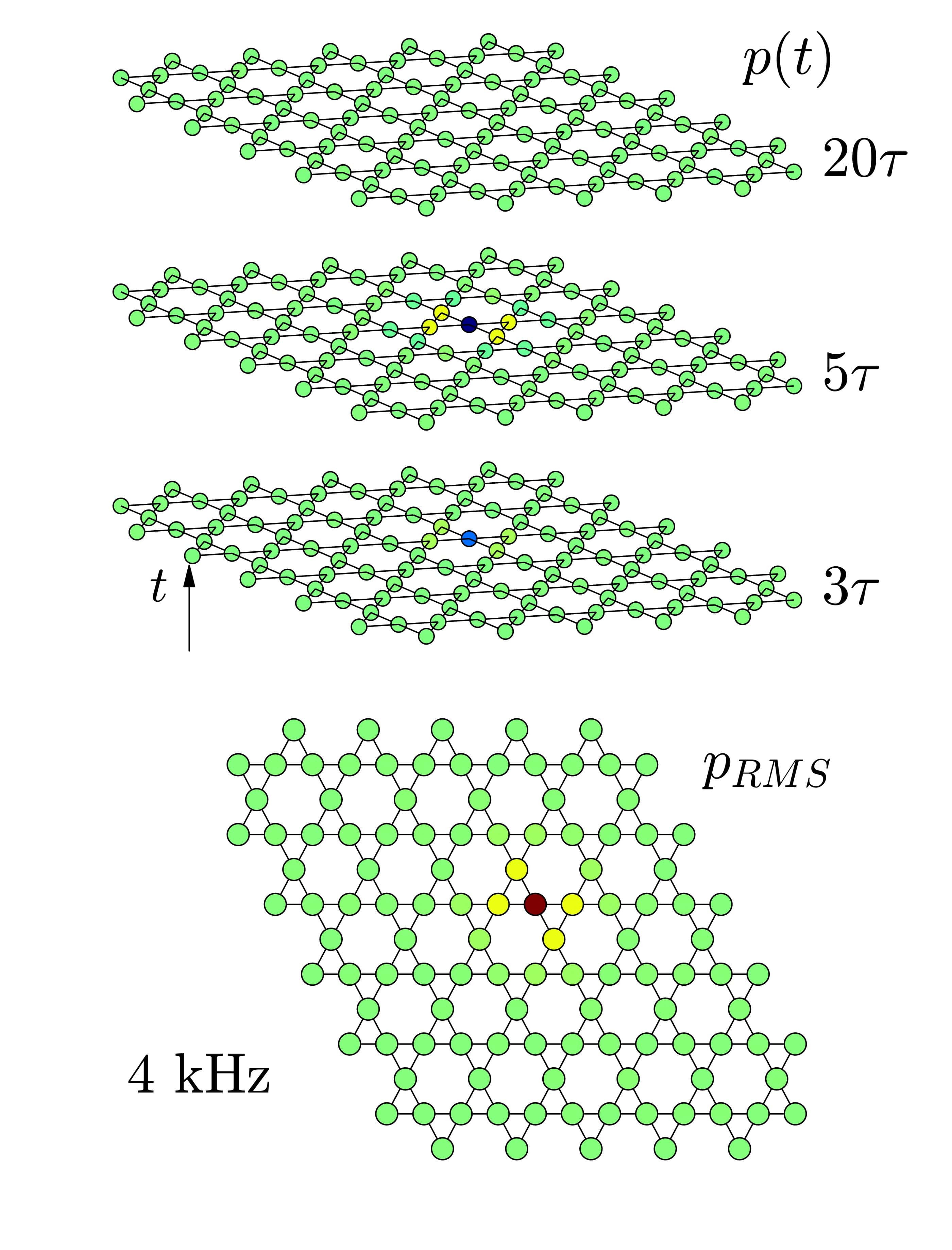}}\hspace{-0.2cm}
	\subfigure[]{\includegraphics[width=0.26\textwidth]{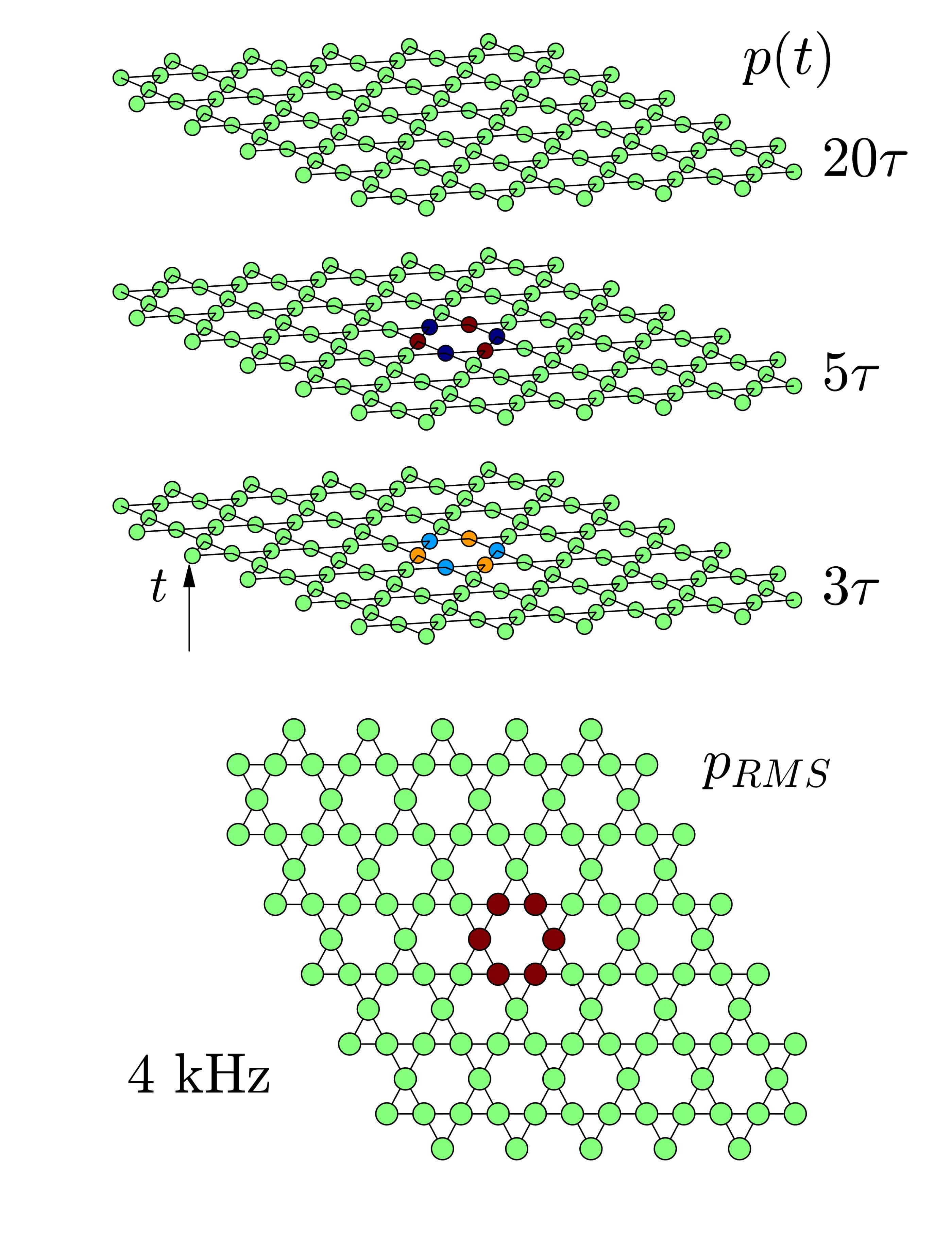}}\hspace{-0.2cm}
	\subfigure[]{\includegraphics[width=0.26\textwidth]{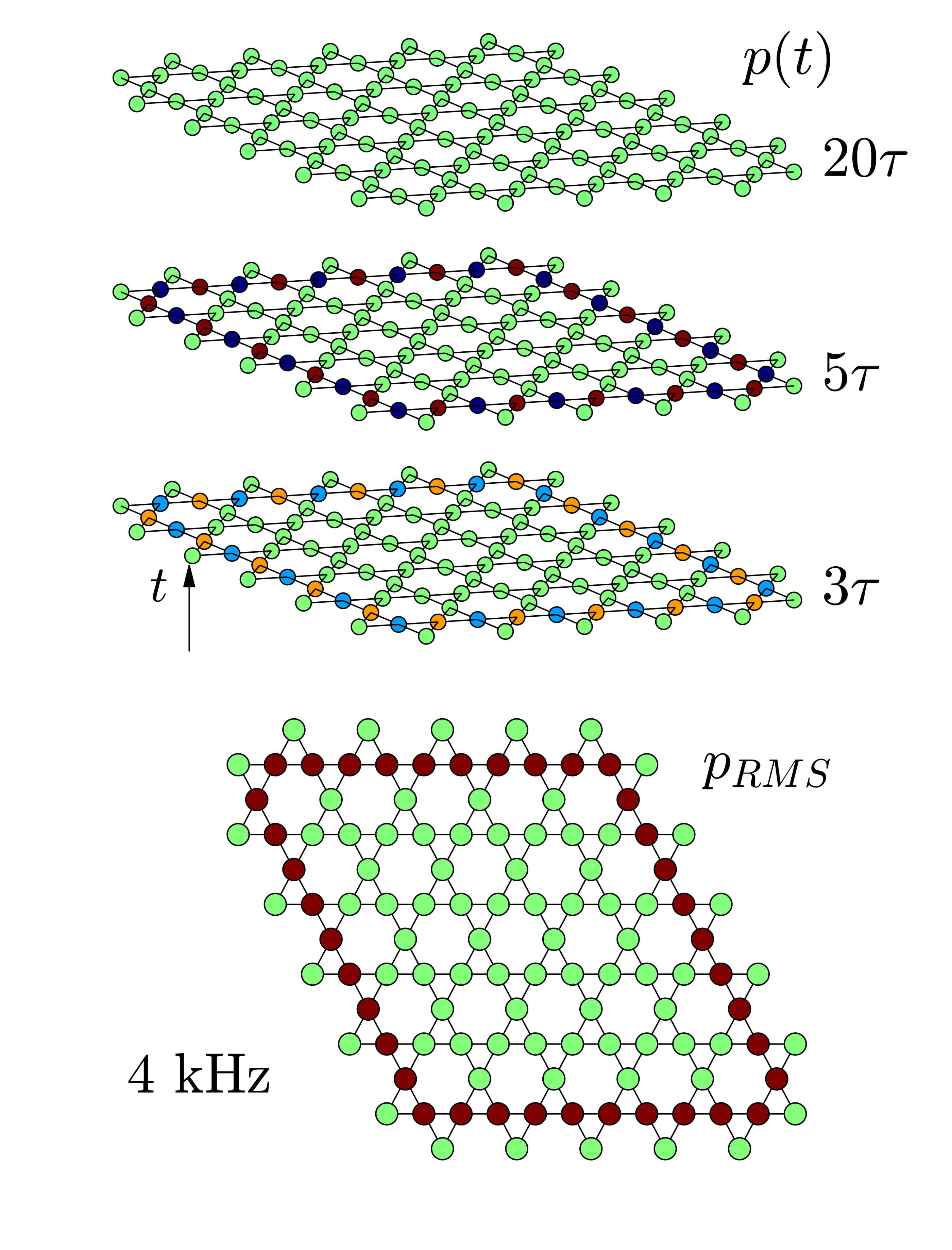}}\hspace{-0.2cm}
	\subfigure[]{\includegraphics[width=0.26\textwidth]{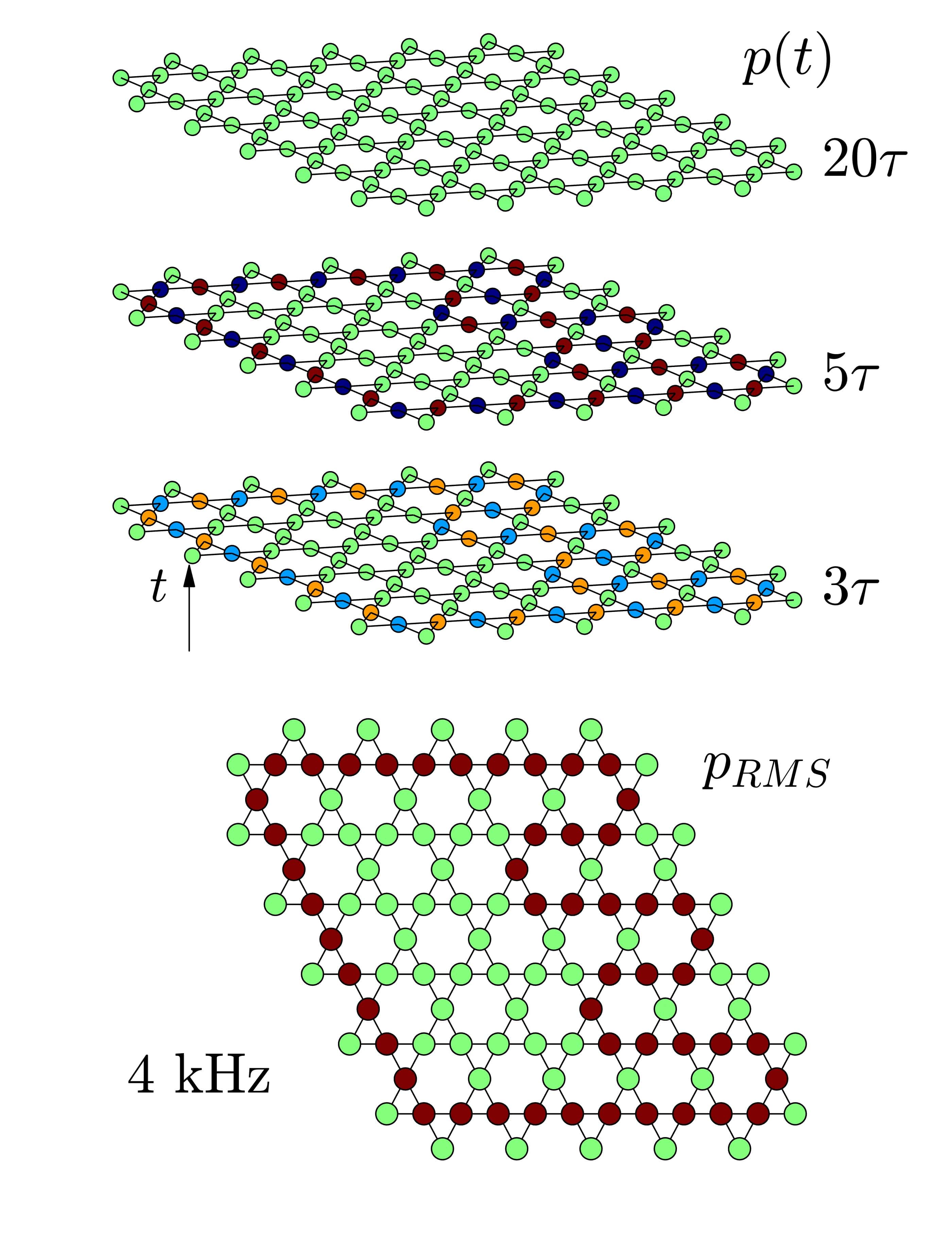}}
	\centering
	\caption{(a-d) Numerical results for the 2D lattice excited with a central frequency of $4kHz$.}
	\label{fig:S3}
\end{figure}

\begin{figure}[!h]
	\centering
	\hspace{-0.75cm}\subfigure[]{\includegraphics[width=0.27\textwidth]{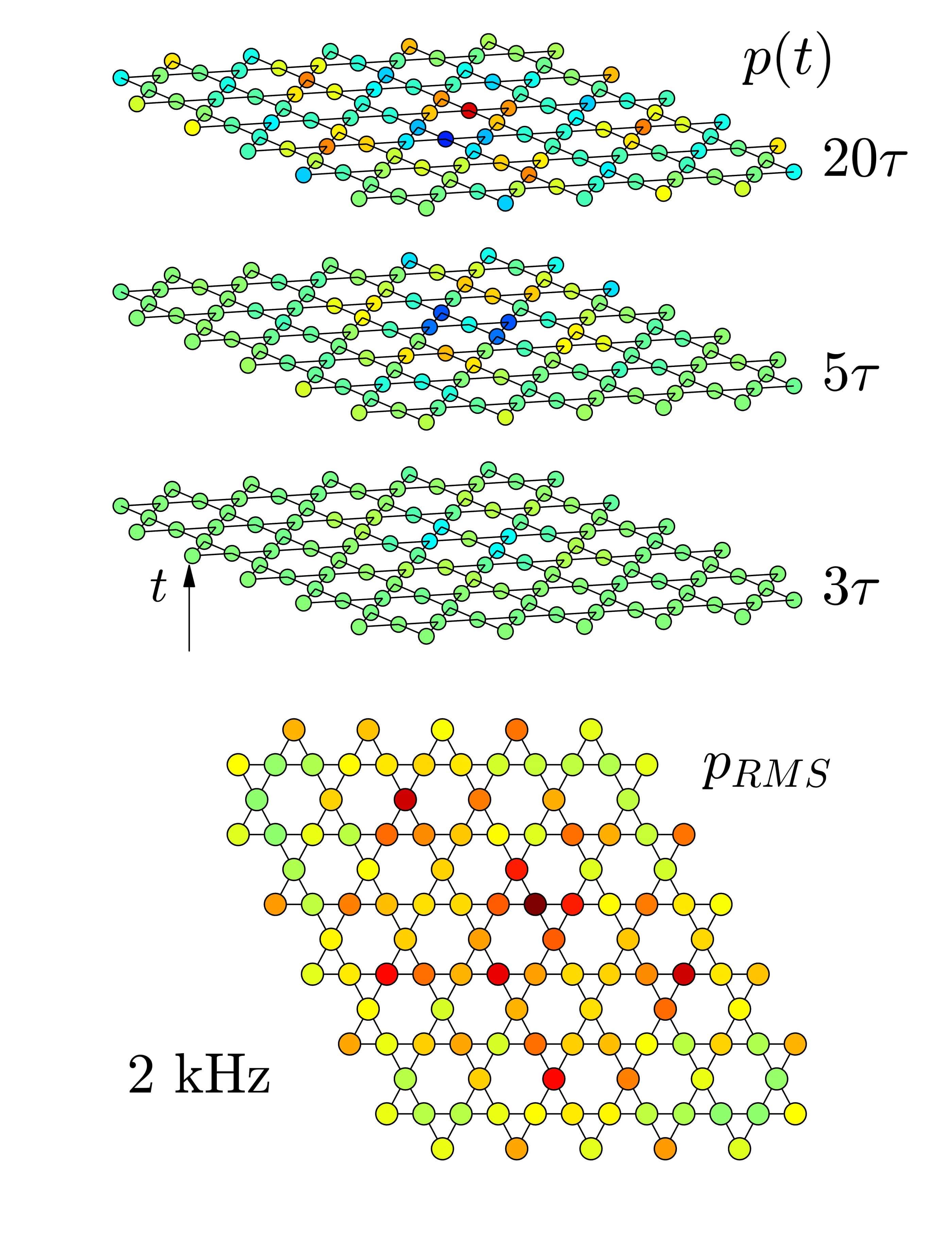}}\hspace{-0.2cm}
	\subfigure[]{\includegraphics[width=0.26\textwidth]{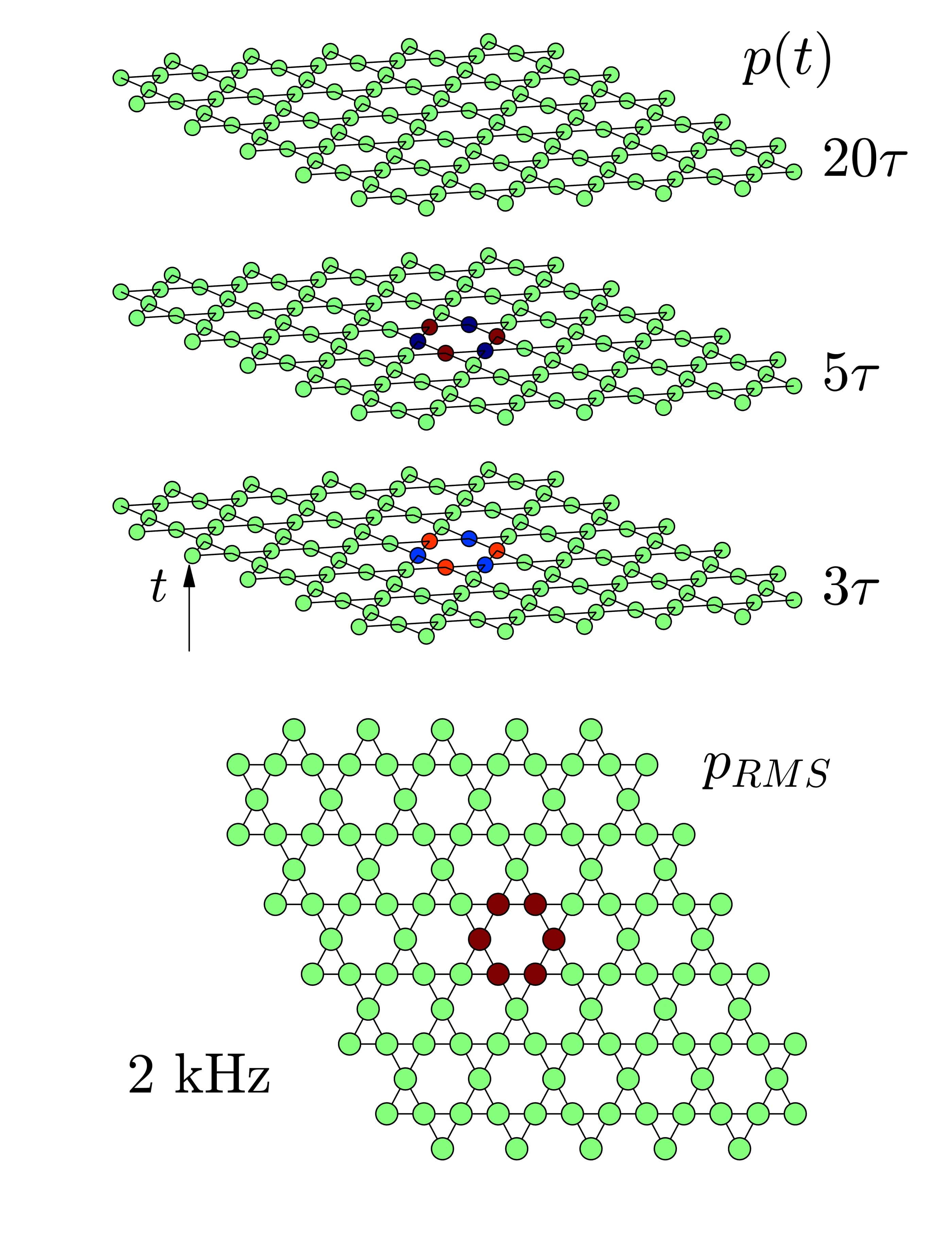}}\hspace{-0.2cm}
	\subfigure[]{\includegraphics[width=0.26\textwidth]{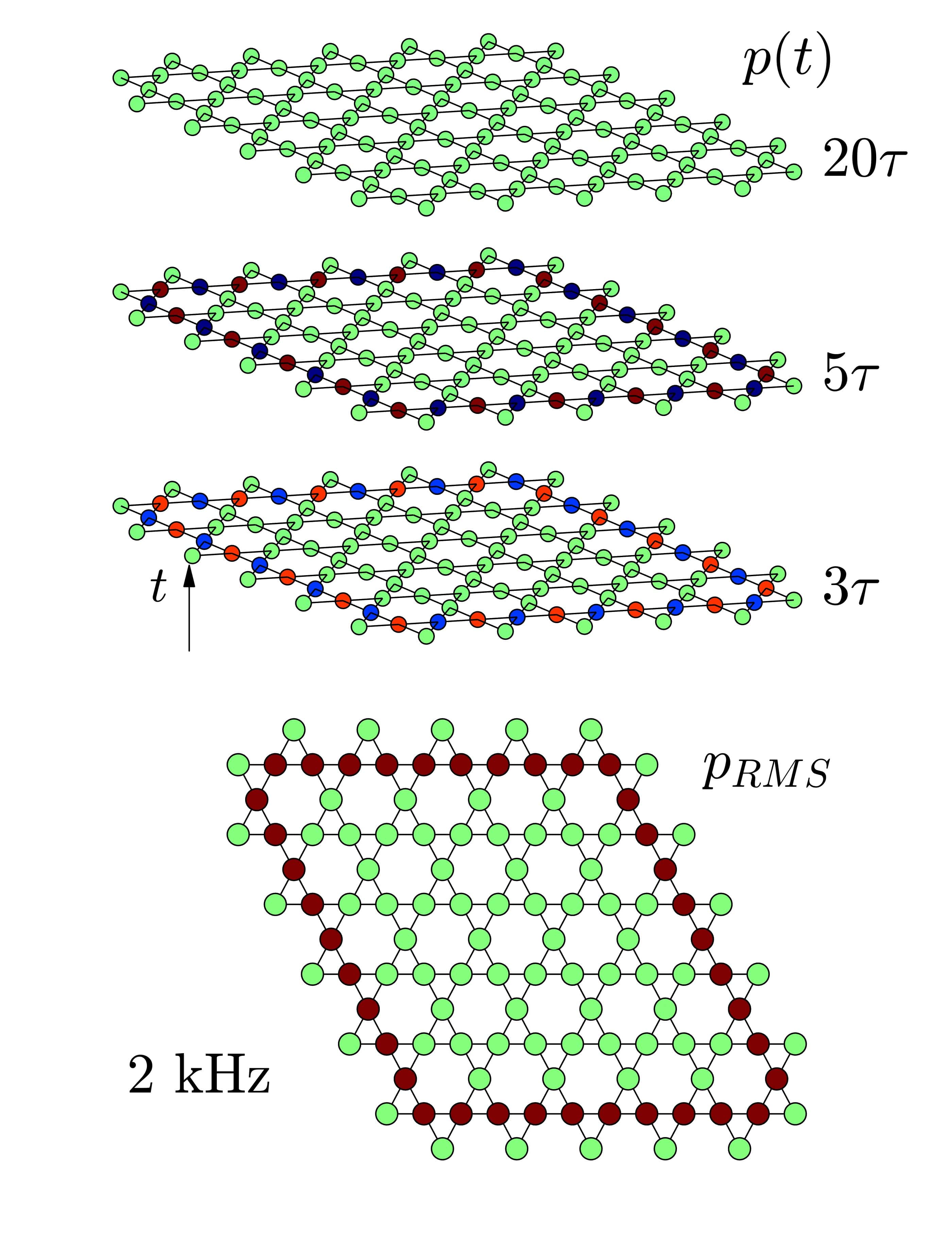}}\hspace{-0.2cm}
	\subfigure[]{\includegraphics[width=0.26\textwidth]{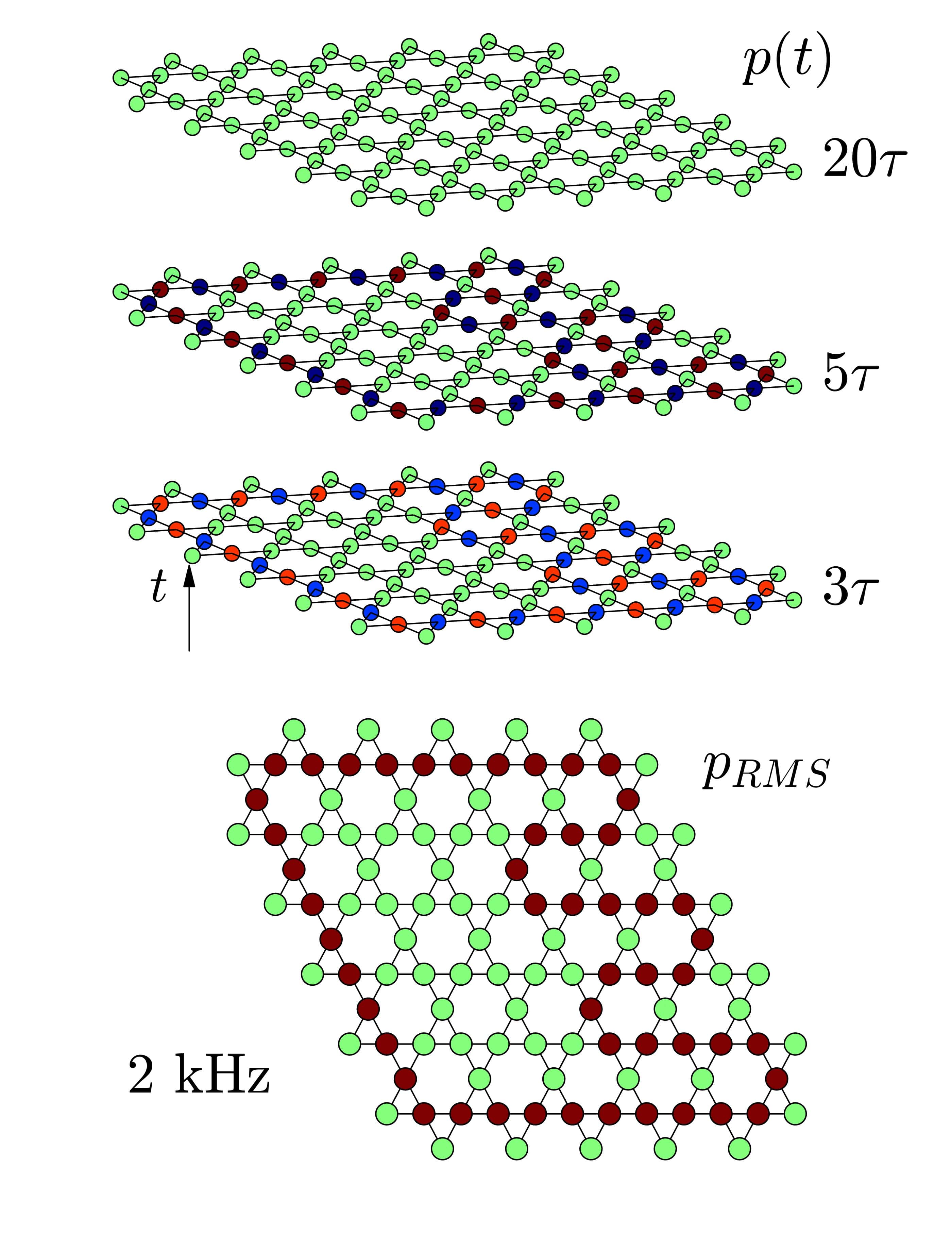}}
	\centering
	\caption{(a-d) Numerical results for the 2D lattice excited with a central frequency of $2kHz$.}
	\label{fig:S4}
\end{figure}

\begin{figure}[!t]
	\centering
	\hspace{-0.75cm}\subfigure[]{\includegraphics[width=0.27\textwidth]{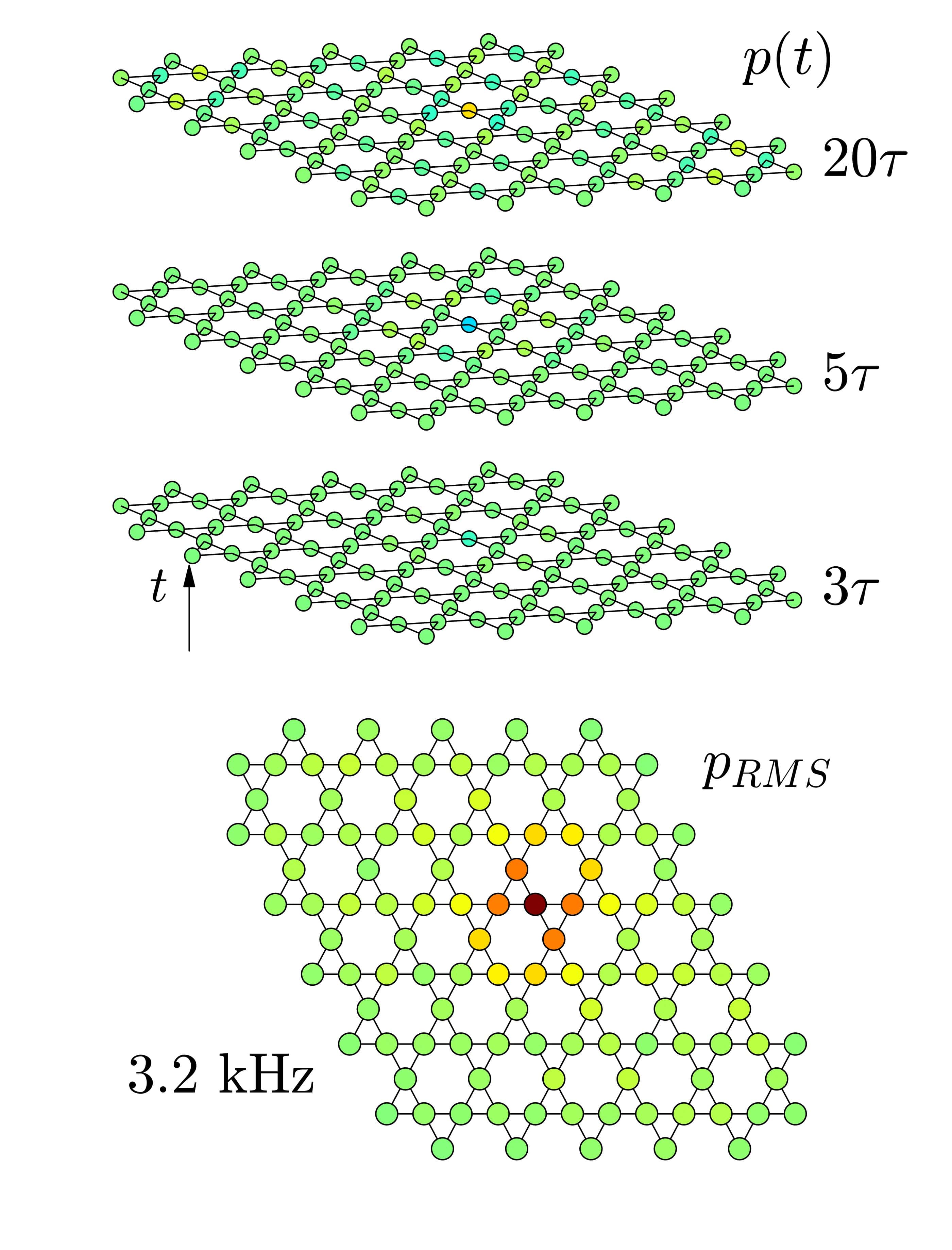}}\hspace{-0.2cm}
	\subfigure[]{\includegraphics[width=0.26\textwidth]{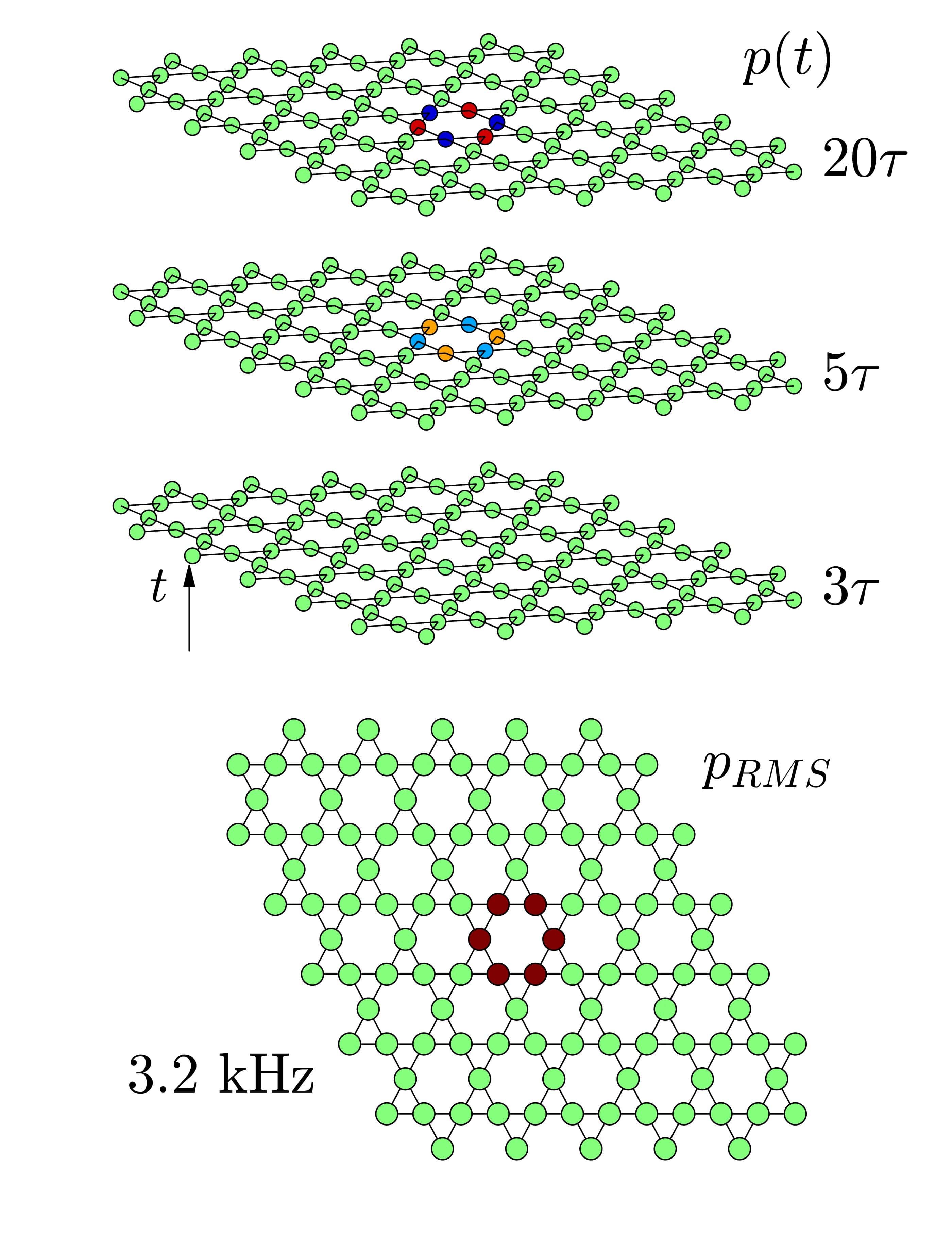}}\hspace{-0.2cm}
	\subfigure[]{\includegraphics[width=0.26\textwidth]{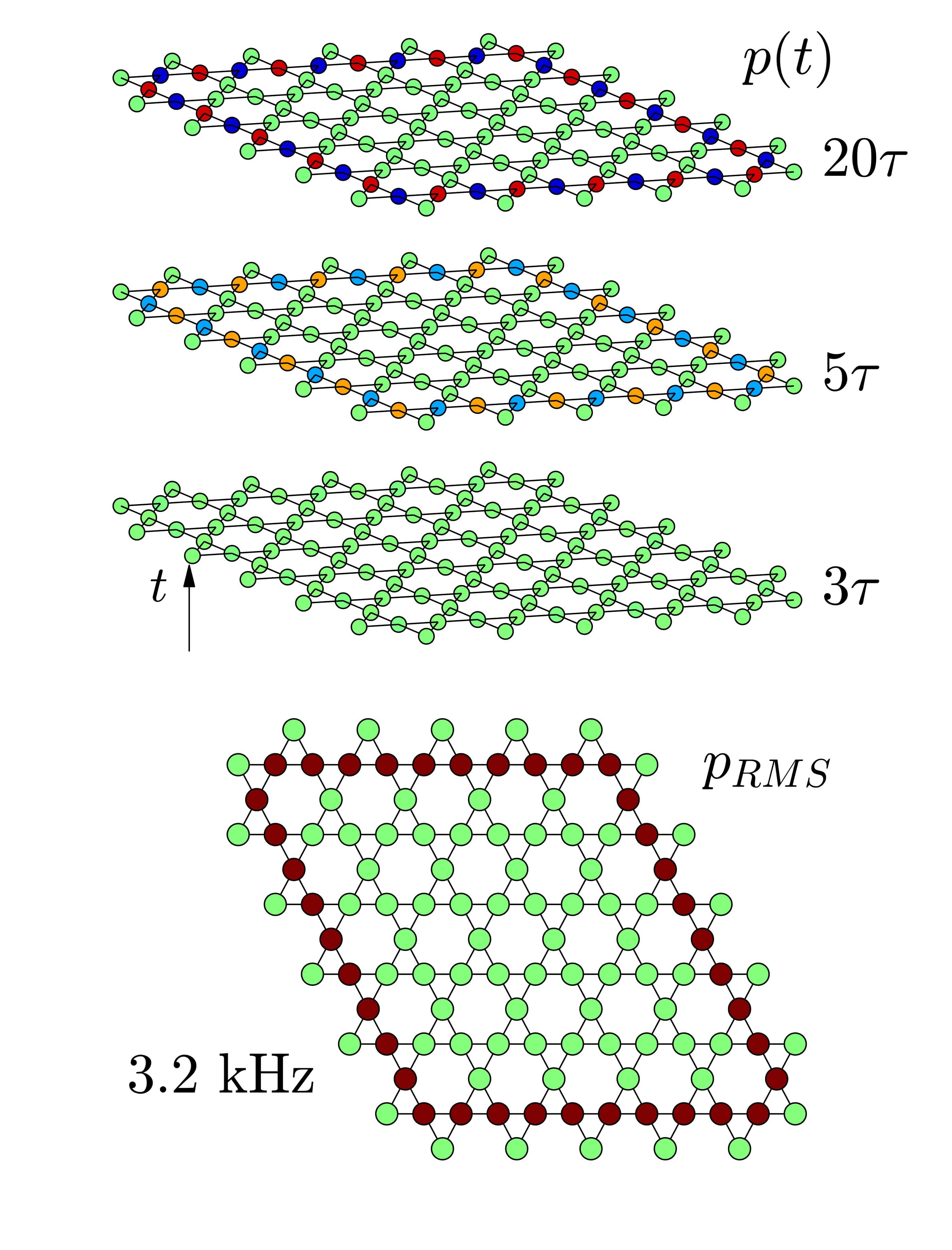}}\hspace{-0.2cm}
	\subfigure[]{\includegraphics[width=0.26\textwidth]{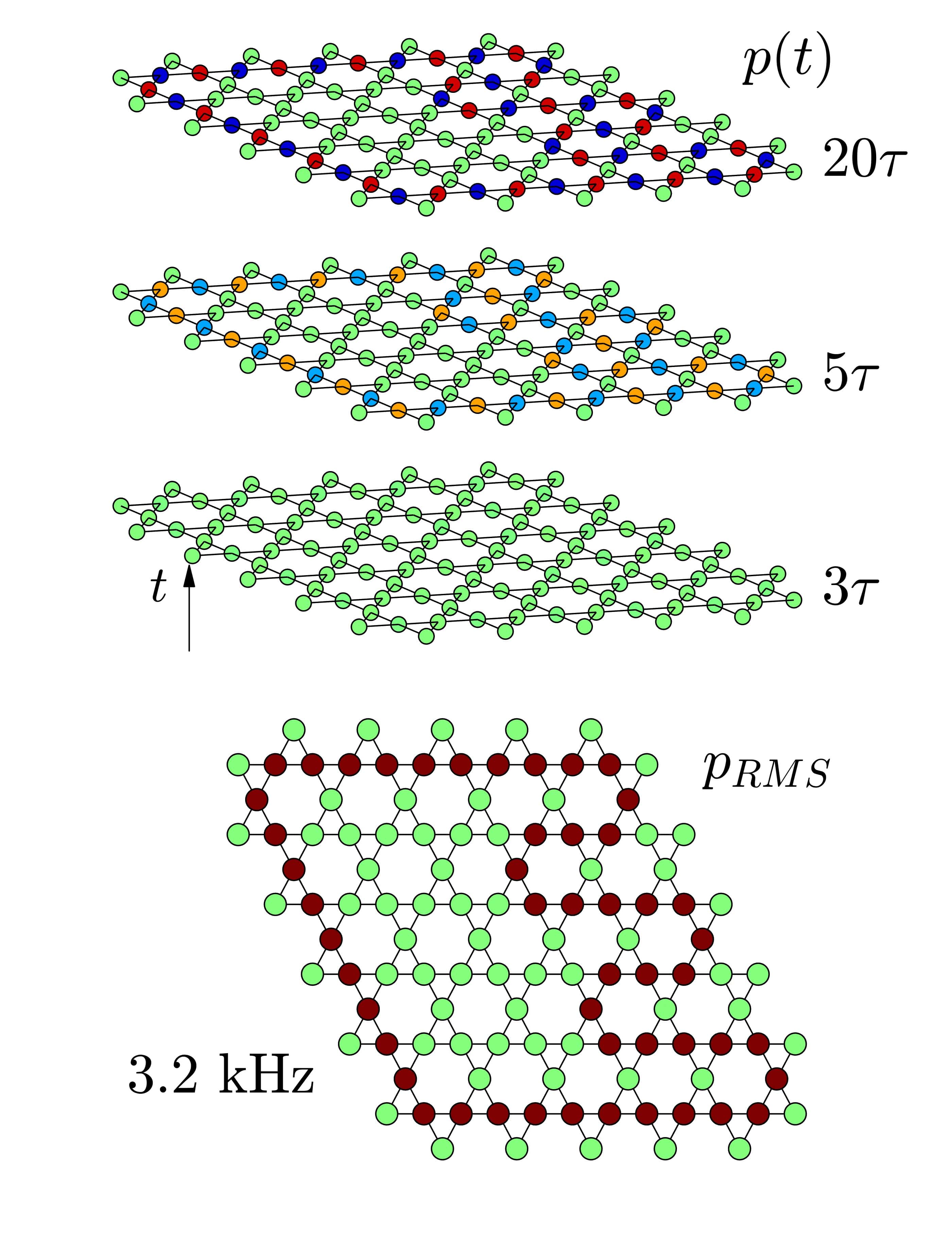}}
	\centering
	\caption{(a-d) Numerical results for the 2D lattice excited at flat band frequency.}
	\label{fig:S5}
\end{figure}

The operational regime of CLSs and ensuing boundary modes is hereafter discussed. In particular, Figs. \ref{fig:S3}-\ref{fig:S5} illustrate the time response for an excitation frequency above (Fig. \ref{fig:S3}), below (Fig. \ref{fig:S4}), and corresponding to the flat band (Fig. \ref{fig:S5}). To better illustrate these results, animations are also provided as separate files. 

We first focus on the point excitation. When the excitation frequency is above the flat band (Fig. \ref{fig:S3}(a)), the energy content is localized close to the excitation site. However, the degree of localization is dictated by the in-gap attenuation, and, as such, the response is not compact localized. In contrast, for excitation frequencies corresponding to the flat band and below (Fig. \ref{fig:S4}(a) and Fig. \ref{fig:S5}(a)), the energy leaks toward the neighboring sites and the response is spread all over the lattice. 

We now discuss the response to a CLS and boundary excitation. Similarly to the results shown in the main text, in the case of excitation frequency either above or below the flat band, the energy remains compact and localized (Fig. \ref{fig:S3}(b-d) and Fig. \ref{fig:S4}(b-d)). However, since the excitation is away from the flat region incident pressure wave is evanescent and, as such, decreases to zero as time elapses. In contrast, a flat band excitation (Fig. \ref{fig:S5}(b-d)) leads to CLSs and boundary modes that oscillate over time without any temporal decay. 

\begin{figure}[!t]
	\centering
	\hspace{-0.75cm}\subfigure[]{\includegraphics[width=0.27\textwidth]{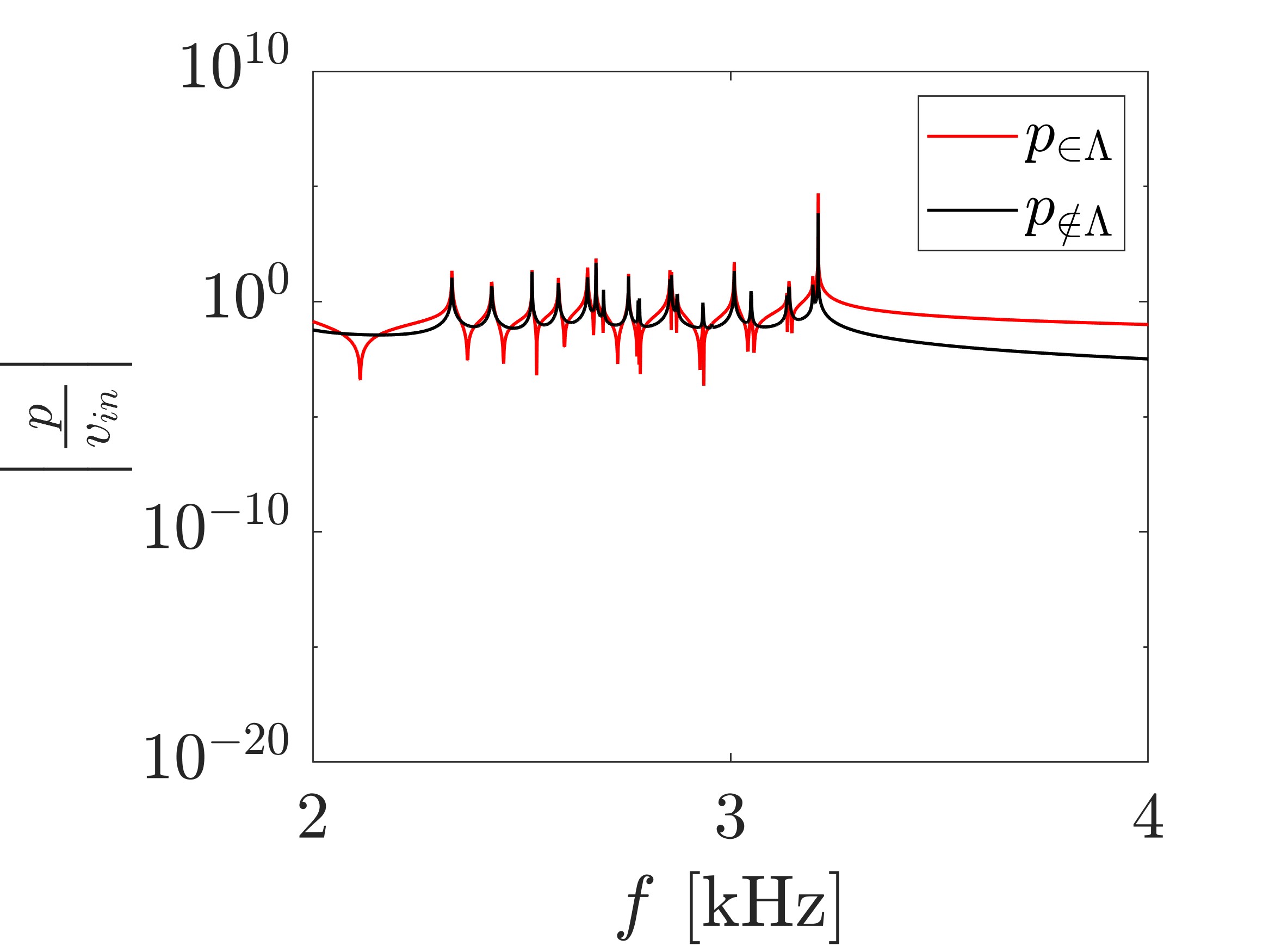}}\hspace{-0.2cm}
	\subfigure[]{\includegraphics[width=0.26\textwidth]{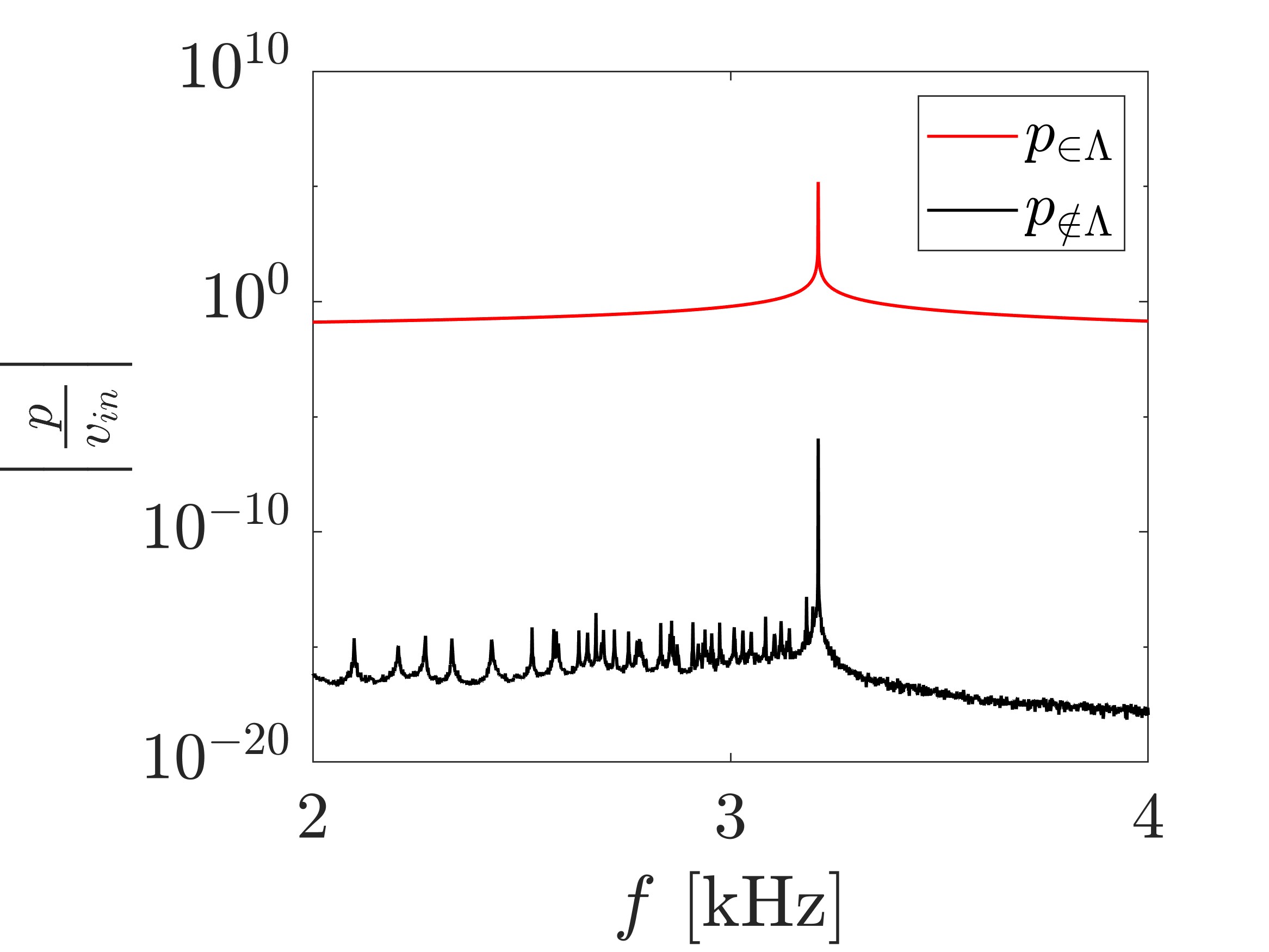}}\hspace{-0.2cm}
	\subfigure[]{\includegraphics[width=0.26\textwidth]{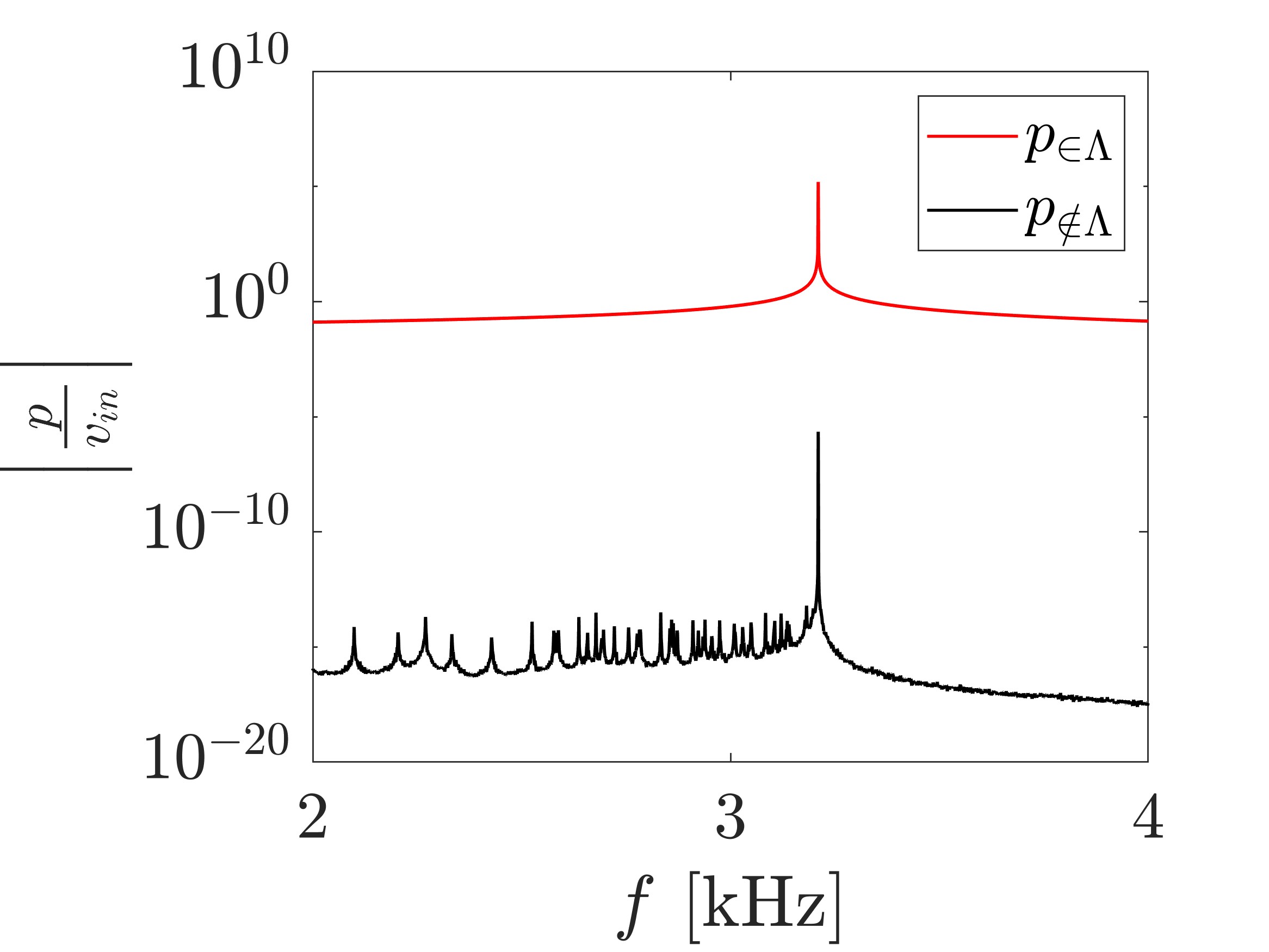}}\hspace{-0.2cm}
	\subfigure[]{\includegraphics[width=0.26\textwidth]{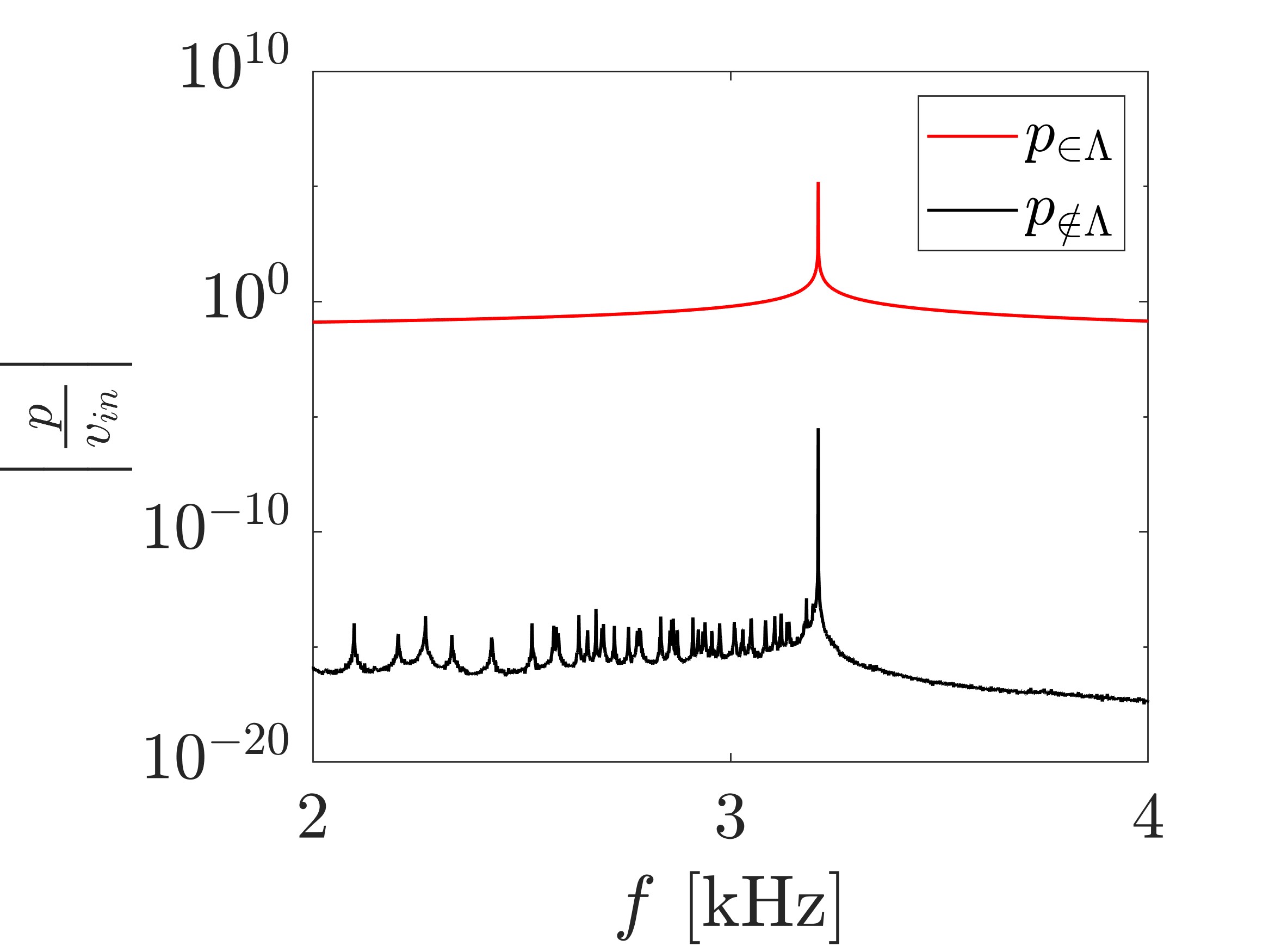}}
	\centering
	\caption{(a-d) Frequency response measured inside the forcing domain $\in\Lambda$ (red curve) and outside the forcing domain $\notin\Lambda$ (black curve) in the case of (a) point excitation, (b) CLS excitation, (c) boundary excitation, and (d) boundary excitation with the removal of two CLSs.}
	\label{fig:S6}
\end{figure}

This is also confirmed in Fig. \ref{fig:S6}, where the frequency response function due to point excitation (Fig. \ref{fig:S6}(a)), CLS excitation (Fig. \ref{fig:S6}(b)), boundary excitation (Fig. \ref{fig:S6}(c)), and boundary excitation with removal of CLSs (Fig. \ref{fig:S6}(d)) is shown. Specifically, the red curve is representative of the RMS response at the excitation sites, while the black curve is representative of the RMS response in all remaining sites. It is straightforward to conclude that, except for very small numerical errors in the order of $\epsilon\approx 10^{-15}$, and in the case of CLS and boundary excitation, the response is compact and localized at the excitation sites. In addition, there is a strong amplification at flat band frequencies, which is dictated by the CLSs excitation and combination of the eigenstates populating the flat dispersion. In contrast, neighboring frequencies are selectively filtered by the excitation mechanism.

\newpage
\section*{Supplementary Note 4: additional numerical results for the 3D lattice}
A similar behavior is observed for the 3D lattice configuration. When extending the dynamics to the additional spatial dimension $z$, the flat-band states propagate across a continuum of frequency-wavenumber pairs $\omega-\kappa_z$ for frequencies above the in-plane flat-band frequency. 

\begin{figure}[!h]
	\centering
	\hspace{-0.75cm}\subfigure[]{\includegraphics[width=0.27\textwidth]{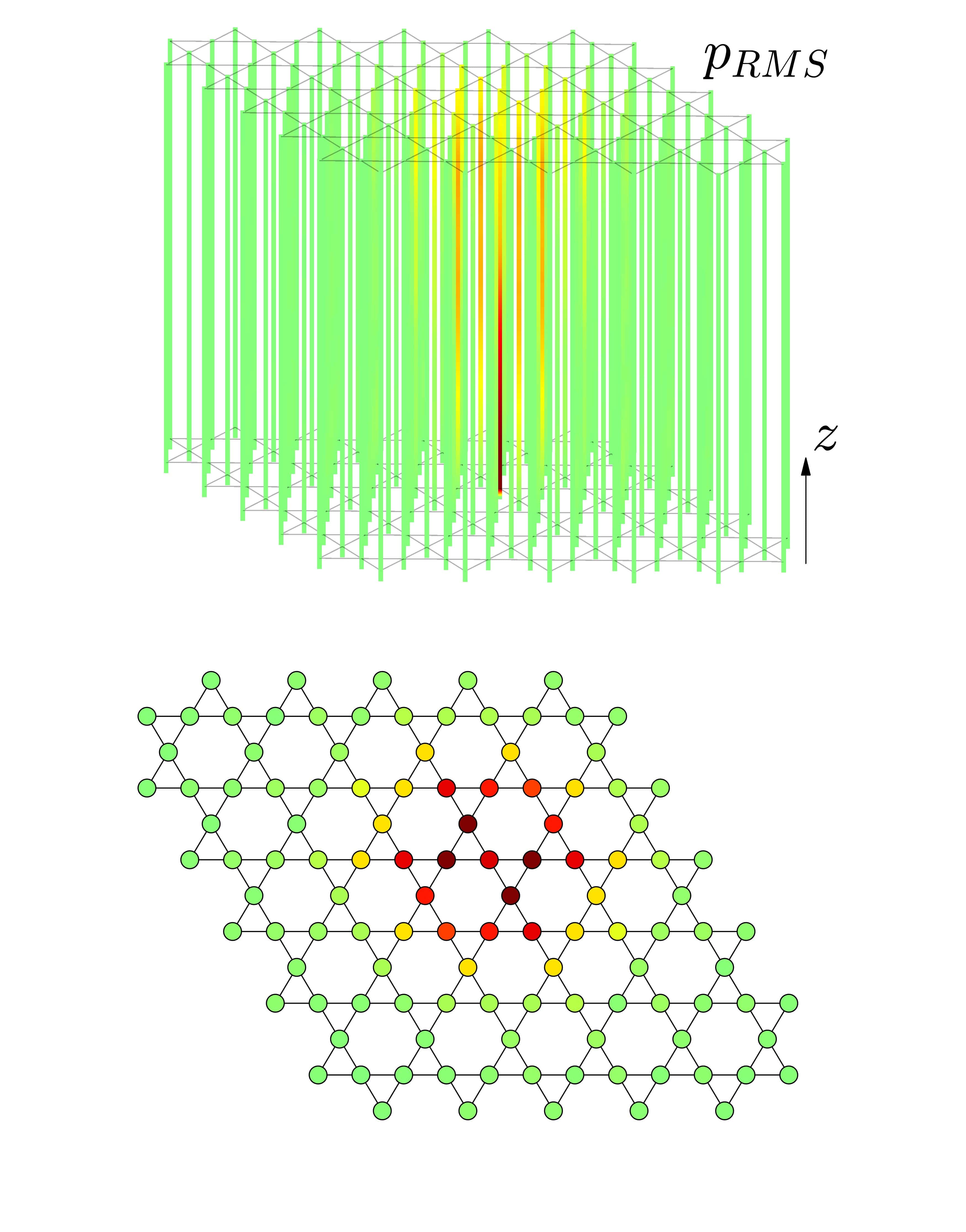}}\hspace{-0.2cm}
	\subfigure[]{\includegraphics[width=0.26\textwidth]{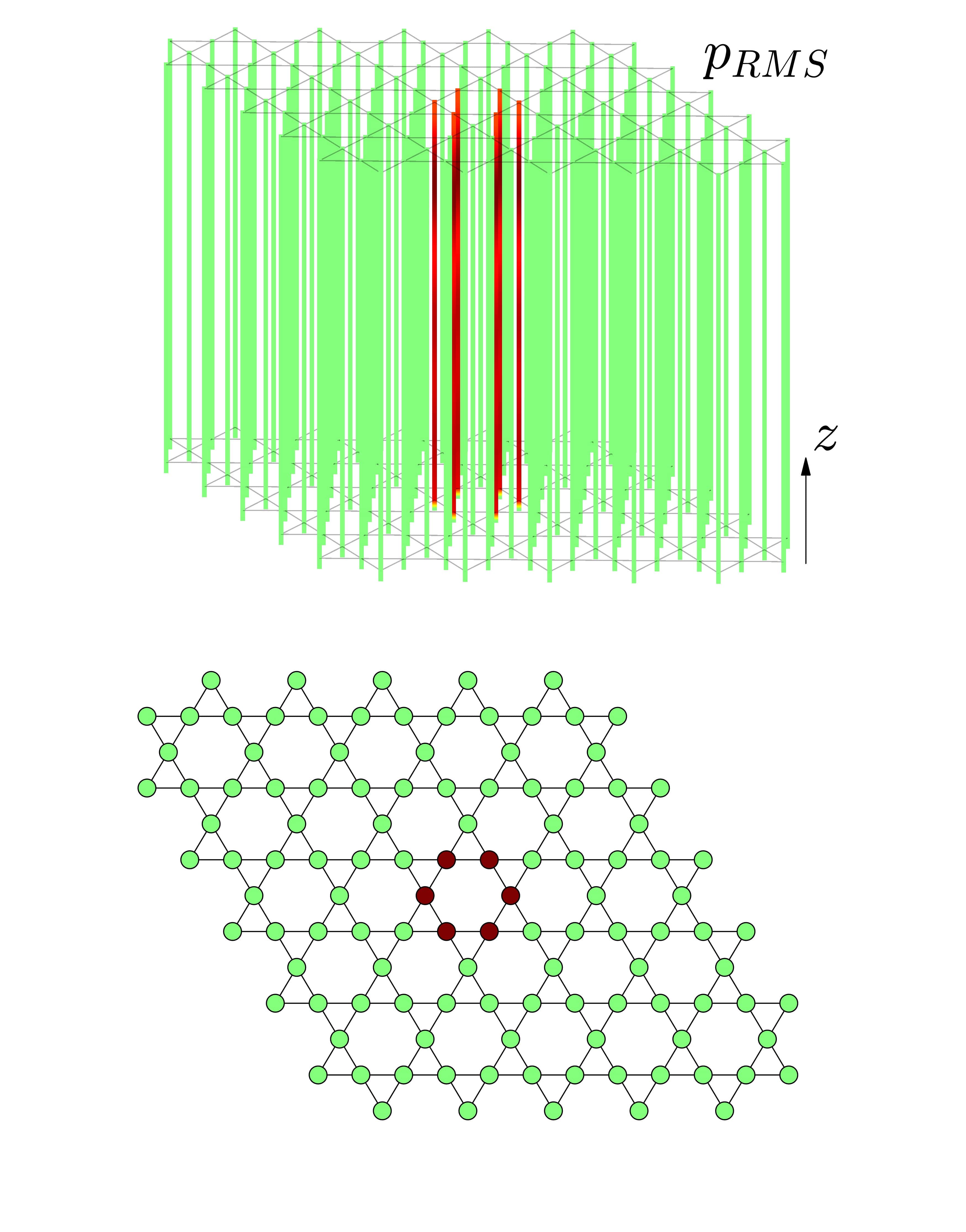}}\hspace{-0.2cm}
	\subfigure[]{\includegraphics[width=0.26\textwidth]{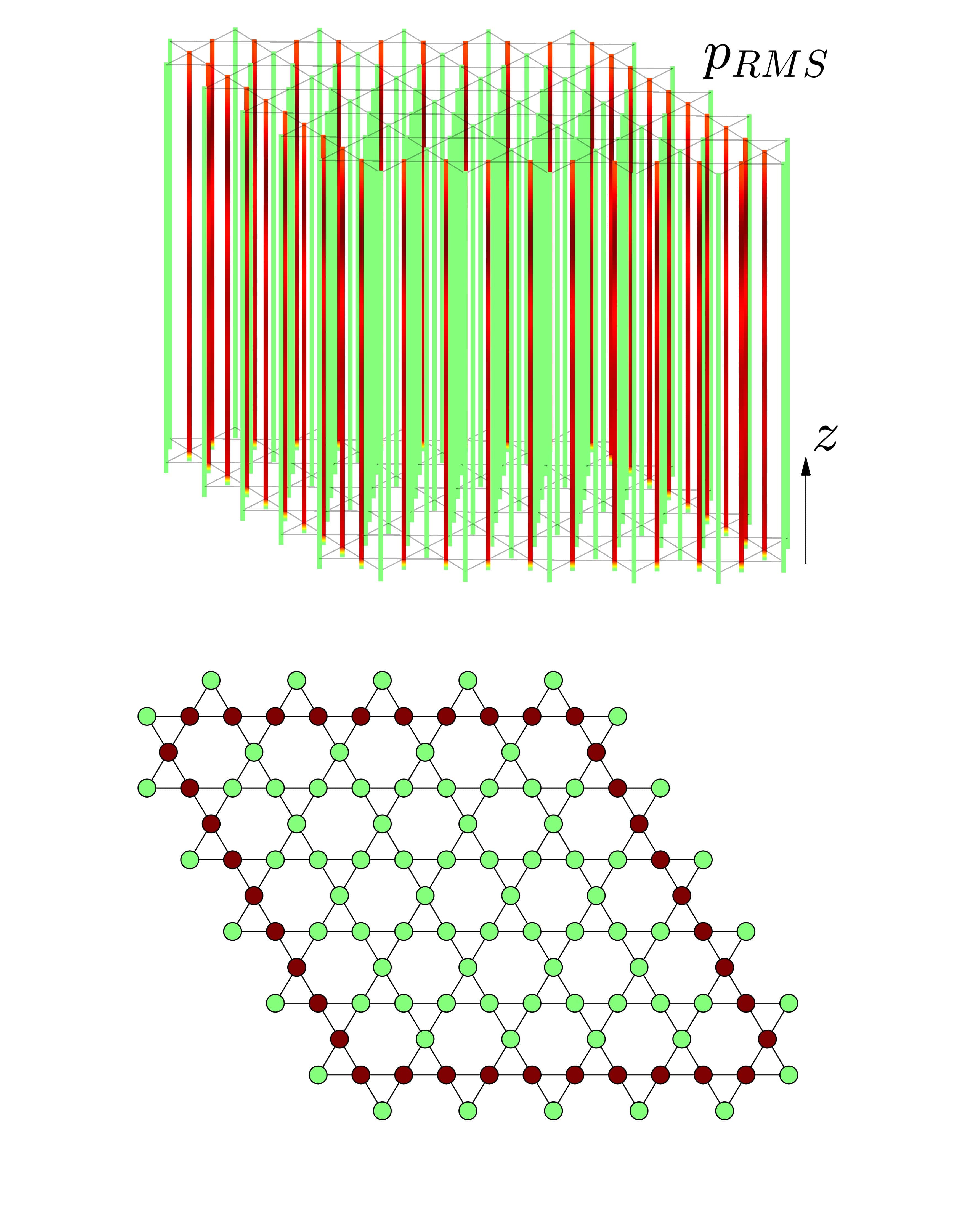}}\hspace{-0.2cm}
	\subfigure[]{\includegraphics[width=0.26\textwidth]{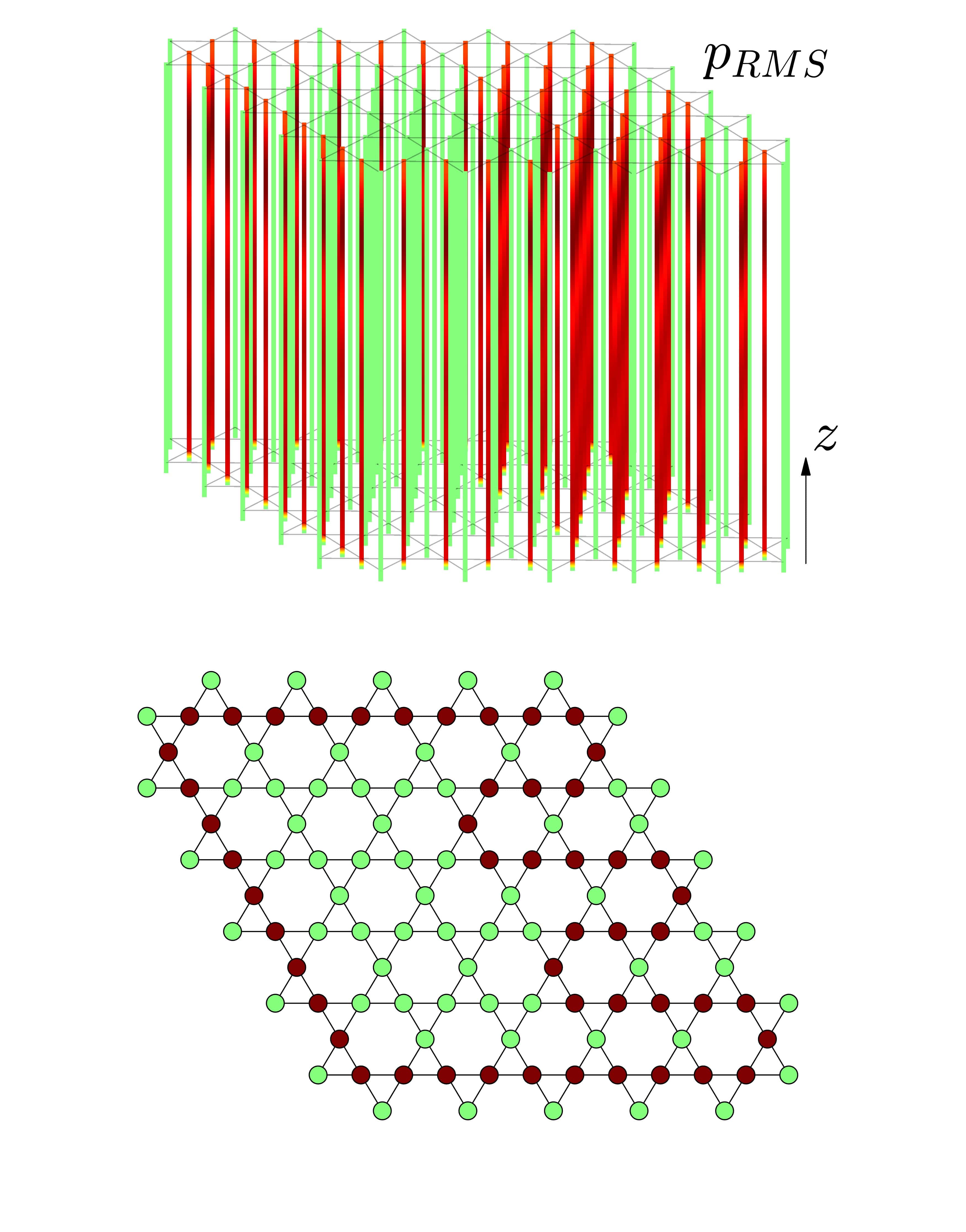}}
	\centering
	\caption{(a-d) Numerical results for the 3D lattice excited above the flat band frequency ($4$ kHz).}
	\label{fig:S7}
\end{figure}

\begin{figure}[!h]
	\centering
	\hspace{-0.75cm}\subfigure[]{\includegraphics[width=0.27\textwidth]{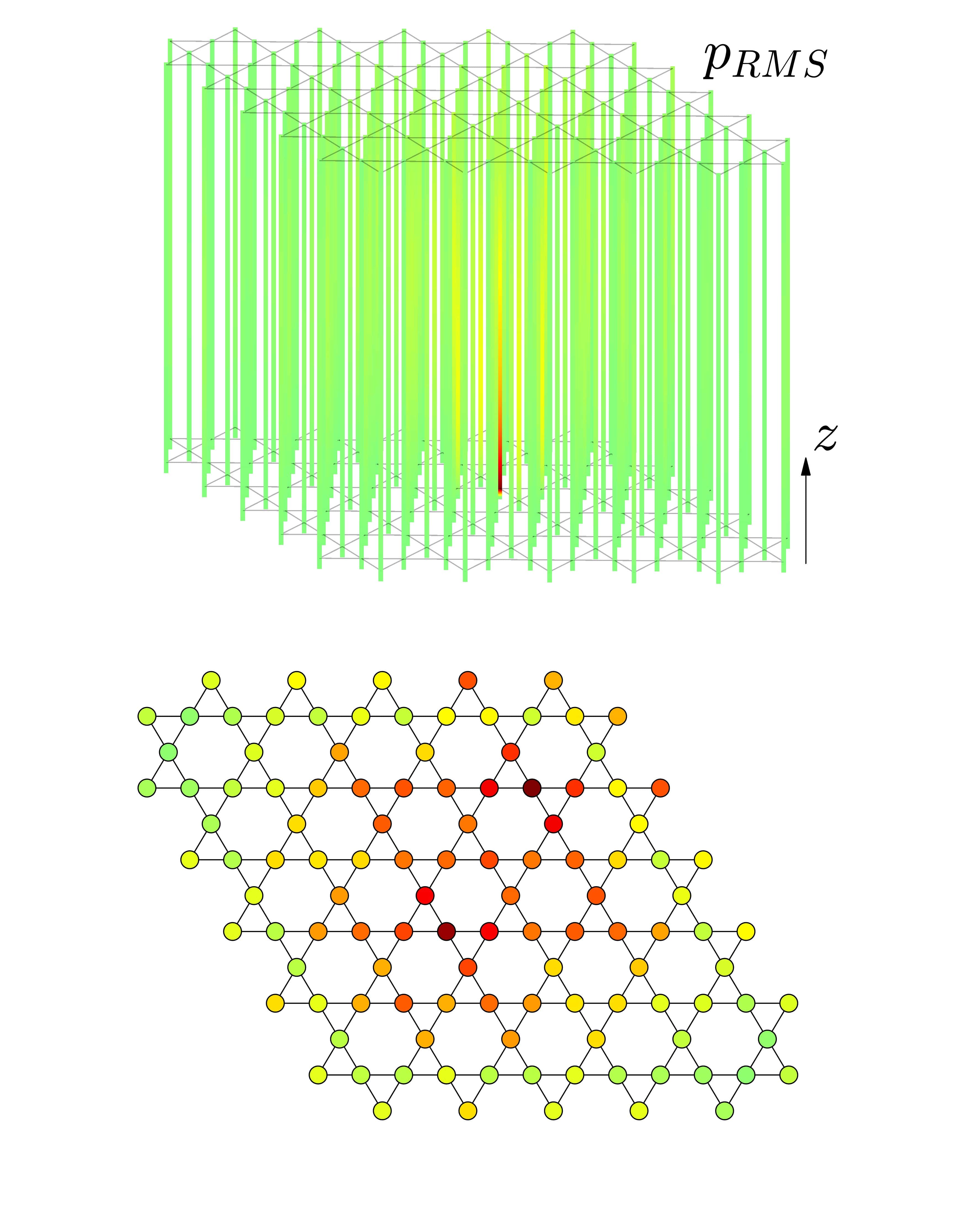}}\hspace{-0.2cm}
	\subfigure[]{\includegraphics[width=0.26\textwidth]{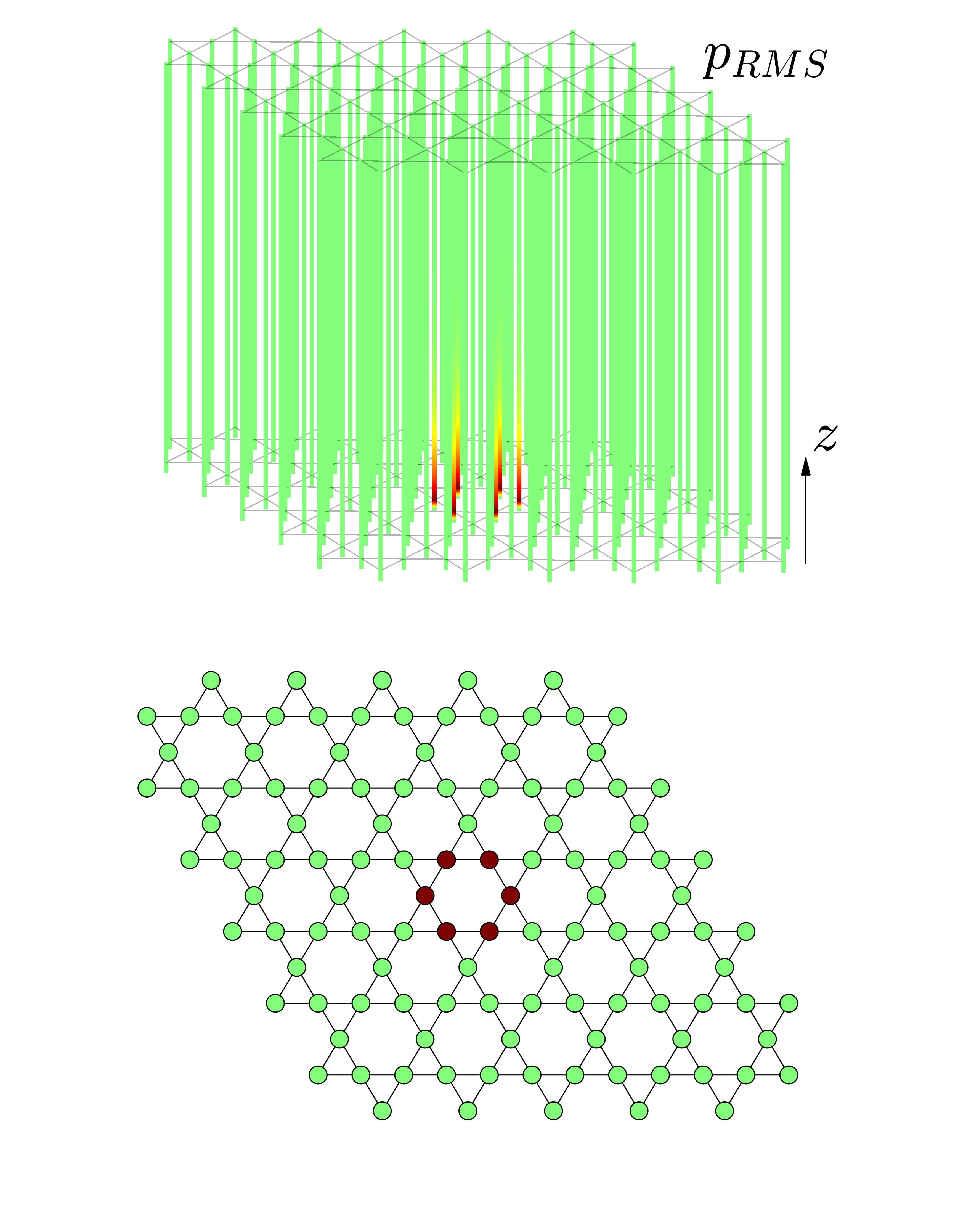}}\hspace{-0.2cm}
	\subfigure[]{\includegraphics[width=0.26\textwidth]{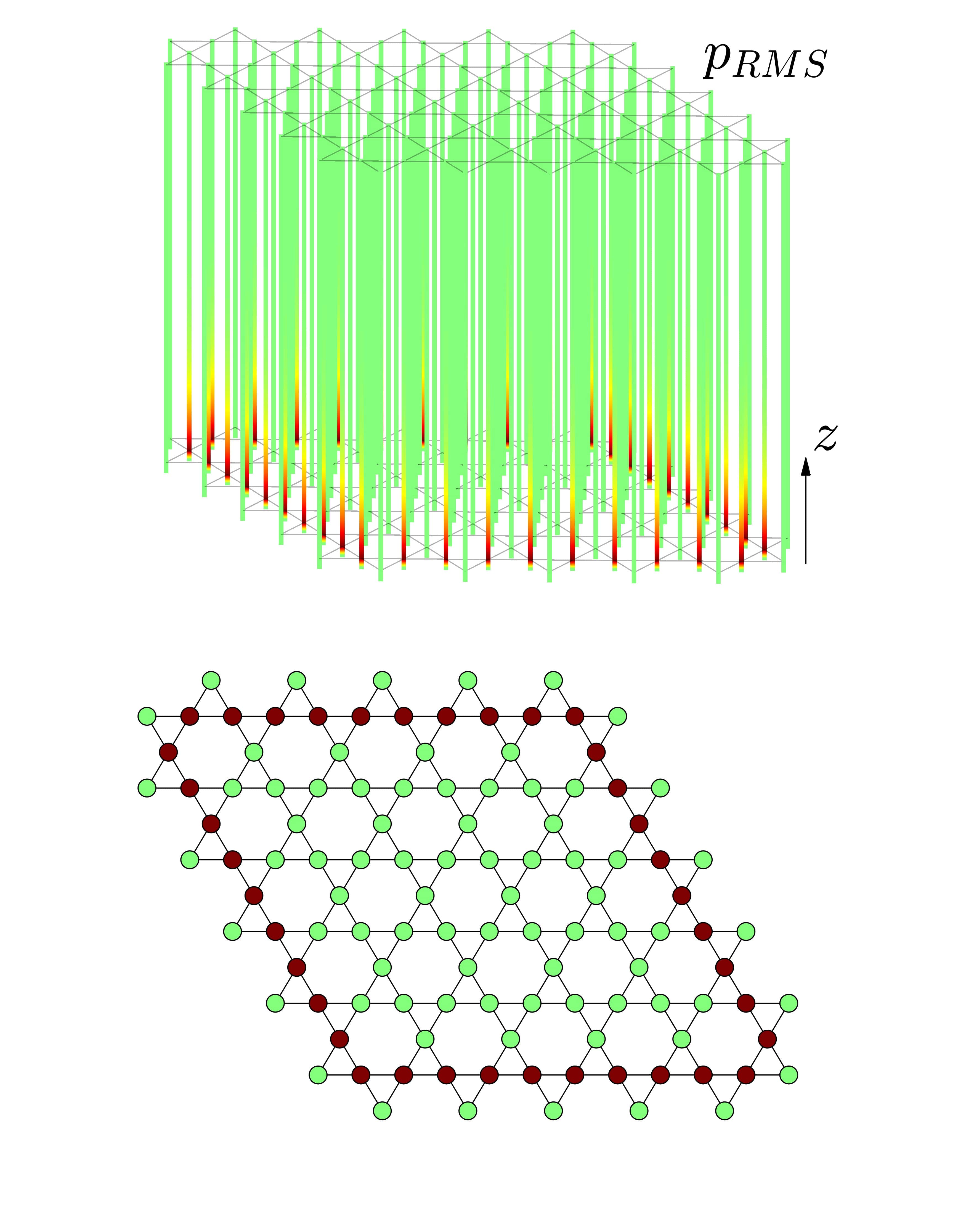}}\hspace{-0.2cm}
	\subfigure[]{\includegraphics[width=0.26\textwidth]{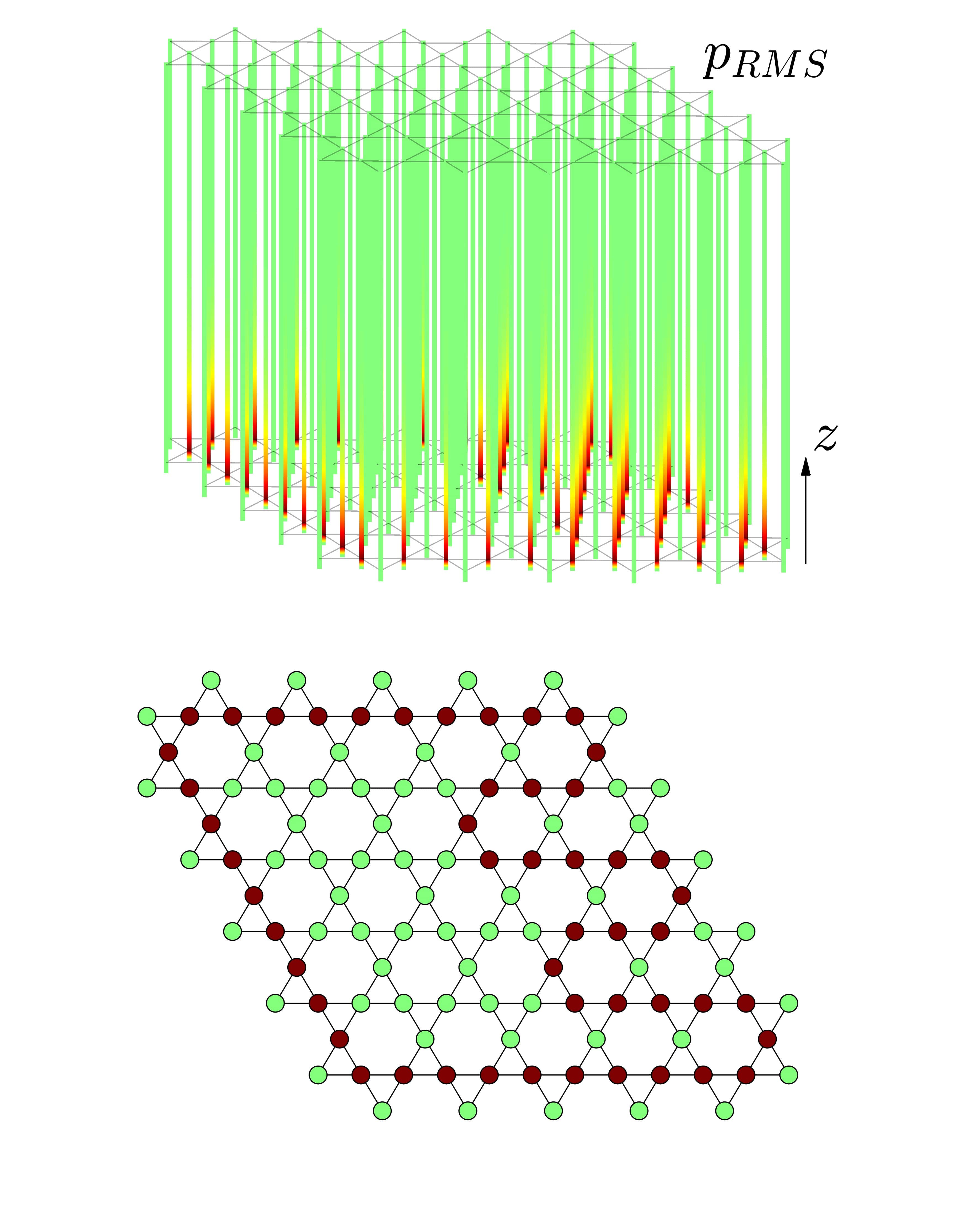}}
	\centering
	\caption{(a-d) Numerical results for the 3D lattice excited below the flat band frequency ($2$ kHz).}
	\label{fig:S8}
\end{figure}

Additional results are shown for excitation frequencies above (Fig. \ref{fig:S7}) and below (Fig. \ref{fig:S8}) this cutoff frequency. Below the cutoff, propagation occurs with spatial attenuation while retaining compact localization.
In contrast, CLSs and boundary modes at frequencies above the flat band can propagate freely. Animations are also provided as separate files. 

\newpage
\section*{Supplementary Note 5: COMSOL model and numerical validation}
All results derived from the simplified model are validated through high-fidelity simulations in a COMSOL Multiphysics environment. These COMSOL models incorporate an infinitely rigid frame to capture the large impedance mismatch between air and the solid lattice material, ensuring an accurate simulation of wave confinement within the lattice. The mesh is composed of quadratic (Lagrange) tetrahedral elements with a maximum size of 1 mm, carefully selected to balance computational efficiency with precision. For excitation, a prescribed velocity field with a 10-period broadband signal is applied at specific lattice sites, matching the experimental configuration.

\begin{figure}[!h]
	\centering
	\hspace{-0.75cm}\subfigure[]{\includegraphics[width=0.24\textwidth]{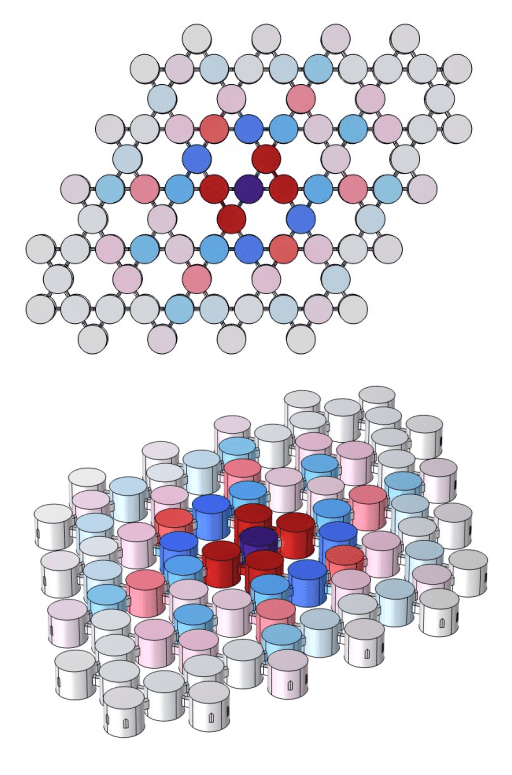}}\hspace{-0.2cm}
	\subfigure[]{\includegraphics[width=0.24\textwidth]{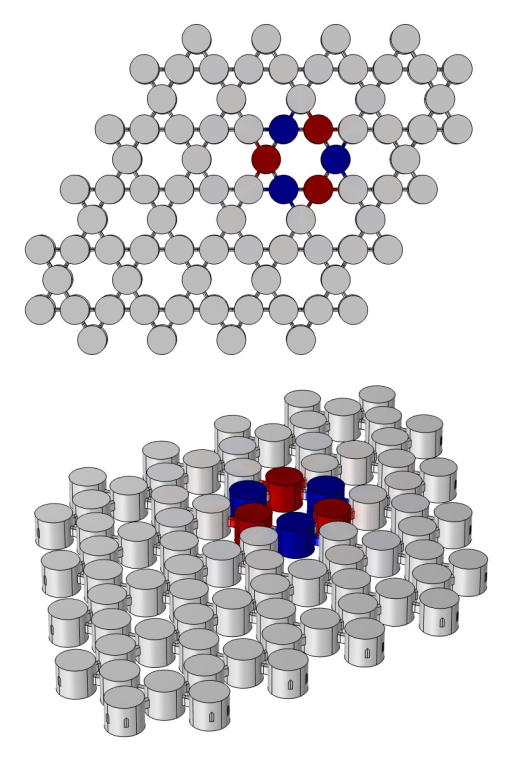}}\hspace{-0.2cm}
	\subfigure[]{\includegraphics[width=0.24\textwidth]{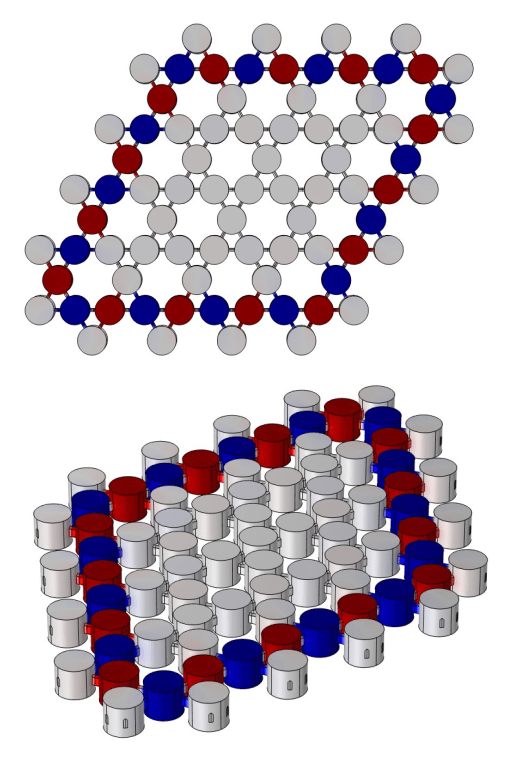}}\hspace{-0.2cm}
	\subfigure[]{\includegraphics[width=0.24\textwidth]{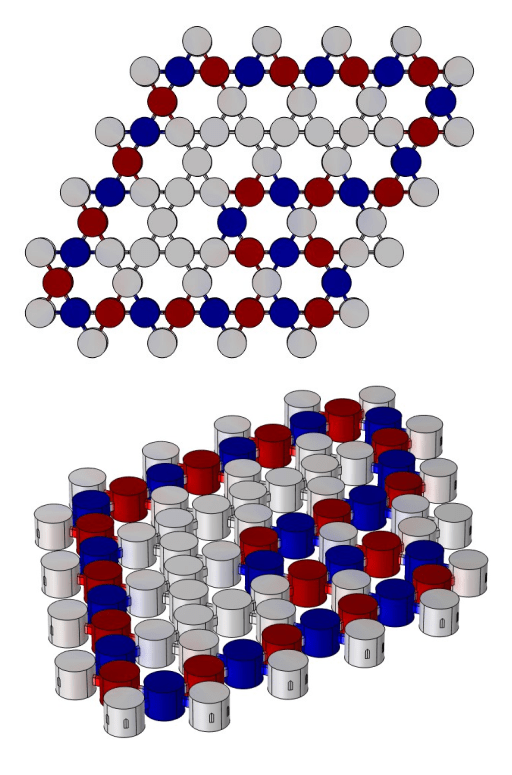}}
	\centering
	\caption{(a-d) Numerical results for the 2D lattice excited at the flat band frequency ($3.2$ kHz).}
	\label{fig:S9}
\end{figure}

\begin{figure}[!h]
	\centering
	\hspace{-0.75cm}\subfigure[]{\includegraphics[width=0.24\textwidth]{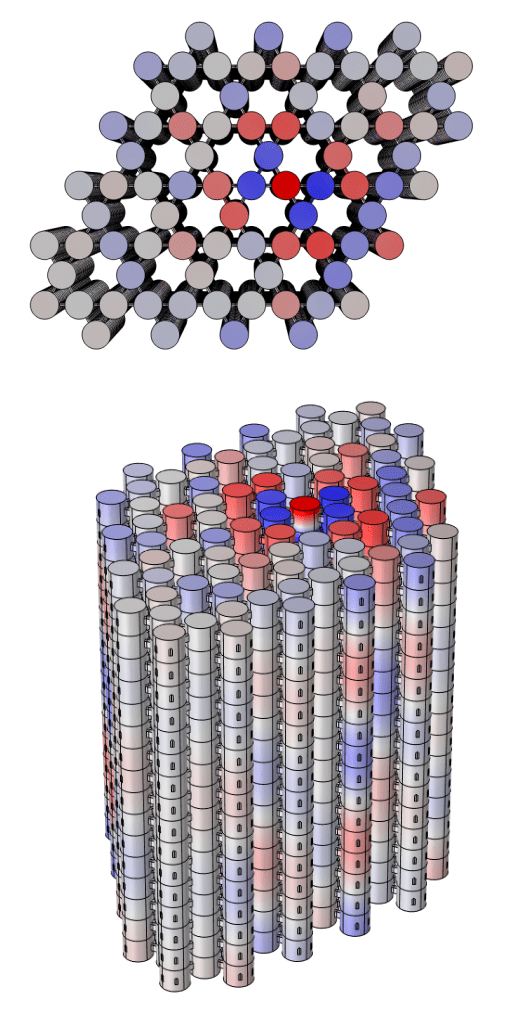}}\hspace{-0.2cm}
	\subfigure[]{\includegraphics[width=0.24\textwidth]{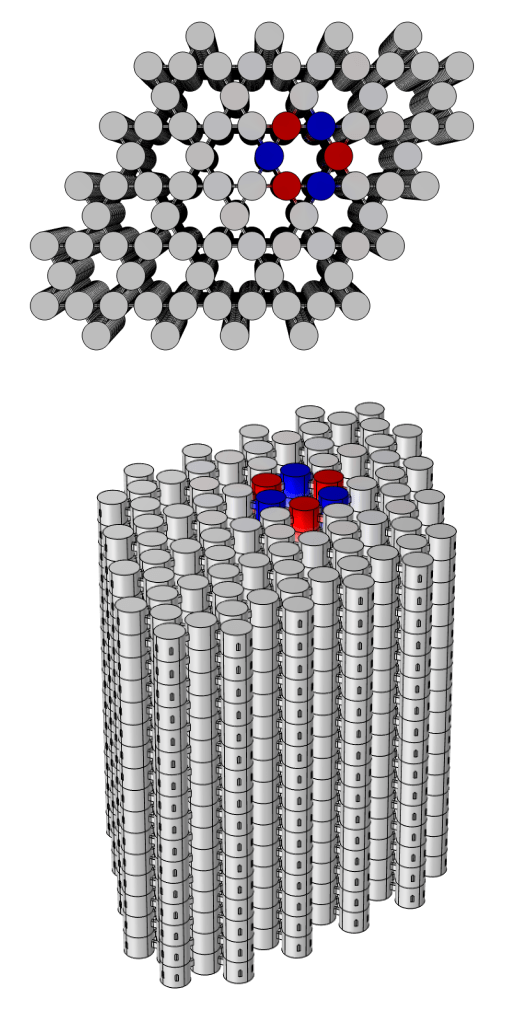}}\hspace{-0.2cm}
	\subfigure[]{\includegraphics[width=0.24\textwidth]{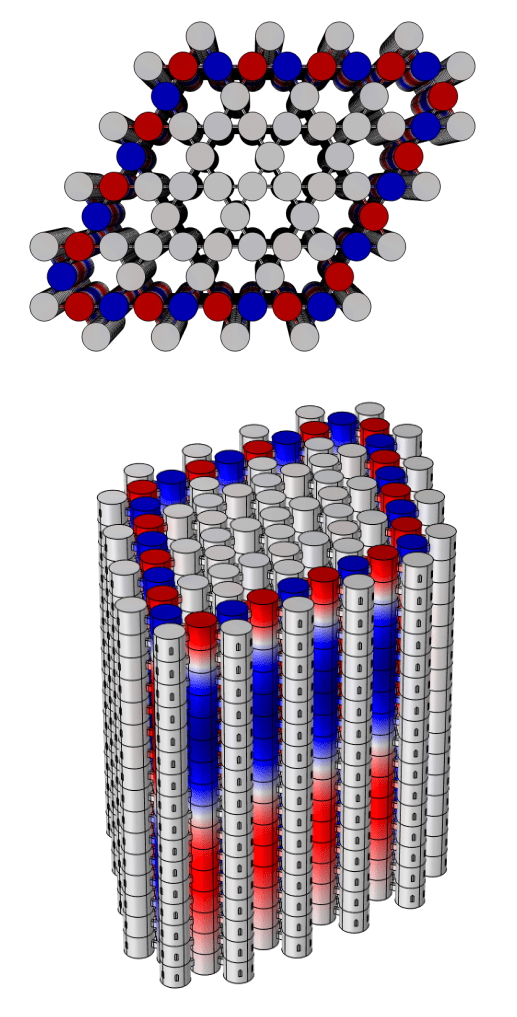}}\hspace{-0.2cm}
	\subfigure[]{\includegraphics[width=0.24\textwidth]{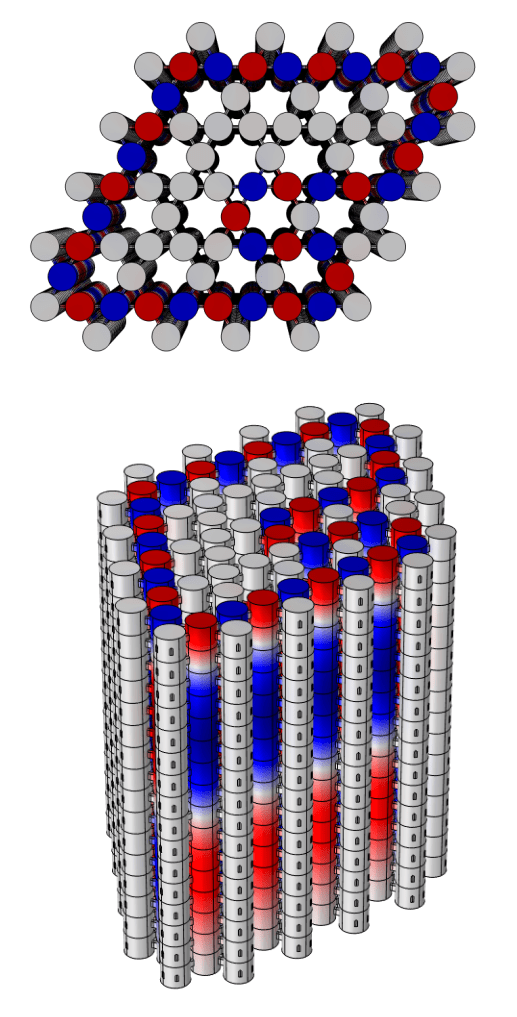}}
	\centering
	\caption{(a-d) Numerical results for the 3D lattice excited at the flat band frequency ($4$ kHz).}
	\label{fig:S10}
\end{figure}

To illustrate the model’s accuracy, two representative case studies are presented, probing the 2D and 3D lattices at frequencies of 3.2 kHz and 4 kHz, respectively, as shown in Fig. \ref{fig:S9} and \ref{fig:S10}. Corresponding animations that highlight the dynamic response in each case are provided as supplementary files.



\bibliography{apssamp}

\end{document}